# A FREE ENERGY PRINCIPLE FOR A PARTICULAR PHYSICS

Karl Friston

*The Wellcome Centre for Human Neuroimaging, UCL Queen Square Institute of Neurology, London, UK WC1N 3AR. Email: k.friston@ucl.ac.uk*

## Abstract

This monograph attempts a theory of every 'thing' that can be distinguished from other 'things' in a statistical sense. The ensuing statistical independencies, mediated by Markov blankets, speak to a recursive composition of ensembles (of things) at increasingly higher spatiotemporal scales. This decomposition provides a description of small things; e.g., quantum mechanics – via the Schrödinger equation, ensembles of small things – via statistical mechanics and related fluctuation theorems, through to big things – via classical mechanics. These descriptions are complemented with a Bayesian mechanics for autonomous or active things. Although this work provides a formulation of every 'thing', its main contribution is to examine the implications of Markov blankets for self-organisation to nonequilibrium steady-state. In brief, we recover an information geometry and accompanying free energy principle that allows one to interpret the internal states of something as representing or making inferences *about* its external states. The ensuing Bayesian mechanics is compatible with quantum, statistical and classical mechanics and may offer a formal description of lifelike particles.

## Contents

# Introduction

This monograph attempts a theory of every 'thing' – in a tongue in cheek way – starting from the premise that a 'thing' is distinguishable from something else and from no 'thing'. Its ambition is to validate a formulation of dynamical systems by appealing to constructs in physics (e.g., quantum, statistical and classical mechanics) and then use the ensuing formulation to derive an account of self-organisation within the same framework[1]. Our starting point is a definition of things in terms of systems that possess an invariant measure; namely, weakly mixing systems that possess an attracting set. The description of such systems usually starts using the formalism of random dynamical systems; for example, the flow or dynamics of systemic states based on random differential equations (e.g., a Langevin equation). This is where the current treatment starts – and then stops. It stops by asking some obvious questions; like, what are states and where do random fluctuations come from? These questions lead to even simpler questions; namely, if we are dealing with the states of something, what is the thing that possesses those states – and how does one distinguish anything from something else? The answers to these questions lead to a theory of everything in a literal sense.

To address the nature of things, we start by asking how something can be distinguished from everything else. In pursuing a formulation of self-organisation, we will call on the notion of conditional independence as the basis of this separation. More specifically, we assume that for something to exist it must possess (*internal* or intrinsic) states that can be separated statistically from (*external* or extrinsic) states that do not constitute the thing. This separation implies the existence of a Markov blanket; namely, a set of states that render the internal and external states conditionally independent. The existence of things (i.e., internal states and their blanket) further implies a partition of the Markov blanket into *active* and *sensory* states – that are not influenced by external and internal states, respectively. This may sound a bit arbitrary; however, this is the minimal set of conditional independencies – and implicit partition of states – that licenses talk about things (that possess states). Specifically, it provides a partition that constitutes the 'self' in self-organisation. The subsequent sections tackle the next obvious question: what are things? At this point, we deploy the Langevin formulation of random dynamical systems as an *ansatz* that is recursively self-verifying, when considered in the light of Markov blankets. In brief, the formulation on offer says that the states of things (i.e., particles) comprise mixtures of blanket states, where the Markov blanket surrounds things at a smaller scale. Effectively, this eludes the question "what is a thing?" by composing things from the Markov blanket of smaller things. By induction, we have Markov blankets all the way down, which means one never has to specify the nature of things.

---

[1] This paper was written as an autodidactic exercise to ensure the author's intuitions played out over complementary formulations in statistical physics. The result is a long, over inclusive paper that tries to adopt conventions from different fields (which the author is not expert in), while emphasizing common themes.

More specifically, we will see that the Langevin formulation of dynamics – at any given spatiotemporal scale – can be decomposed into an ensemble of Markov blankets. These blanket states have a dynamics at a higher scale with exactly the same (Langevin) form as the dynamics of the original scale. When lifting the dynamics from one scale to the next, internal states are effectively eliminated, leaving only slow, macroscopic dynamics of blanket states. These become the states of things at the next level, which have their own Markov blankets and so on. The endpoint of this formalism is a description of everything at progressively higher spatial and temporal scales. The implicit separation of temporal scales is used in subsequent sections to examine the sorts of dynamics, physics or mechanics of progressively larger things.

This monograph comprises 12 sections organised into three parts. The first part establishes some basic results, the second part applies these results to limiting cases of dynamical systems to recover quantum, statistical and classical mechanics. The third part considers the special case of active or autonomous systems, in terms of a Bayesian mechanics for particles with internal states that 'matter' for their behaviour.

**Part One**: The first section is a foundational treatment that introduces some constraints on the dynamics of Markov blankets that possess measurable characteristics. The constraint of measurability – or possessing an invariant measure over sufficiently long periods of time – allows one to express the flow of states as a function of their non-equilibrium steady-state (NESS) density[2]. The relationship between flow and the NESS density follows in a straightforward way from the Fokker Planck formulation of density dynamics and, in particular, its eigensolution. The interesting results here are the dependencies – implicit in the system's equations of motion – that inherit from Markov blankets at nonequilibrium steady-state. The ensuing, relatively straightforward lemma and corollaries concerning marginal flows and conditional independencies then form the basis for emergent behaviours in subsequent sections. The second section looks at various ways in which one can characterise density dynamics in terms of symmetry breaking and self-organisation. This section uses information theory and geometry to characterise different sorts of self-organisation to nonequilibrium steady-state. The third section provides an illustration of self-organisation, using numerical analyses of a particular system (a synthetic primordial soup based on ensemble of Lorenz systems). This system is used throughout the monograph to illustrate how one can take complementary perspectives on the same dynamics. The fourth section considers the behaviour of this (Langevin) formulation of Markov blankets at nested scales. In brief, we assume that as one ascends to higher scales, random and intrinsic fluctuations are progressively suppressed, resulting in a move from dissipative dynamics – that are dominated by random fluctuations – through to large systems whose conservative dynamics are dominated by divergence-free flow.

---

[2] NESS could also be an acronym for Nearly Ergodic Steady-State in weakly mixing systems. As observed by my young colleague Brennan Klein, nonequilibrium steady-state puts the "ness" in "thingness" (From Middle English *-nes*, *-nesse*: appended to adjectives to form nouns meaning "the state of being"). We will argue later that any 'state of being' rests upon a NESS.

**Part Two**: Section 5 considers the very small in terms of quantum mechanics. This section derives the Schrödinger wave equation using the relationship between a particle's flow and the NESS density established in the first section. The trick here is to express or factorise the NESS density in terms of (complex) roots that play the role of a wave function. Section 6 then considers the collective behaviour of small things in terms of ensemble dynamics and stochastic thermodynamics. Our focus here is on linking the dissipative dynamics of ensembles to established results in statistical mechanics; namely, the laws of thermodynamics and related fluctuation theorems such as the Jarzynski equality. We then turn to the physics of big things in the limit of small amplitude random fluctuations. This limit allows us to write down equations of motion in terms of a classical Lagrangian or Hamiltonian, leading to classical mechanics, Newtonian laws of motion and Maxwell's equations.

**Part Three**: Having cast quantum, statistical and classical mechanics as limiting cases of the density dynamics of *inert particles*, we turn to the ontology of big things – whose internal states cannot be ignored – that show autonomous behaviour (e.g., large *active particles* like ourselves). Section 8 asks why one might attribute representational or inferential capacities to biological self-organisation. In other words, how notions like the good regulator theorem (Conant and Ashby, 1970) and the Bayesian brain hypothesis (Helmholtz, 1878 (1971); Knill and Pouget, 2004) could be substantiated in terms of a sentient physics. The argument here is fairly straightforward: namely, that the internal states of a system encode probabilistic beliefs about external states that cause sensory impressions on the Markov blanket – and are caused by the influence of active states on external states. This section provides a formal basis for an information geometry and attending free energy principle that describes autonomous things (e.g., cells or brains) as inferring the causes of actively sampled sensations. Here, we pursue a variational theme by showing how variational Bayes (Beal, 2003) is an emergent property of certain kinds of particles, leading to a form of Bayesian mechanics. Section 9 illustrates a particular inference using numerical analyses of the synthetic soup from Part One (and a virus like denizen). Section 10 then considers active states and agency in terms of corollaries of the free energy principle based upon an integral fluctuation theorem and expected free energy. The penultimate section considers the ensuing active inference in light of previous (thermodynamic) treatments. We conclude with a brief discussion of the relationships between quantum, stochastic, classical and Bayesian mechanics.

# Part One: the setup

## Something or nothing

The "*Siphonaptera*" is a nursery rhyme, sometimes referred to as Fleas:

*Big fleas have little fleas,*
*Upon their backs to bite 'em,*
*And little fleas have lesser fleas,*
*and so, ad infinitum.*

This nicely frames one approach to the question of 'what is a thing?' by appealing to an infinite regress in which the question goes away. This deflationary account[3] says that the states of things are constituted by their Markov blanket, while the Markov blanket comprises the states of smaller things with Markov blankets within them – and so on *ad infinitum*. This appeal to blankets 'all the way' down offers a recursive definition of everything – at separable spatial and temporal scales – that are unpacked in Section 4, using notions from the renormalisation group. The idea of blankets all the way down (and up) suggest that there is no privileged scale, other than the scale that 'matters' for a thing in question. In the final sections, we will see that to 'matter' means there is an information geometry in play at certain scales, which afford autonomous and itinerant dynamics but are sufficiently large to suppress random fluctuations.

To foreshadow a more technical description, the basic story can be illustrated with a common-sense example. Imagine a solar system whose physics (i.e., dynamics or flow) is described sufficiently by the position, velocity and irradiation of heavenly bodies. These quantities constitute ensemble averages of each body's surface (i.e., Markov blanket) with internal states fluctuating beneath. Now let us descend a scale and focus on a particular planet (e.g., Earth). At this scale, the meteorological flows and geography of the planet's surface now constitute an appropriate level of description with many (internal) microscopic states lying below. Now assume we have zoomed in to a city and can access these microscopic states that transpire to be the ebb and flow of commuters during the daily cycle of a metropolis. Now we come down a further level and appreciate that each element, previously contributing to the average behaviour of commuters, is an individual or entity with its own Markov blanket; namely, an embodied brain. At this level of description, the delicate and structured fluctuations of the inner workings of the brain are hidden behind the Markov blanket and yet, if we zoom in further, now become the Markov blankets of neuronal elements and processes that have been coordinating commuter behaviour. Here, the Markov blanket corresponds to a cell surface that itself surrounds intrinsic or internal intracellular processes;

---

[3] An account that might also dissolve the **prime mover** (Latin: *primum movens*), advanced by Aristotle as a primary cause of all motion in the universe. In Book 12 (Greek "Λ") of his *Metaphysics*, Aristotle describes the prime mover as being perfectly beautiful, indivisible, and contemplating only the perfect contemplation: itself contemplating.

namely, exchange among intracellular organelles with their own Markov blankets. One could imagine going further and further through macromolecules down to atomic and subatomic levels. At each stage, we find a sufficient level of description according to the states of an ensemble that present themselves for engagement with – and coupling to – the Markov blanket of things at the level in question. Crucially, beneath (or sequestered behind) each Markov blanket are intrinsic or internal states that themselves are constituted by (mixtures of) ensembles of blanket states. In what follows, we will retell this story (from the bottom up), trying to show why these hierarchical levels of description are a necessary consequence of any (weakly mixing) random dynamical system that possess Markov blankets. However, first, we consider some preliminaries and background material that will be necessary to connect the different perspectives adopted in subsequent sections.

## Some preliminaries

This section can be read as a foundational (introductory) treatment of physics. It is not rigorous but is sufficient to convey the basic ideas. To unpack some of the assertions and lemmas, each section is accompanied by numerical examples and comprehensive descriptions in the figure legends[4]. The numerical analyses illustrate various phenomena using stochastic chaos based on a single Lorenz system (Lorenz, 1963) – or an ensemble of Lorenz systems, dressed with blanket states, to simulate active matter. These examples try to emphasise that, although the maths may look complicated, it describes sensible, emergent phenomena.

The mathematical notation is largely standard: the section on quantum mechanics will occasionally use the Dirac notation and the section on statistical mechanics follows (Seifert, 2012). Occasionally, we will use the (Einstein) summation convention when dealing with tensors. An exception to standard notation is the use of boldface variables; where $x \in X$ denotes a (generalised) coordinate in phase or state-space, while $\mathbf{x} \in X$ denotes the expected or most likely value. $\mathbf{x}(a)$ will denote an expectation, conditioned on a variable $a$. Boldface capital letters $\mathbf{X}$ will denote operators. For clarity, functional derivatives and integrals involving time are expressed in terms of orbits, trajectories or paths $x[\tau] = \{x(\tau) : \tau \in [0, t]\}$ , where a value at time $\tau$ is denoted by $x(\tau) \equiv x_\tau$ . We will also be dealing with time-dependent probability densities $p(x, \tau) \equiv p_\tau(x_\tau)$ that have stationary or steady-state solutions $p(x, \infty) \equiv p(x)$ in the limit $\tau \to \infty$ ; similarly, for their negative log density or surprisal $\Im(x, \tau) \equiv \Im_\tau(x_\tau) = -\ln p(x, \tau)$ . For ease of reference, a glossary of terms and expressions is provided at the end of the monograph. Most of what follows rests on three equivalent and complementary descriptions of stochastic dynamics; the Langevin equation, path integral formulation and Fokker Planck equation.

---

[4] These numerical analyses serve two purposes. First, they show how one can characterise the same system from complementary perspectives. For example, one can treat a system as a small particle (e.g., an electron), in terms of quantum mechanics; or we can treat it as an ensemble of particles (e.g., a gas), to examine its statistical mechanics; or we can treat it as a blob of mass (e.g., a ball) in some active medium and describe its response in terms of classical mechanics. In Part Three, we will take this further and look at autonomous behaviour; namely, how one part of a system actively infers or 'measures' another. The second purpose is more pedagogical (for biological readers); in the sense that the simulations dispel any mysticism surrounding high end physics. In short, all the mechanics considered in this monograph lend themselves to straightforward and intuitive numerical analyses that allow one kind of mechanics to be understood in relation to the others.

**Langevin dynamics:** this formulation expresses the dynamics of systemic states $x(\tau)$ (i.e., states of some system) in terms of a state-dependent flow and some random fluctuations $\omega(\tau)$:

$$
\left.
\begin{aligned}
\dot{x}(\tau) &= f(x,\tau) + \omega \\[6pt]
E[\omega(\tau)] &= 0 \\
E[\omega(\tau) \cdot \omega(\tau - t)] &= 2\Gamma\rho(t) = 2\Gamma\delta(t)
\end{aligned}
\right\} \Rightarrow p(\dot{x}\,|\,x,\tau) = \mathcal{N}(f, 2\Gamma)
\tag{1.1}
$$

Here, the random fluctuations are normally distributed with a covariance of $2\Gamma$, under the assumption that they fluctuate sufficiently quickly, in relation to states *per se*, that we can ignore temporal correlations. This formulation underwrites everything that follows. In section 4, we will look more closely at where the Langevin formulation comes from – and why random fluctuations are Gaussian and uncorrelated.

**The path integral formulation:** this formulation deals with paths or trajectories $x[\tau]$, from $x(0) \equiv x_0$ that are generated by the Langevin dynamics above:

$$
\begin{aligned}
\Im(x[\tau]) &\triangleq -\ln p(x[\tau]) = \mathcal{A}(x[\tau]) \\
\mathcal{A}(x[\tau]) &= \int_0^t \mathcal{L}(x,\dot{x})\,d\tau \\
\mathcal{L}(x,\dot{x}) &= \tfrac{1}{2}[(\dot{x}-f)\cdot\tfrac{1}{2\Gamma}(\dot{x}-f) + \nabla\cdot f] \\
&= \tfrac{1}{4\Gamma}\dot{x}\cdot\dot{x} - \tfrac{1}{2\Gamma}f\cdot\dot{x} + \tfrac{1}{h}V(x) \\
V(x) &= \tfrac{h}{4\Gamma}f\cdot f + \tfrac{h}{2}\nabla\cdot f
\end{aligned}
\tag{1.2}
$$

This formulation expresses the probability of a path in terms of the *action* associated with a trajectory. It says that the surprisal (i.e., negative log probability) of a path (i.e., action) is the surprisal accumulated along its trajectory, based upon the difference between the path's motion and the flow expected at each point in state-space. Under Gaussian assumptions about the random fluctuations, the surprisal at each point (i.e., *Lagrangian*) has a simple quadratic form, with an additional divergence term that arises from the implicit use of Stratonovich integrals (Seifert, 2012). See Appendix A for an explanation of this term. Here, we have expressed the Lagrangian in terms of a *Schrödinger potential* that will figure later in quantum mechanics. This potential depends on, and only on, the flow. For non-quantum treatments, Planck's constant is usually set to $h = 1$.

It will be useful to introduce the Legendre transform of the Lagrangian called a *Hamiltonian* that will arise in the characterisation of how things behave:

$$\mathcal{H}(x, \dot{x}) = \dot{x} \cdot \frac{\partial \mathcal{L}}{\partial \dot{x}} - \mathcal{L}(x, \dot{x}) = \dot{x} \cdot \mathrm{p} - \mathcal{L}(x, \dot{x})$$
$$= \frac{1}{4\Gamma} \dot{x} \cdot \dot{x} - \frac{1}{\hbar} V(x)$$

$$\mathrm{p} \triangleq \frac{\partial \mathcal{L}}{\partial \dot{x}} = \frac{1}{2\Gamma}(\dot{x} - f) \qquad (1.3)$$

Here, the last equality defines the generalised *momentum*. Note that the most likely path obtains when the random fluctuations take their most likely value of zero, giving:

$$\dot{\mathbf{x}} = f(\mathbf{x}) \Rightarrow \mathcal{H}(\mathbf{x}, \dot{\mathbf{x}}) = -\mathcal{L}(\mathbf{x}, \dot{\mathbf{x}}) = -\frac{1}{2}\nabla \cdot f(\mathbf{x}) \qquad (1.4)$$

This means that the Hamiltonian along the most likely path reduces to the divergence of the flow. Furthermore, in conservative systems with divergence-free flow (i.e., with negligible random fluctuations) the Hamiltonian is zero everywhere. This speaks to the importance of the Hamiltonian in characterising conservative (i.e., classical) mechanics. Finally, path-dependent phase measurements $\Omega(x)$ can then be averaged in the following path integral, given an initial density, $p_0(x_0) \equiv p(x, 0)$:

$$E_{p(x[\tau], x_0)}[\Omega(x)] = \int dx_0 \int dx[\tau] [\Omega(x[\tau]) \, p(x[\tau] \,|\, x_0) \, p_0(x_0)] \qquad (1.5)$$

This concludes the key results from the path integral formulation.

**The Fokker Planck equation:** this formulation deals with the probability density over the states, which describes the probability of finding a system in state $x$ at time $\tau$. Given a Langevin system, one can describe the density dynamics as follows:

$$\dot{p}(x, \tau) = \mathbf{L} p(x, \tau) = -\nabla \cdot j(x, \tau)$$
$$\mathbf{L} = \nabla \cdot (\Gamma \nabla - f)$$
$$j(x, \tau) = f(x, \tau) p(x, \tau) - \Gamma \nabla p(x, \tau) \qquad (1.6)$$

Here, $\mathbf{L}$ is the Fokker Planck operator and $j(x, \tau)$ is the probability current that provides a convenient summary of the flow of probability mass. This comprises a flow-dependent term and (a usually opposite) part, generated by random fluctuations over probability gradients.

## Nonequilibrium steady states

Equipped with the Fokker Planck formulation of density dynamics, we can now consider the nonequilibrium long-term behaviour of any random dynamical system. Because the system is weakly mixing it will, after a sufficient amount of time, converge to an invariant set of states called a *pullback* or *random global attractor*. The attractor is random because it is itself a random set (Crauel, 1999; Crauel and Flandoli, 1994). The associated NESS density $p(x)$ is the solution to the Fokker-Planck equation (Frank, 2004). Equation (1.6) shows that the NESS density depends upon flow, which can always be expressed in terms of curl and divergence-free components. This is the Helmholtz decomposition (a.k.a., the fundamental theorem of vector calculus) and can be formulated in terms of an anti-symmetric matrix $Q = -Q^T$ and a scalar potential $\Im(x)$ (Ao, 2004)[5],

$$f = (Q - \Gamma)\nabla\Im \tag{1.7}$$

Using this standard form (Yuan et al., 2010), it is straightforward to show that $p(x) = \exp(-\Im(x))$ is the solution to the Fokker Planck equation (Friston and Ao, 2012). In information theory, the scalar potential $\Im(x) = -\ln p(x)$ is known as self-information, *surprisal* or more simply *surprise* (Jones, 1979; Tribus, 1961). This means we can express the flow in terms of the NESS density or surprisal, according to the *NESS lemma* in Appendix B and (Friston, 2013):

$$f = (\Gamma - Q)\nabla \ln p(x)$$
$$\Rightarrow \frac{j(x)}{p(x)} = -Q\nabla \ln p(x) \Rightarrow \nabla \cdot j(x) = 0 \Rightarrow \dot{p}(x) = 0$$

$$\Im(x) = -\ln p(x)$$
$$f(x) = (Q - \Gamma)\nabla\Im(x)$$
$$\nabla \cdot f(x) = -\Gamma\nabla^2\Im(x)$$

$$\tag{1.8}$$

This is the key result upon which most of this monograph rests. It says that the flow of any random dynamical system, at nonequilibrium steady-state, comprises orthogonal components: a dissipative flow that ascends the gradients established by the logarithm of the nonequilibrium steady-state density and a conservative (*divergence-free*), solenoidal flow circulating on the corresponding isocontours. Heuristically, the dissipative (*curl-free*) flow counters the dispersion of the density that would otherwise be caused by random fluctuations. This means the only remaining probability current is solenoidal. We will see later that this simple result has some remarkable implications, when we consider the flow of various subsets of states that are conditionally independent. This particular structure rests on the notion of a Markov blanket that is the second cornerstone of all that follows.

The next move is to substitute the NESS solution to the Fokker Planck equation into the path integral formulation to express the probability of any trajectory in terms of an action, expressed in terms of surprisal. From (1.2) and (1.8) this gives:

---

[5] For simplicity, we will assume that $Q = -Q^T$ does not depend on $x$, at least locally. Furthermore, the amplitude of random fluctuations is assumed to be spherical, so that we can treat $\Gamma$ as a scaled identity matrix or a scalar quantity.

$$\mathcal{A}(x[\tau]) = \int_0^t \mathcal{L}(x, \dot{x}) d\tau$$
$$\mathcal{L}(x, \dot{x}) = \tfrac{1}{2}[\tfrac{1}{2\Gamma}(\dot{x} - Q\nabla\mathfrak{I}) \cdot (\dot{x} - Q\nabla\mathfrak{I}) + \dot{x} \cdot \nabla\mathfrak{I} + \Gamma(\tfrac{1}{2}\nabla\mathfrak{I} \cdot \nabla\mathfrak{I} - \nabla^2\mathfrak{I})] \qquad (1.9)$$
$$\mathcal{H}(x, \dot{x}) = \tfrac{1}{2}[\tfrac{1}{2\Gamma}(\dot{x} - Q\nabla\mathfrak{I}) \cdot (\dot{x} - Q\nabla\mathfrak{I}) - \Gamma(\tfrac{1}{2}\nabla\mathfrak{I} \cdot \nabla\mathfrak{I} - \nabla^2\mathfrak{I})]$$

This result uses the fact that the solenoidal and gradient flows are orthogonal $Q\nabla\mathfrak{I} \cdot \Gamma\nabla\mathfrak{I} = 0$. Equation (1.9) is essentially the path integral formulation of nonequilibrium steady-state dynamics. It expresses the probability of any trajectory through state-space as the path integral of a Lagrangian; where the Lagrangian can be expressed in terms of motion and surprise (i.e., the NESS potential).

Crucially, the three terms in (1.9) depend in different ways on the amplitude of random fluctuations. This dependency can be seen more clearly if we consider the expression for a one-dimensional system, where there is no solenoidal flow:

$$\mathcal{A}(x[\tau]) = \underbrace{\tfrac{1}{2}\int_0^t \tfrac{1}{2\Gamma}\dot{x}^2 d\tau}_{kinetic} + \underbrace{\tfrac{1}{2}\int_0^t \dot{\mathfrak{I}} d\tau}_{path\text{-}independent} + \underbrace{\tfrac{1}{\hbar}\int_0^t V(x) d\tau}_{path\text{-}dependent}$$
$$\int_0^t \dot{\mathfrak{I}} d\tau = \mathfrak{I}(x_t) - \mathfrak{I}(x_0) \qquad (1.10)$$
$$V(x) = \tfrac{\hbar}{2}\Gamma(\tfrac{1}{2}\nabla\mathfrak{I} \cdot \nabla\mathfrak{I} - \nabla^2\mathfrak{I})$$

This implies that the action of any path can be expressed in terms of a motion-dependent (*kinetic*) term, a (*path-independent*) term – that depends upon the change in surprisal – and a (*path-dependent*) term that scales with the amplitude of random fluctuations. Appendix C provides a brief treatment of the expected Lagrangian and associated Hamiltonian.

The key thing to note here is that the first term in (1.10) has the form of a kinetic energy, in which the amplitude of random fluctuations plays the role of an inverse mass. The second term is simply a (NESS) potential difference, while the third (Schrödinger potential) term increases with the amplitude of random fluctuations. This means that when random fluctuations are large, the state behaves as if it has negligible mass and the Schrödinger potential dominates. Conversely, when the amplitude of random fluctuations is negligible, the first two terms predominate enabling a decomposition of action into kinetic and potential terms. This dialectic will appear later as the distinction between quantum and classical mechanics. Equation (1.10) suggests $\hbar = 0$ in the classical limit, when the contribution of the Schrödinger potential disappears (Feynman, 1948). However, in this monograph, Planck's constant is treated as a constant of proportionality (that endows the amplitude of random fluctuations with certain units), such that the classical limit is attained when their amplitude tends to 0; i.e., $\Gamma = 0$.

In this limit, the classical path is the most likely path that can be described with a variational principle of least action:

$$\delta_x \mathcal{A}(\mathbf{x}[\tau]) = 0$$
$$\Rightarrow \dot{\mathbf{x}}(\tau) = f(\mathbf{x}(\tau)) \qquad (1.11)$$
$$\Rightarrow \mathbf{x}[\tau] = \arg\min_{x[\tau]} \mathcal{A}(x[\tau])$$

This means the most likely path minimises action, rendering its variation with respect to the path zero. Crucially, at nonequilibrium steady-state, path-dependent and independent contributions to action can be expressed in terms of the surprisal and its gradients. We will call upon this consequence of nonequilibrium steady-state dynamics in several different settings. In treatments of synergetics and pattern formation, this least action principle is sometimes expressed as the destruction of energy gradients (Tschacher and Haken, 2007).

## Fluctuations and information length

"*Time is designed in such a way that given the present, the future is independent of the past*" (Caticha, 2015b) p6116

We will be concerned with the 'measure' of things; both in terms of probability measures and metrics that inherit from differential geometry. This section provides a brief background to the notion of *length* and *information geometry* – that arises when applying differential geometry to probability theory. The main message here is that all these measures depend, in a deep way, on time.

As part of this preamble, it is useful to consider the nature of random fluctuations. We will see later that these fluctuations are mixtures of states that fluctuate very quickly, in relation to states *per se* that play the role of slow variables. In this sense, random fluctuations are just fast states that are (by definition) not correlated with slow states. The implicit statistical independence of fast and slow states licences us to talk about 'random' fluctuations.

The form of the Langevin dynamics in (1.1) speaks to this separation of temporal scales. For example, the units of $\Gamma$ (per second) suggest it plays the role of a rate constant. Indeed, on one view, the amplitude of fluctuations corresponds to the rate at which the variance or dispersion of states – due to fluctuations – accumulates over time (Cox and Miller, 1965). Furthermore, at steady state, the amplitude is effectively a rate constant that couples the flow of (slow) states to the gradients of surprisal (1.7). In other words, for a given NESS density, the flow increases in proportion to the amplitude of fluctuations. One can formalise this by introducing the notion of *length*:

$$\ell = \int_0^t \sqrt{\dot{x}(\tau)^i \, g_{ij} \, \dot{x}(\tau)^j} \, d\tau \tag{1.12}$$

We have used the (Einstein) summation convention here, with an implicit summation over repeated superscripts and subscripts. Equation (1.12) expresses the length along a path in terms of a Riemannian metric supplied by the metric tensor $g_{ij}$. Paths of locally minimal distance are called *geodesics*, and are the analogues of straight lines in Euclidean space. From (1.10), the action of a path in the classical limit of low amplitude fluctuations can be interpreted as an upper bound on path length (by the Cauchy-Schwarz inequality):

$$\lim_{Q,\Gamma \to 0} \mathcal{A}(x[\tau]) = \int_0^t \dot{x}(\tau)^i \, g_{ij} \, \dot{x}(\tau)^j \, d\tau \geq \ell^2$$
$$g = \frac{1}{4\Gamma}$$

(1.13)

For simplicity, we have ignored solenoidal flow. This suggests that the most likely (classical) path will be the shortest, if length in measured in terms of the *precision* (i.e., inverse covariance) of random fluctuations. Equivalently, the precision furnishes a Riemannian metric that equips state-space with a geometry in which points are close together when fluctuations have a large amplitude. The equivalence between the amplitude of random fluctuations and the metric tensor underlies our simplifying assumption that the amplitude of random fluctuations is spherical (i.e., looks the same from all directions). This is equivalent to working in a symmetric state-space, whose Riemannian metric is invariant to the choice of coordinates.

One can take this metric treatment further and equip spaces of the sufficient statistics (i.e., parameters) of a density with an *information geometry*. In brief, information geometry rests on Riemannian metrics that can be used to measure distances on *statistical manifolds* (Amari, 1998; Ay, 2015). A statistical manifold is a metric space in which each point represents a probability density; i.e., a parameter space whose points correspond to the sufficient statistics of a probability density, such that nearby points on the statistical manifold correspond to similar densities. For example, the two-dimensional space spanned by the statistical moments (i.e., mean and precision) of a Gaussian density constitutes a ubiquitous statistical manifold. The special thing about statistical manifolds is that they are always equipped with a metric tensor, supplied in the form of a Fisher information metric.

In the present setting, following (Crooks, 2007), one can characterise systemic density dynamics in terms of *information length*, where the metric is a Fisher information. Consider the decomposition of the density into time-independent collective variables or modes $\zeta_i(x)$ and time-dependent conjugate variables that play the role of sufficient statistics $\lambda(\tau)$ [6]:

$$\Im(x,\tau) = \lambda^i(\tau)\zeta_i(x) + \ln Z$$

(1.14)

Under this parameterisation, the Fisher information metric $\mathbf{I}(\lambda)$ is:

$$\ell = \int_0^t \sqrt{g_{ij}\dot{\lambda}^i\dot{\lambda}^j} \, d\tau$$
$$g = \mathbf{I}(\lambda) \Leftrightarrow g_{ij} = \mathrm{cov}(\zeta_i(x),\zeta_j(x)) = E\left[\frac{\partial \Im}{\partial \lambda^i}\frac{\partial \Im}{\partial \lambda^j}\right] = \frac{\partial D[\,p_{\lambda'}(x)\,\|\,p_{\lambda}(x)]}{\partial \lambda''\partial \lambda'^j}\bigg|_{\lambda'=\lambda}$$

(1.15)

Notice that the information geometry is in the space of the conjugate variables that parameterise the density over states, as opposed to state-space *per se*. This space is a statistical manifold and its geometry will become a central aspect of Bayesian mechanics later. At the moment it serves to foreshadow the intimate relationship between information, geometry and statistical mechanics (Crooks, 2007; Kleeman, 2014).

---

[6] We will use $Z$ to denote a partition function or normalising constant throughout.

The final equality in (1.15) can be confirmed by straightforward calculation of the second derivative of the divergence between two probability distributions that are infinitesimally close (Caticha, 2015a); where $\lambda' = \lambda + d\lambda$. This means information length on the statistical manifold accumulates more quickly when the parameterised density changes quickly. This interpretation also licences a formulation of information length as the accumulation of (the square root of) divergences between successive densities over small displacements $d\lambda \to 0$ on the statistical manifold,

$$
\begin{aligned}
\ell &= \int d\ell \\
d\ell^2 &= g_{ij} d\lambda^j d\lambda^i \\
\tfrac{1}{2} d\ell^2 &= D[p_{\lambda'}(x) \parallel p_{\lambda}(x)] = \tfrac{1}{2} \frac{\partial D[p_{\lambda'}(x) \parallel p_{\lambda}(x)]}{\partial \lambda''^{i} \partial \lambda'^{j}} \bigg|_{\lambda' = \lambda} d\lambda^i d\lambda^j
\end{aligned}
\tag{1.16}
$$

The last equality follows from a Taylor expansion of the divergence, where the first non-vanishing term is the second derivative in (1.15). This follows because the divergence and its first derivative are zero when $\lambda' = \lambda$. Equation (1.16) means that information length can be construed as the number of distinct configurations the density passes through when moving through parameter space. These results provide a graceful way to connect thermodynamic variables to different configurations of a system, by treating them as parameters of a probability distribution (Crooks, 2007; Kleeman, 2014). However, our focus will be on a special parameter – time – that endows density dynamics with an information geometry.

An important information length follows from the fact that time itself parameterises an evolving density, $\lambda(\tau) = \tau \Rightarrow \dot{\lambda} = 1$ and therefore *time has an information geometry*. Consider a system prepared in some initial state with a density $p_0 \equiv p(x, 0)$ such that

$$
\begin{aligned}
p(x, \tau) &= p_0 + \tau \dot{p}_0 + \dots \\
&= p_0 + \tau \mathbf{L} p_0 + \dots
\end{aligned}
$$

$$
\begin{aligned}
\ell(t) &= \int_0^t \sqrt{g(\tau)} \, d\tau \\
g(\tau) &= \mathbf{I}(\tau) = E_{p_0}[(\partial_\tau \mathfrak{I}_0)^2] = \int \frac{(\mathbf{L}_0 p_0)^2}{p_0} \, dx \\
\mathbf{L}(\tau) &= \nabla \cdot \Gamma \nabla - \nabla \cdot f(\tau)
\end{aligned}
\tag{1.17}
$$

In this context, the Fisher information metric furnishes a temporal scaling that reflects the rate at which the density evolves. From (1.16):

$$
\begin{aligned}
d\ell(\tau)^2 &= g(\tau) d\tau^2 \\
\tfrac{1}{2} d\ell(\tau)^2 &= D[p(x, \tau + d\tau) \parallel p(x, \tau)]
\end{aligned}
\tag{1.18}
$$

In other words, the metric for time plays the role of a (squared) rate constant, with units of per second (squared). Heuristically, as time progresses, information length depends upon the amplitude of random fluctuations and (the divergence of) flow via the Fokker Planck operator in (1.17). This means that metric time can proceed slowly from one part of state-space and quickly from elsewhere. Equation (1.18) also suggests that information length can be regarded as an accumulation of (squared) divergences over infinitesimally short time steps. Note that this accumulation enables a metric to be assembled from (pre-metric) divergences. This should be contrasted with the divergence between the initial density and the density at some later point in time. We will see an example of this in the next section.

The information length in (1.18) provides a useful characterisation of convergence to equilibrium or nonequilibrium steady-state (Kim, 2018). As time goes by, the future density from any initial state converges to the NESS density and the information length converges to an asymptotic limit. This limit scores the distance from the initial density to the NESS density. At convergence, the divergence in (1.18) disappears and there is no further increase in information length

$$\lim_{\tau \to \infty} D[\, p(x, \tau) \,\|\, p(x)\,] = 0 \Rightarrow d\ell(\tau) = 0 \tag{1.19}$$

Effectively, from the point of view of information length, time slows down in the future. Heuristically, imagine what you will be doing in an hour – and how it differs from what you are doing at the moment. Now, repeat the exercise but imagine yourself in a decade's time and a decade plus one hour. In the sense that it is difficult to distinguish between evolving versions of yourself in the distal future, time has effectively stopped.

In summary, the information length of time-dependent densities, parameterised by time, provides a metric that complements the use of divergence between initial and final densities, which does not depend upon the path or evolution of the density from its initial state. Both play an important role in characterising convergence to nonequilibrium steady-state. Later, we will see the divergence in (1.19) arises in the form of free energy in both stochastic and Bayesian mechanics. We will also see that information length is closely related to stochastic entropy production in thermodynamic formulations. So far, everything we have previewed applies to any random dynamical system. In the next section, we will revisit these characterisations, when the system possesses a Markov blanket.

## Random dynamical systems and Markov blankets

A Markov blanket is a set of states that separates two other sets in a statistical sense. The term Markov blanket was introduced in the context of Bayesian networks or graphs (Pearl, 1988) and refers to the children of a set (the set of states that are influenced), its parents (the set of states that influence it) and the parents of its children. The existence of a Markov blanket induces a partition of states into *internal* and *external* states, where external states are hidden (insulated) from the internal (insular) states by the Markov blanket. In other words, external states can only be seen vicariously by internal states, through blanket states. Furthermore, the Markov blanket can itself be partitioned into two sets that are, and are not, children of external states. We will refer to these as *sensory states* and *active states* respectively: $b = \{s, a\} \in B$. Put simply, the existence of a Markov blanket implies a partition of states into *external*, *sensory*, *active* and *internal* states: $x = \{\eta, s, a, \mu\} \in X$. External states cause sensory states that influence – but are not influenced by – internal states, while internal states cause active states that influence – but are not influenced by – external states. Crucially, the dependencies induced by Markov blankets create a circular causality that is reminiscent of the action-perception cycle: see Figure 1 and (Fuster, 2004). Circular causality here means that external states cause changes in internal states, via sensory states, while the internal states couple back to the external states through active states, such that internal and external states influence each other in a vicarious and reciprocal fashion.

## Markov blankets and marginal flows

In the next section, we will unpack the implications of Markov blankets for self-organisation in terms of information theory. This treatment rests upon the conditional independencies among the partition of states that result from precluding an influence of external states on active and internal states – and an influence of internal states on sensory and external states. This dynamical architecture is summarised in terms of a *marginal flow lemma* and its corollaries below. In brief, these results express the flow at nonequilibrium steady-state under the conditional independencies implied by a Markov blanket and *vice versa*. In other words, they connect the sparse influences mediated by Langevin flow to conditional independencies among subsets of states. Effectively, this generalises the standard form for flow in (1.8) to a partition of states that contains a Markov blanket. Appendix B contains the accompanying proofs and considers the complementary perspectives on how sparse influences underwrite conditional independence and *vice versa*.

The generalisation of the NESS density laminar to Markov blankets rests on the notion of *marginal flow*; namely, the flow of certain states averaged (i.e., marginalised) over other states. We will use the ~ notation to denote the complement of a subset of states; for example, $x = (\mu, \tilde{\mu})$.

**Lemma** (marginal flow): *for any weakly mixing random dynamical system at nonequilibrium steady-state, the marginal flow $f_\eta(\mu)$ of any subset of states $\eta \in X$, averaged under the complement of another $\mu \in X$ can be expressed in terms of the gradients of the logarithm of the corresponding marginal density:*

$$f_\eta(\mu) \triangleq E_{p(\tilde{\mu}|\mu)}[f_\eta(\mu, \tilde{\mu})] = (Q_{\eta\eta} - \Gamma_{\eta\eta})\nabla_\eta \mathfrak{I}(\mu) + Q_{\eta\tilde{\eta}}\nabla_{\tilde{\eta}} \mathfrak{I}(\mu) \qquad (1.20)$$

**Corollary** (conditional independence): *if the flow of one subset of states does not depend on another, then it becomes the flow expected under the second subset. For example, in terms of the Markov blanket:*

$$
\begin{bmatrix} f_\eta(x) \\ f_s(x) \\ f_\mu(x) \\ f_a(x) \end{bmatrix} = \begin{bmatrix} f_\eta(\eta,b) \\ f_s(\eta,b) \\ f_\mu(\mu,b) \\ f_a(\mu,b) \end{bmatrix} = \begin{bmatrix} (Q_{\eta\eta} - \Gamma_{\eta\eta})\nabla_\eta \Im(\eta,b) \\ (Q_{ss} - \Gamma_{ss})\nabla_s \Im(\eta,b) + Q_{sa}\nabla_a \Im(\eta,b) \\ (Q_{\mu\mu} - \Gamma_{\mu\mu})\nabla_\mu \Im(\mu,b) \\ (Q_{aa} - \Gamma_{aa})\nabla_a \Im(\mu,b) + Q_{as}\nabla_s \Im(\mu,b) \end{bmatrix} \tag{1.21}
$$

In short, the conditional independencies induced by the Markov blanket mean that the flow of external states is the same for every internal state, which is just its average over the internal states (similarly for other partitions).

**Corollary** (expected flow): *the marginal flow of any subset $\eta \subset x$ averaged over all other states depends only on the gradients of its marginal density, provided there is no solenoidal coupling with its complement:*

$$
f_\eta(\eta) = (\Gamma_{\eta\eta} - Q_{\eta\eta})\nabla_\eta \ln p(\eta) = (Q_{\eta\eta} - \Gamma_{\eta\eta})\nabla_\eta \Im(\eta) \tag{1.22}
$$

This is a special case of the marginal flow lemma, when $\eta = \mu$ and $Q_{\eta\bar{\eta}} = 0$. It implies that the expected flow of any state or subset of states, averaged over all other states, will behave in exactly the same way as all states considered together. In other words, it will ascend the gradients of its (marginal) density.

The marginal flow lemma allows us to express the conditional independencies implicit in the structural or probabilistic graphical model of Figure 1 in terms of the flows in (1.21). In other words, if a system maintains a Markov blanket at nonequilibrium steady-state, then it must possess flows that depend only upon certain states. This structured dynamics underwrites everything that follows.

## Summary

This section has introduced the technical foundations that we will call upon later to characterise dynamics in various settings. Its focus was on self-organisation to nonequilibrium steady-state, which can be characterised as the solution to the Fokker Planck formulation of density dynamics. Crucially, this enables one to express the flow of states in terms of a nonequilibrium steady-state density, surprisal or potential. We have looked briefly at the geometry of density dynamics in terms of information length. Finally, the conditional independencies implied by a Markov blanket have been expressed in terms of how the (marginal) flow of certain states depends on other states. The marginal flows induced by a Markov blanket will become important later, when we interpret gradient flows in relation to information geometry – in Part Three.

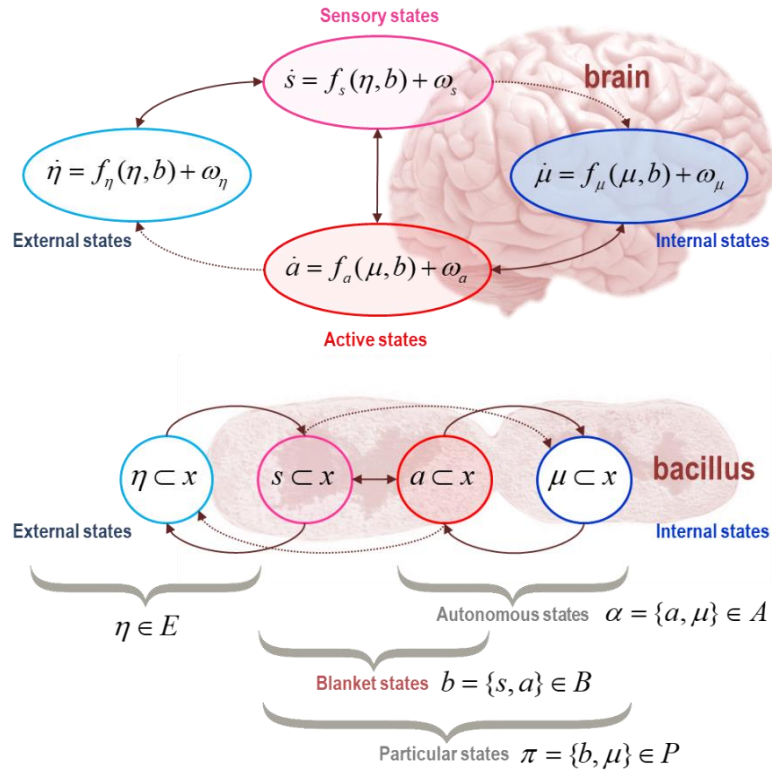

**FIGURE 1**

*Markov blankets.* This probabilistic graphical model illustrates the partition of states into *internal* states (blue) and hidden or *external* states (cyan) that are separated by a Markov blanket – comprising *sensory* (magenta) and *active* states (red). The upper panel shows this partition as it would be applied to action and perception in a brain. In this setting, self-organisation of internal states then corresponds to perception, while active states couple brain states back to external states. The lower panel shows the same dependencies but rearranged so that the internal states are associated with the intracellular states of a Bacillus, where the sensory states become the surface states or cell membrane overlying active states (e.g., the actin filaments of the cytoskeleton). Note that the only missing influences are between internal and external states – and directed influences from external (respectively internal) to active (respectively sensory) states. The surviving directed influences are highlighted with dotted connectors. *Autonomous* states are those states that are not influenced by external states, while *particular* states constitute a particle; namely, autonomous and sensory states – or blanket and internal states. The equations of motion in the upper panel follow from the marginal flow lemma.

## Symmetry breaking and self-organisation

"*How can the events in space and time which take place within the spatial boundary of a living organism be accounted for by physics and chemistry?*" (Schrödinger, 1944)

The introduction of Markov blankets – and the distinction between the external and internal states of a particle – changes the game somewhat, in terms of ensemble densities. In the absence of a partition, we can only talk about the entropy of a density and how it changes with time. However, in the setting of a partition, we can consider the entropy of particular states in relation to hidden states (or *vice versa*). This relative entropy is known as mutual information. So, are we interested in systems with a high or a low mutual information? It transpires that the answer is both, in the sense that we are interested in particles that explore their state-space but have a well-defined attracting manifold with low measure (i.e., low entropy). This speaks to a dialectic between opposing constraints. In brief, if the NESS entropy of particular states is small, then the average uncertainty about particular states, given external states must be small. In other words, knowing the external state resolves *ambiguity* about particular states. However, at the same time, the mutual information or coupling between external and particular states must also be small; otherwise, there will be a *risk* of being unable to disambiguate external from particular states; i.e., the particle will dissipate or dissolve. Heuristically, this allows for the fact that we can identify Markov blankets that are distinct from their external milieu (e.g., disambiguating a fish from the water in which it is swimming), while – at the same time – observing an intricate and self-organised coupling between particular dynamics and external states (e.g., a particular fish swimming in water). Even more simply, a fish remains a fish, despite the myriad of delicate, context-sensitive behaviours that preserve its integrity (Clarke et al., 2015). In what follows, we consider how this dialectic emerges from a straightforward statistical treatment using information theory.

Having established a partition of systemic states, we are now in a position to define the sort of self-organising systems we want to characterise. In brief, these are systems with space-filling random dynamical attractors with low measure. In other words, their probability mass is concentrated in small volumes that are connected in a way that permits itinerant (i.e. wandering) *percolation* of trajectories through state-space, from one manifold of the attractor to another: c.f., the percolation produced by phase transitions in deterministic systems (Vespignani and Zapperi, 1998). The implicit symmetry breaking (i.e., divergence of nearby trajectories to different regimes of phase-space) is a hallmark of nonequilibrium dynamics (Evans and Searles, 2002) and is intimately related to phenomena like self-organised criticality in dynamical systems (Bak et al., 1988; Vespignani and Zapperi, 1998). Indeed, much of complexity science addresses the problem of how to formalise multiscale, itinerant and chaotic dynamics. This is a vast field, encompassing renormalisation group theory, scale-invariance, criticality and universality (Kwapien and Drozdz, 2012; Nicolis and Prigogine, 1977; Schwabl, 2002). In this monograph, we will elude many of the finer details (and phenomena such as bifurcations, frustration and phase transitions) and suppose that the interesting behaviour of self-organising systems can be captured by nonequilibrium steady-state densities with the right sort of shape.

So, what is the right sort of shape? We start by considering the marginal (NESS) density over particular states. Given a partition into external and particular states, it is straightforward to characterise a simple form of self-organisation in terms of entropy production. This follows because there is a separation between autonomous $\alpha = \{a, \mu\} \in A$ and sensory states $s \in S$. Crucially, by definition, the flow of autonomous states does not depend on external states $\eta \in E$. This means that autonomous states will appear to suppress the self-information or surprisal of particular states $\pi \in P$ and its long-term average; namely, their entropy. From the marginal flow lemma (1.21), we have (ignoring solenoidal coupling between active and sensory states):

$$f_\alpha(\pi) = (Q_{\alpha\alpha} - \Gamma_{\alpha\alpha})\nabla_\alpha \mathfrak{I}(\pi)$$
$$E_{p(\pi)}[\mathfrak{I}(\pi)] = \lim_{\tau \to \infty} \tfrac{1}{\tau}\int_0^\tau \mathfrak{I}(\pi(t))dt \qquad (2.1)$$
$$= H(P)$$

We will refer to the entropy of particular states as *particular* or *self-entropy*. The flow in (2.2) will make it look as if autonomous states are trying, on average, to minimise the entropy of particular states. From elementary information theory, it follows that autonomous states will also look as if they are trying to minimise the mutual information between external states and particular states, while – at the same time – minimising the entropy of particular states, given external states. This is because mutual information is the uncertainty about particular states, minus the uncertainty, given external states (i.e., when there is no reduction in uncertainty afforded by knowing the external states, the mutual information is zero).

The decomposition of (self) entropy into mutual information and conditional entropy can also be expressed, from a statistical perspective, as a decomposition of *surprisal* into in*accuracy and complexity*:

$$\mathfrak{I}(\pi) = E_{p(\eta|\pi)}[\mathfrak{I}(\pi)]$$
$$= E_{p(\eta|\pi)}[\ln p(\eta \mid \pi) - \ln p(\eta, \pi)]$$
$$= E_{p(\eta|\pi)}[\ln p(\eta \mid \pi) - \ln p(\eta) - \ln p(\pi \mid \eta)]$$
$$= \underbrace{E_{p(\eta|b)}[\mathfrak{I}(\pi \mid \eta)]}_{inaccuracy} + \underbrace{D[p(\eta \mid \pi) \parallel p(\eta)]}_{complexity} \qquad (2.3)$$

$$E_{p(\pi)}[\mathfrak{I}(\pi)] = H(P) = \underbrace{H(P \mid E)}_{ambiguity} + \underbrace{I(E,P)}_{risk}$$
$$\underbrace{H(P \mid E)}_{ambiguity} = \underbrace{H(P)}_{entropy} - \underbrace{I(E,P)}_{info\ gain}$$

Complexity here is used in a statistical sense, where it scores the divergence between a posterior and prior distribution over external (hidden) states, while accuracy is the expected log probability of particular states, under the posterior. On this view, the conditional entropy is expected inaccuracy (i.e., *ambiguity*), while mutual information becomes the expected complexity cost (i.e., *risk*). The last equality evinces the dialectic that attends mutual information: on the one hand, minimising entropy requires the minimisation of mutual information, where it plays the role of *risk*, while minimising ambiguity requires a maximisation of mutual information, where it plays the role of *information gain*. These complementary roles are easily reconciled by noting that ambiguity and risk (or inaccuracy and complexity) are just two sides of the same coin; namely, self-entropy.

*Complexity* is a ubiquitous cost function in optimal control theory and Bayesian statistics. In optimal control, it scores the divergence between predicted external states, given the sensory and active (control) states and some desired states (Kappen, 2005; Kappen et al., 2012). In economics, this is called *risk-sensitive control* (Fleming and Sheu, 2002; van den Broek et al., 2010). In Bayesian statistics, the complexity scores the degree to which a posterior density over hidden states diverges from its prior; in other words, the degrees of freedom required to encode posterior beliefs about hidden states (Spiegelhalter et al., 2002). Reducing complexity cost underwrites Occam's principle; i.e., the best explanation provides an accurate account with the smallest change in posterior beliefs relative to prior beliefs (Penny et al., 2004). Formally, this is closely related to the notion of *causal entropic forces* in the modelling of adaptive behaviour in nonequilibrium systems (Wissner-Gross and Freer, 2013). Finally, the causal entropic forces can, themselves, be related to the maximum entropy principle of Jaynes (Jaynes, 1957).

The *ambiguity* term has epistemic, uncertainty reducing, interpretations; in which the marginal flow of autonomous states will appear to minimise the uncertainty about sensory states, given external states. In other words, self-organisation will appear to seek out regimes of phase-space in which external states cause unambiguous sensory states – much like searching under a lamp post for lost keys (Demirdjian et al., 2005). This dynamics is self-organising in the sense that (on average) autonomous states will appear to reduce the entropy of particular states. This particular entropy is the mutual information between blanket and external states plus their conditional uncertainty, conditioned on the external states. In other words, autonomous states will appear to minimise the statistical coupling (i.e., mutual information) with external states while, at the same time, resisting their dispersion, under any given hidden states.

We can express this active resistance to dissipation in terms of entropy production (a full treatment of entropy production can be found in the thermodynamics section below). The entropy production due to the flow of autonomous states can be expressed as:

$$
\begin{aligned}
\dot{H}^{\alpha} &= \int p(\pi) f_{\alpha}(\pi) \cdot \nabla_{\alpha} \Im(\pi) d\pi \\
&= -\int p(\pi) \nabla_{\alpha} \Im(\pi) \cdot \Gamma_{\alpha\alpha} \nabla_{\alpha} \Im(\pi) d\pi \leq 0
\end{aligned}
\tag{2.4}
$$

It follows that autonomous entropy production is always zero or less (because the covariance of random fluctuations is positive definite – and the solenoidal flow cancels). In other words, at nonequilibrium steady-state, autonomous flow resists the dispersion of particular states due to random fluctuations and the influences of external states. We can drill down on this entropy reducing behaviour by considering the marginal flow of autonomous states expected under sensory states. By the marginal flow lemma (1.22), we have:

$$
f_{\alpha}(\alpha) = (Q_{\alpha\alpha} - \Gamma_{\alpha\alpha}) \nabla_{\alpha} \Im(\alpha) \Rightarrow p(\dot{\alpha} \mid \alpha) = \mathcal{N}(f_{\alpha}(\alpha), 2\Gamma_{\alpha})
\tag{2.5}
$$

Following (1.11), when random fluctuations dominate, the most likely (marginal) path of autonomous states minimises their action:

$$
\begin{aligned}
\delta_{\alpha} \mathcal{A}(\boldsymbol{\alpha}[\tau]) &= 0 \\
&\Rightarrow \dot{\boldsymbol{\alpha}} = f_{\alpha}(\boldsymbol{\alpha}) = (Q_{\alpha\alpha} - \Gamma_{\alpha\alpha}) \nabla_{\alpha} \Im(\boldsymbol{\alpha}) \\
&\Rightarrow \boldsymbol{\alpha}[\tau] = \arg\min_{\alpha[\tau]} \mathcal{A}(\alpha[\tau])
\end{aligned}
\tag{2.6}
$$

So, what does this entail? To build an intuition about autonomous dynamics, we can use the same inflationary device as in (2.3) to express the surprisal of autonomous states in terms of complexity cost and the information gained by conditioning on sensory states:

$$
\begin{aligned}
\Im(\alpha) &= E_{p(\tilde{\alpha}|\alpha)}[\Im(\alpha)] \\
&= E_{p(\tilde{\alpha}|\alpha)}[\ln p(\eta\,|\,\pi) - \ln p(\eta) - \ln p(\eta\,|\,\pi) + \ln p(\eta\,|\,\alpha) - \ln p(a\,|\,\eta)] \\
&= E_{p(\eta|\alpha)}[\Im(\alpha\,|\,\eta)] + E_{p(\tilde{\alpha}|\alpha)}[\underbrace{D[\,p(\eta\,|\,\pi)\,\|\,p(\eta)]}_{complexity} - \underbrace{D[\,p(\eta\,|\,\pi)\,\|\,p(\eta\,|\,\alpha)]}_{information\ gain}]
\end{aligned}
\tag{2.7}
$$

$$
\begin{aligned}
E_{p(\alpha)}[\Im(\alpha)] &= H(A) = I(A,E) + H(A\,|\,E) \\
&= H(A\,|\,E) + \underbrace{I(E,P)}_{risk} - \underbrace{I(E,S\,|\,A)}_{active\ information}
\end{aligned}
$$

Here, $\tilde{\alpha} = \{\eta, s\}$ is the complement of autonomous states. This decomposition means that the marginal flow of autonomous states minimises its surprisal, which can be decomposed into terms that reflect the dialectic between trying to couple to external states and yet resist their dispersive effects. In this decomposition, ambiguity reduction is expressed in terms of information gain:

$$
\underbrace{E_{p(\pi)}[D[\,p(\eta\,|\,\pi)\,\|\,p(\eta\,|\,\alpha)]}_{expected\ information\ gain} = I(E,S\,|\,A) = I(E,S) - I(E,S,A)
\tag{2.8}
$$

This equality has been introduced to establish a connection with characterisations of self-organisation in terms of higher-order mutual information below. Information gain is sometimes referred to as *epistemic value* or *intrinsic motivation* in artificial intelligence and robotics (Friston et al., 2015b; Oudeyer and Kaplan, 2007; Schmidhuber, 2010). It corresponds to change in the probability density over external states afforded by sensory states: i.e., the Kullback-Leibler (KL) divergence between the posterior density with and without sensory states, conditioned upon autonomous states. This is also the mutual information between the sensory and hidden states, afforded by autonomous activity. In the life sciences (e.g., cognitive neuroscience), this measure is often referred to as *Bayesian surprise* or *salience* (Itti and Baldi, 2009; Mirza et al., 2016; Sun et al., 2011). We will return to these interpretations in Part Three. The expression for expected information gain in (2.8) shows that it comprises the mutual information between external and sensory states minus the third order mutual information (among external, sensory and autonomous states).

In summary, self-organisation can be cast as an autonomous suppression of self-entropy (resp. surprisal). In turn, self-entropy can be decomposed into risk (resp. complexity) and ambiguity (resp. inaccuracy) resolving components that look as if they are mediated by the flow of autonomous states. Clearly, in one sense, all interesting systems that possess a random dynamical attractor will show some degree of self-organisation. Expressing self-organisation in terms of self-entropy, risk and ambiguity just means that one can talk about – and quantify – self-organisation in terms of sentient, epistemic behaviour.

## Self-organization and self-evidencing

In statistics, the surprisal of particular states is known as the negative logarithm of the marginal likelihood or *evidence*. This means self-organisation can be construed as *self-evidencing*, in the sense that the most likely flow of autonomous states reduces surprisal and therefore increases evidence. This interpretation will play an important role in Part Three, in terms of understanding behaviour – of the sort considered in computational and cognitive neuroscience. The position taken here is not to ask how self-organisation emerges; rather, what properties do self-organising systems exhibit? This may seem as if we are avoiding a hard problem; however, nearly every system encountered in the real world is self-organising to a greater or lesser degree – suggesting that self-organisation is, in itself, unremarkable. Put another way, if systems did not self-organise they would have dissipated before we had a chance to observe them. This means that the interesting questions are what self-organisation looks like and what sort of mechanics does it entail?

In what follows, we look at how self-organisation is manifest and, in Part Three, turn to the apparent teleology that emerges from a Bayesian mechanics. From the point of view of a physicist, this means we are starting from the assumption that any interesting system that has a random dynamical attractor will, in the course of settling onto its attracting set, *reduce* its entropy. This stands in stark contrast with statistical thermodynamics that normally appeals to an *increase* in the entropy of a closed system. However, the very closure of the system – in terms of its insulation from external states by a Markov blanket – may be the more interesting problem. In other words, how did the heat bath or container emerge and what explains its persistence? This is not to say that classical (and equilibrium) statistical mechanics go away: we will see in subsequent sections how they emerge as special cases, when we consider ensembles of Markov blankets contained within blankets.

## Self-organisation, frustration and supersymmetry

The preceding characterisation of self-organisation was introduced heuristically; however, it has some construct validity in relation to various characterisations of complexity. Perhaps the most direct is the relationship between high-order mutual information and the itinerant dynamics associated with (geometrical) *frustration* in dynamical systems (Kaluza and Meyer-Ortmanns, 2010). In particular, the correlations among ensembles with negative third-order mutual information can be considered frustrated, in the sense that "two-body preferences are simultaneously unsatisfied" (Matsuda, 2000) p3099. These correlations are especially important in frustrated statistical systems such as spin glasses. In these systems, frustration – due to competing interactions or geometrical constraints – can induce complicated phase transitions, partial disorder, and non-exponential relaxation (Fierro et al., 1999; Matsuda, 2000). Interestingly, high-order mutual information also underpins the measures of neural complexity introduced by (Tononi et al., 1994): see (Ay, 2015).

The current formulation (in terms of random dynamical systems) is probably best considered in relation to the supersymmetry theory of stochastic systems (Parisi and Sourlas, 1982). All stochastic differential equations for continuous time dynamical systems – of the sort we are dealing with here – possess *topological supersymmetry* (Ovchinnikov, 2016). Topological supersymmetry (TS) refers to the preservation of phase-space continuity; in other words, infinitely close points remain close during continuous time evolution, even in the presence of noise. Spontaneous TS breaking underpins ubiquitous dynamical phenomenon; such as *chaos*, *turbulence* and *self-organized criticality* (Ovchinnikov, 2016). Symmetry breaking of this sort entails the emergence of long-range dynamical behaviour by the *Goldstone theorem* (Goldstone et al., 1962). This manifests as 1/f noise and the scale-free statistics of sudden (instantonic) processes that conform to *Zipf's law*; e.g., earthquakes, neuronal avalanches, solar flare *etc.*, (Beggs and Plenz, 2003; Kauffman and Johnsen, 1991; Newman, 2005; Plenz and Thiagarajan, 2007; Shew et al., 2011).

To explain this kind of dynamical behaviour, it has been suggested that some random dynamical systems are attracted to critical points. This constitutes the phenomenological approach of self-organized criticality – SOC (Bak et al., 1987; Bak et al., 1988). Spontaneous TS breaking offers an alternative perspective that eschews critical phenomena and regards SOC as noise-induced symmetry breaking. Heuristically, this can be thought of as noise-induced tunnelling among different attracting manifolds. Figure 2 helps build an intuition about symmetry breaking by illustrating its emergence as self-entropy is reduced. A complementary illustration is provided in Figure 3, which shows the emergence of dynamical instability (i.e., criticality) as self-entropy falls.

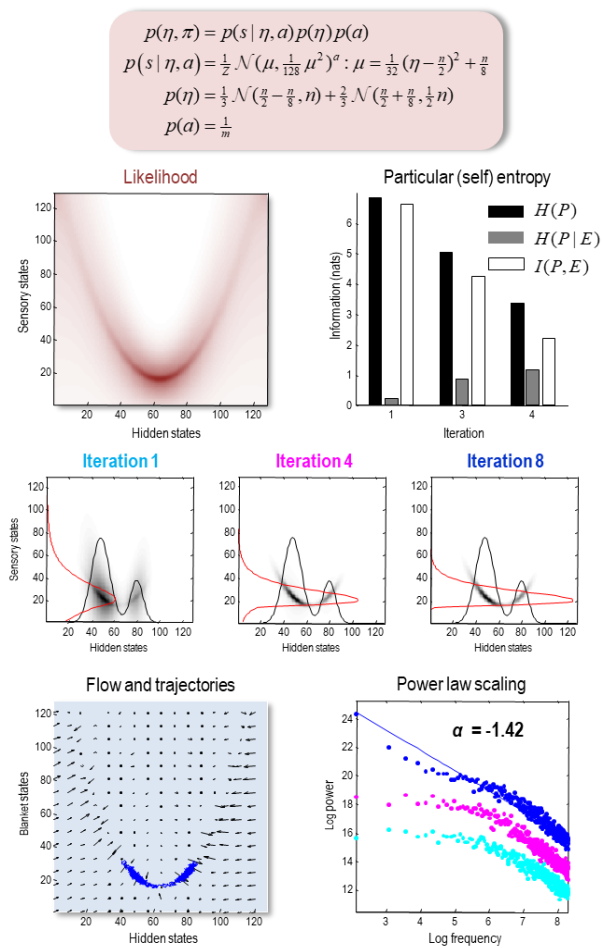

**FIGURE 2**

*Self-evidencing, self-organisation and scale-free dynamics*. This figure illustrates the itinerant, scale-free dynamics that emerge with the reduction of self-entropy – and the mutual information between external and blanket states. This simple example starts with a fixed-form likelihood (upper left panel) mapping from external or hidden states to a particle's states. For simplicity, there are no internal states in this simulation, so the particle is constituted by blanket states. Here, one hidden state maps to a blanket state via a nonlinear (quadratic) function with a state-dependent dispersion (modelled with a quadratic function of the hidden state). The self-entropy was minimised with respect to the marginal density over hidden states (using a gradient descent). The ensuing joint distributions are shown in the middle row, as mutual information decreases over iterations of the gradient descent. The black and red lines correspond to the marginal densities over hidden and blanket states, respectively. The key observation here is that the joint distribution progressively concentrates probability mass in regions that reduce mutual information, while avoiding regions with a high conditional uncertainty over sensory states. This is reflected in the mutual information measures shown in the upper right panel. In this example, decreases in *self-entropy* entail a decrease in mutual information (i.e., expected complexity or *risk*), with slight increases in conditional uncertainty (i.e., *ambiguity*). Given the joint distribution, one can derive the flow and solve for particular trajectories (here, over $2^{20}$ time steps). The lower left panel shows the flow in terms of a quiver plot and part of the trajectory (over $2^{10}$ time steps, assuming the amplitude of random fluctuations was unity with solenoidal flow of one quarter). This segment was chosen to illustrate noise-induced tunnelling; i.e., a trajectory that connects the two regimes of the attracting set. Technically, this is known as an *instanton* (Ginzburg, 1987). The associated dynamics show itinerancy as the trajectory wanders within the attracting set. The ensuing scale-free behaviour is illustrated in the lower right panel in terms of a power law. Here, the log spectral density has been plotted against log frequency; showing a roughly linear relationship with a power law exponent of -1.42. Technically, this corresponds to a noised induced (*N*–phase) breaking of topological supersymmetry or, more colloquially, noise induced tunnelling from one regime of the attracting set to another. Although this scale-free dynamics resembles anomalous diffusion – usually associated with non-extensive (or non-Gaussian) dynamics; e.g., (Pavlos et al., 2012) – it emerges from extensive (Gaussian) dynamics, under a joint density that was optimised to induce self-organising flow. The cyan and magenta dots correspond to the equivalent simulation at earlier iterations, to illustrate the emergence of power law scaling. These simulations used a discretised state-space of 128 bins for each of the two states. The trajectory was integrated using a simple Euler scheme.

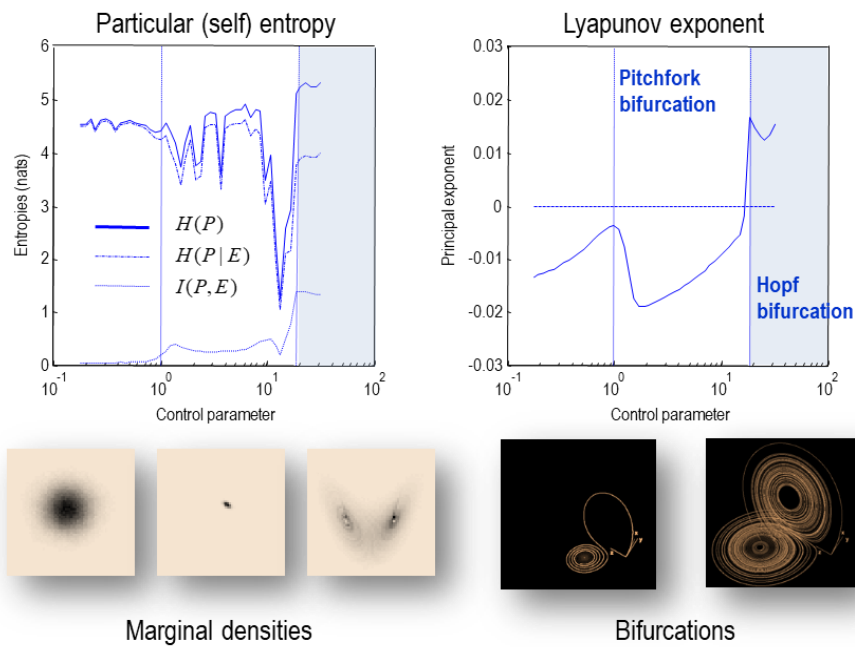

**FIGURE 3**

*Symmetry breaking and bifurcations*. This figure illustrates the relationship between particular (i.e., self) entropy and the exponential divergence of trajectories that underlies symmetry breaking in the (stochastic) Lorentz system (Agarwal and Wettlaufer, 2016; Lorenz, 1963). This example illustrates how entropy changes with bifurcations and their associated (random) attracting sets. Here, we integrated a Lorenz system (for $2^{18}$ time steps of 1/64 units of time) with increasing values of the Rayleigh (control) parameter that, in deterministic systems, induces a pitchfork bifurcation (at $\rho = 1$) and a subsequent (subcritical) Hopf bifurcation (at $\rho = 24.74$). Because we added small (standard deviation of 1/128) random fluctuations to the flow, noise induced topological symmetry breaking emerges around $\rho = 14$ (typically, in a regime of chaotic transients). The exponential divergence of trajectories was measured with the maximal Lyapunov exponent (averaged over the orbits). The resulting changes in the Lyapunov exponent (right panel) cause the attracting set to become space-filling (see insets beneath), with associated changes in self-entropy and mutual information (left panel). Here, we exploited the fact that the flow of the Lorenz system has one missing link (from the third to the first state). This means we can associate the first state with an active state, the second with a sensory state and the last with an external or hidden state (note there are no internal states here and the particular states reduce to blanket states). The remarkable result in this simulation is that the highest degree of self-organisation

– as reflected in the entropy of the (active and sensory) blanket states shows a profound dip just before the onset of stochastic chaos – in a regime associated with critical slowing in deterministic systems. The insets illustrate the bifurcations and attracting sets, in terms of illustrative trajectories (right) and associated ensemble densities (left), arbitrarily rescaled to the minimum and maximum values of the trajectory. Cardinal bifurcations are illustrated with vertical lines, while the horizontal line indicates when the principal Lyapunov exponent first crosses zero at the onset of (stochastic) chaos. The mutual information measures (between the first and remaining states) were evaluated using a discretisation of state-space into 32 bins in each of the three dimensions. This example is offered as a numerical study to illustrate that a simple definition of self-organisation – in terms of the entropy of blanket (i.e., particular) states – has some construct validity in relation to self-organised criticality in stochastic chaos: a validation that may or may not generalise. Note that the dynamics illustrated here are not deterministic because each state is equipped with random fluctuations. We will see later that this means each state of the Lorenz attractor is modelled as a mixture of blanket states at lower scale – that inherit fast fluctuations from their internal states.

## Self-organisation and information length

So far, we have considered self-organisation in terms of particular entropy, where a low entropy appears to go hand-in-hand with the phenomenology of symmetry breaking and self-organised criticality. This begs the question: how can one quantify itinerancy of this sort? One approach borrows the notion of information length; namely, the number of discernible probabilistic configurations a system passes through *en route* to nonequilibrium steady-state. In other words, one can associate itinerant symmetry breaking (of the sort seen in biological systems) with long information lengths from a particular state to nonequilibrium steady-state. The use of information length eludes difficult questions about the meaning of high versus low self-entropy, which is only defined to within an additive constant (Jones, 1979). Conversely, the information length is a *metric* that can be applied to any density dynamics to score the wandering, itinerant dynamics we are trying to quantify.

To build an intuition about information length, Figure 4 shows three examples that illustrate the role of flows and random fluctuations in generating itinerant but structured dynamics. In this illustration, we use the Lorentz system of Figure 3 to show the different ways in which an initial density – given a particular state – converges to the nonequilibrium steady-state density. The upper panel shows the familiar self-organisation induced by the Lorentz attractor, using low amplitude random fluctuations. In this regime, the system has undergone a Hopf bifurcation, guaranteed by using a Rayleigh parameter of 28. The evolution of the initial density was evaluated in terms of: (i) its KL divergence from the NESS density $D(\tau)$ – and (ii) the difference in information length between the density over time and the final (steady-state) density $\Delta(\tau)$.

$$D(\tau) = D[\,p(x, \tau \mid \pi_0)\,\|\,p(x, \infty \mid \pi_0)\,]$$
$$\delta(\tau)^2 \leq \tfrac{1}{2} D(\tau)$$

$$\Delta\ell(\tau) = \ell(\tau) - \ell(\infty) \tag{2.9}$$
$$\tfrac{1}{2} d\ell(\tau)^2 = D[\,p(x, \tau + d\tau \mid \pi_0)\,\|\,p(x, \tau \mid \pi_0)\,]$$

$$D(\tau) = 0 \Leftrightarrow \Delta\ell(\tau) = 0 \Leftrightarrow \delta(\tau) = 0 \Leftrightarrow d\ell(\tau) = 0$$

The inequality above is known as Pinsker's inequality, where, $\delta(\tau)$ is called *total variation distance* (Rényi, 2007) and is upper bounded by $D(\tau)$, which we will refer to as *divergence length*. Recall from (1.18) that the characterisations of statistical distances $D(\tau)$ and $\Delta\ell(\tau)$ are related but differ in their use of KL divergences. The path length is an accumulation of divergences over small increments over time – to ensure the information length is a (Riemannian) measure of distance. Conversely, the divergence length between initial and final densities is not. The final expression in (2.9) says that after a sufficient period of time the density 'forgets' about the particular state it started from; rendering increments in divergence and information length zero. Conversely, particles that 'remember' their initial state have a long information length with itinerant density dynamics.

A long information length means, effectively, the initial density is a long way away from the final density and therefore convergence takes longer (indicated by the small blue arrows in Figure 4). In this example, it takes about eight seconds before convergence to steady state. This can be contrasted with the lower panels that illustrate examples of fast convergence; meaning that the initial densities have a short information length.

As one might intuit from inspection of (1.1), there are two ways to reduce information length. First, one can increase the amplitude of random fluctuations while keeping the flow fixed. This enables trajectories to explore state-space quickly and find their attracting set from any initial density. Panel B illustrates this by increasing the amplitude of random fluctuations – by decreasing their log precision from 8 to 0. This markedly attenuates information length, such that convergence to nonequilibrium steady-state takes less than a second. Alternatively, one can change the flow, without changing the amplitude of random fluctuations. The example in panel C features the same reduction of information length – and accompanying KL divergence – when the Rayleigh parameter was reduced from 28 to 1. In this flow regime, the Lorentz attractor becomes a point attractor and the itinerancy due to stochastic chaos is lost (see Figure 3).

Generally speaking, the information length preserves the linearity of the system's dynamics. For example, with linear flow we have a familiar Ornstein–Uhlenbeck process, where information length decreases with the amplitude of random fluctuations. Following (Kim, 2018):

$$f(\pi) = -\gamma\pi + \omega$$
$$\Rightarrow \ln\ell(\infty) = \ln|\pi_0| + \tfrac{1}{2}\ln\gamma - \tfrac{1}{2}\ln\Gamma \tag{2.10}$$
$$\Rightarrow \frac{\partial\ln\ell(\infty)}{\partial\ln|\pi_0|} = \nu = 1$$

In contrast, nonlinear flow changes the linear scaling of geometric structure to produce power-law scaling $\nu \neq 1$ characteristic of symmetry breaking, itinerancy and self-organised criticality[7]. An interesting aspect of the numerical analyses of chaotic systems in (Kim, 2018) is the dependency of information length on the initial state, where unstable or critical points have the shortest information length. From the numerical analysis in Figure 3, one might picture noise-induced tunnelling from unstable points as mediating 'shortcuts' to nonequilibrium steady-state. The linear case suggests that as random fluctuations attain large amplitudes, all initial conditions are drawn close to steady-state and, by implication, self-organisation to nonequilibrium steady-state is (almost) instantaneous (c.f., Panel B of Figure 4). Later, we will associate this sort of behaviour with small (quantum) particles.

In Part Three, we will look more closely at self-organisation in systems that have a long information length – and contrast these particles with short information length systems, such as quantum and other small particles (e.g., viruses). On this view, information length distinguishes between the *simple*, fast, 'hot' self-organisation of small (quantum) particles and the *itinerant*, slow, 'cold' behaviour of large (classical) particles.

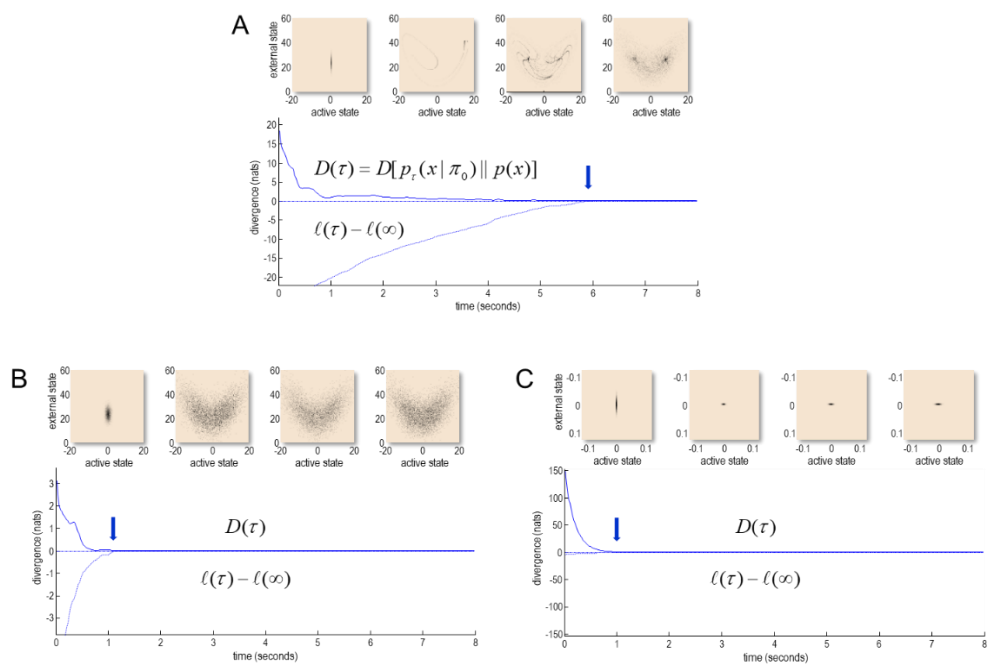

*Convergence to nonequilibrium steady-state*. **A**: this is a simple demonstration of (nearly) deterministic convergence to nonequilibrium steady-state, using the Lorenz system of Figure 3. Deterministic solutions (with a Rayleigh parameter of 28) were obtained for 8192 initial states, integrating over eight seconds (with a time step of 1/64 seconds and low amplitude random fluctuations with a log precision of eight). The initial particular states were the same for each solution and yet their final density converges to non-equilibrium steady-state over time. This manifests as a collapse in the divergence between the sample densities and final (NESS) density – as evaluated using a Gaussian approximation to the ensemble densities at each point in time. The upper insets show the propagated sample densities at four points in time. As time progresses, this density comes to assume the familiar butterfly form of the Lorenz attractor. However, these solutions are not trajectories through state-space, they are the endpoints of paths from an ensemble of starting locations (shown in the right plot). For comparison, convergence is also shown in terms of relative information length. One can see that information length effectively stops increasing when the divergence is zero. The lower panels show the same simulations but using random fluctuations with a log precision of zero (i.e., a variance of one) – panel **B** – and a Rayleigh parameter of one – panel **C**. In these illustrations, we treated the first state of the Lorenz system as the active state, the second state constituted the sensory state and the third state played the role of an external or hidden state. This designation is based upon the fact that the first state is not influenced by the third. This numerical example shows how uncertainty about external states is propagated over time to induce uncertainty about a particle's state; even when the initial (particular) state is known.

---

[7] See Kim, E.-j., 2018. Investigating Information Geometry in Classical and Quantum Systems through Information Length. Entropy 20, 574. for further discussion and an (entertaining) application of information length to music.

## Summary

The treatment in this section suggests that self-organising systems reduce their self-entropy, to the extent allowed by coupling to external states (or other Markov blankets) and random fluctuations. Using information theory, it is possible to interpret this kind of behaviour in terms of statistical imperatives; namely, the avoidance of complexity cost (i.e., *risk*) and inaccuracy (i.e., *ambiguity*). The Nirvana of *simple* self-organisation is a complete resolution of particular or self-entropy: a trivial solution here would be when a particular density collapses to a point mass (i.e., a delta function). Although we are not interested in these simple solutions, it is interesting to reflect that small particles may be trying to get back to how the universe started. A more interesting example of *itinerant* self-organisation is provided in Figure 4. In Part Three, we will return to the emergence of self-organisation and symmetry breaking. In brief, we will see that systems with these characteristics can always be construed as engaging in something called *active inference* (a.k.a. *self-evidencing*) via a minimisation of a variational free energy.

Table 1 provides a summary of the information measures introduced in the section, which we will refer to later. However, before dealing with the sentient foundations of self-organisation, we will spend some time unpacking the NESS lemma (Appendix B) in terms of Markov blankets (in the remainder of Part One) and its relationship to quantum, statistical and classical mechanics (in Part Two) – to contextualise later treatments of active inference (in Part Three).

Information measures of particular states that characterise self-organisation

| Measure | Definition | Comments |
|---------|-----------|----------|
| Self-information | $\Im(\pi) = -\ln p(\pi)$ | A.k.a. surprise, surprisal or negative log-evidence, where evidence is also known as the marginal likelihood |
| Self-entropy | $H[P] = E_{p(\pi)}[\Im(\pi)]$ | The entropy of particular states |
| Complexity | $D[p(\eta \mid \pi) \parallel p(\eta)]$ | The divergence between the posterior and prior over external (i.e., hidden) states |
| Risk (expected complexity) | $I(E,P) = E_{p(\pi)}[D[p(\eta \mid \pi) \parallel p(\eta)]]$ | The expected complexity or mutual information between external and particular states |
| Accuracy | $E_{p(\eta,\pi)}[\ln p(\pi \mid \eta)]$ | The expected log likelihood of particular states |
| Ambiguity (expected inaccuracy) | $H(P \mid E) = E_{p(\eta,\pi)}[\Im(\pi \mid \eta)]$ | Negative expected accuracy or log likelihood. This is the conditional entropy of particular states given external states |
| Information gain | $D[p(\eta \mid \pi) \parallel p(\eta \mid \alpha)]$ | A relative entropy, a.k.a. intrinsic value, salience and epistemic value. |
| Expected information gain | $I(E,S \mid A) = E_{p(\pi)}[D[p(\eta \mid \pi) \parallel p(\eta \mid \alpha)]]$ | Expected information gain or mutual information between sensory and external states, conditioned on active states |

# Synthetic soups and active matter

In this section, we describe an exemplar system that will be used to illustrate crosscutting themes in subsequent sections. Here, it is used to simulate a primordial soup – to illustrate the emergence of self-organisation in terms of Markov blankets and internal states. This soup or *active matter* (Ramaswamy, 2010), comprises an ensemble of particles that are coupled through short range interactions. Each particle corresponds to the Lorentz system of the previous section that has been 'dressed' with blanket states to create an internal state – and enable interactions among particles. The resulting simulations are similar to those used to characterise pattern formation in dissipative systems; for example, Turing instabilities (Turing, 1952) and other dissipative structures in nonequilibrium systems, such as turbulence and convention in fluid dynamics (e.g., Bénard cells) or percolation in reaction-diffusion systems such as the Belousov-Zhabotinsky reaction (Belousov, 1959). In our case, we can treat our system is an ensemble of macromolecules; however, the details of the simulation are not important, similar results would be obtained with any coupled random dynamical system. The description below summarises the material in (Friston, 2013), where interested readers can find more details.

## An active soup

To simulate the emergence of a Markov blanket, each constituent of the ensemble or $i$-th macromolecule was equipped with notional Newtonian and electrochemical states, $\{b_n^{(i)}, b_e^{(i)}\}$. Here, $b_n^{(i)} = \{a_n^{(i)}, s_n^{(i)}\}$ can be considered coordinates of motion; e.g., position and velocity, while $b_e^{(i)} = \{a_e^{(i)}, s_e^{(i)}, \mu^{(i)}\}$ could correspond to electrochemical states; e.g., concentrations or electromagnetic states. The electrochemical dynamics of each macromolecule was chosen to have a Lorenz attractor, which provides a ubiquitous model of itinerant systems; e.g. in electrodynamics, lasers and chemical reactions (Poland, 1993). Figure 5 provides the summary of the dynamics. Specifically, the Langevin equation for the $i$-th macromolecule is:

$$\begin{bmatrix} \dot{s}_e^{(i)} \\ \dot{a}_e^{(i)} \\ \dot{\mu}^{(i)} \end{bmatrix} = \begin{bmatrix} 10(a_e^{(i)} - s_e^{(i)}) + \mathbf{s}_e^{(i)} \\ 32 \cdot s_e^{(i)} - a_e^{(i)} - \mu^{(i)} s_e^{(i)} \\ s_e^{(i)} a_e^{(i)} - \frac{8}{3} \mu^{(i)} \end{bmatrix} \cdot \kappa^{(i)} + \omega_e \tag{3.1}$$

$$\mathbf{s}_e^{(i)} = \sum_{j=\{j:\Delta_{ij}<1\}} s_e^{(i)}$$

$$\Delta_{ij} = |a_n^{(j)} - a_n^{(i)}|$$

Here, changes in electrochemical states are coupled through the local average $\mathbf{s}_e^{(i)}$ of the states of other macromolecules that lie within a distance of one unit. This means $\Delta$ can be regarded as an adjacency matrix that encodes the dependencies among the electrochemical states of the ensemble. Crucially, this means electrochemical coupling depends upon the spatial relationships among the macromolecules. The corresponding rate parameters $\kappa^{(i)} = \frac{1}{32}(1 - \exp(-4 \cdot u))$; where $u \in (0,1)$ was selected from a uniform distribution to ensure topological symmetry breaking.

Similarly, the (Newtonian) motion of each macromolecule depends upon the electrochemical state of its neighbours

$$\dot{a}_n^{(i)} = (1 + \tfrac{1}{64}\mu^{(i)})s_n^{(i)} + \omega_n$$

$$\dot{s}_n^{(i)} = 2F^{(i)} - 8s_n^{(i)} - a_n^{(i)} + \omega_n$$

$$(3.2)$$

$$F^{(i)} = \sum_{j=\{j:\Delta_{ij}<1\}} \Delta_{ij}\left( \frac{8\exp(-|a_e^{(j)} - a_e^{(i)}|) - 4}{\Delta_{ij}^2} - \frac{1}{\Delta_{ij}^3} \right)$$

This motion rests on forces $F^{(i)}$ exerted by other macromolecules that comprise a strong repulsive force (with an inverse square law) and a weaker attractive force that depends on electrochemical states. This force was chosen so that macromolecules with coherent electrochemical states are attracted to each other but repel otherwise. The remaining two terms in the second equality represent viscosity that depends upon velocity and an exogenous force that attracts all macromolecules to the origin – as if they were moving in a simple (quadratic) potential energy well. This ensures the synthetic soup falls to the bottom of the well. We now take a closer look at the self-organisation that emerges under these equations of motion.

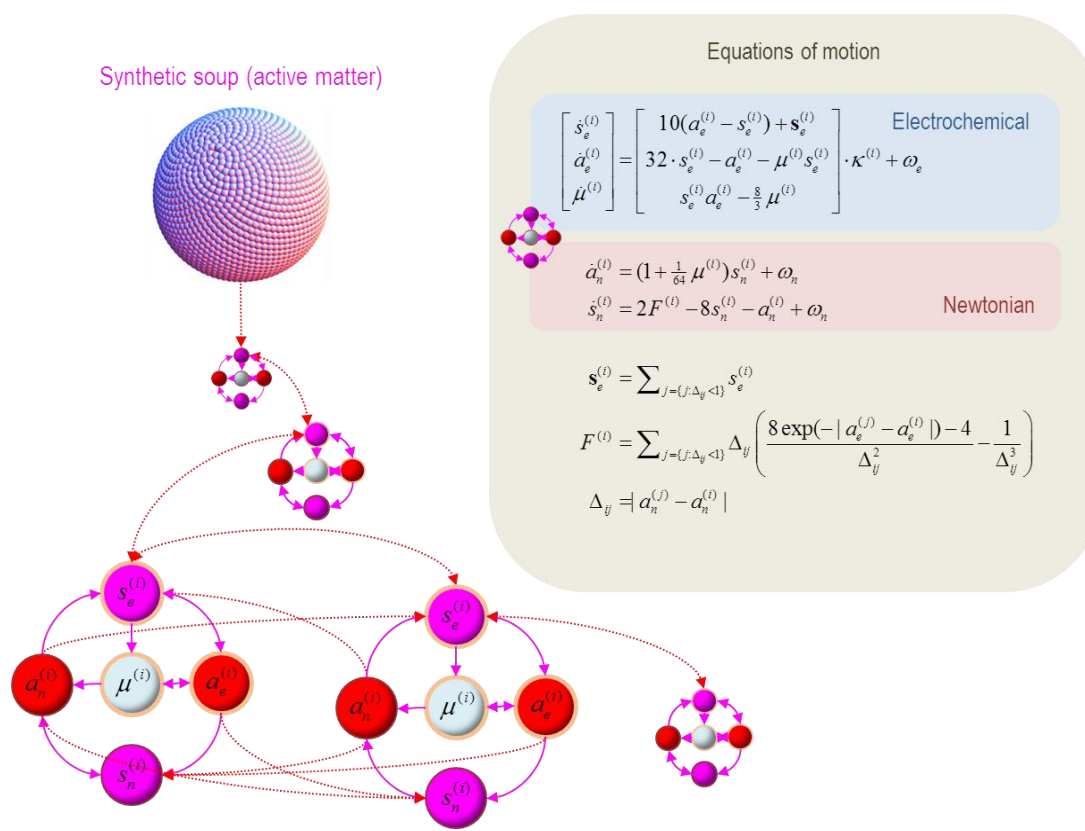

**FIGURE 5**

*Synthetic soups and active matter.* This graphic describes the equations of motion used to simulate coupled (random) dynamical systems (i.e., particles) to illustrate self-organisation. The equations describe the dynamics (that have been separated into electrochemical and Newtonian components). The schematics illustrate the conditional dependencies among particles, where each particle comprises its Markov blanket and internal states. The states with orange outlines are electrochemical states and the remaining pair constitute the Newtonian states. Note that active states (red circles) play the role of position, while sensory states (magenta circles) become velocity that depends on active states. These roles of active and sensory states will figure later, when we consider classical mechanics.

## A random dynamical attractor and its Markov blankets

In the simulations below an ensemble of 128 particles (i.e., macromolecules) were integrated using Euler's (forward) method with step sizes of 1/512 seconds and initial conditions sampled from a normal distribution. By adjusting the parameters in the equations of motion (3.1) and (3.2), one can produce a repertoire of plausible and interesting behaviours (the code for these simulations and the figures in this monograph are available as part of the SPM academic software – see software note). These behaviours range from gas-like behaviour (where particles occasionally get close enough to interact) to a cauldron of activity, when particles are forced together at the bottom of the potential well. In this regime, macromolecules are sufficiently close for the inverse square law to blow them apart. In other regimes, a more crystalline structure emerges with muted interactions.

However, for most values of the parameters, weakly mixing behaviour emerges, as the ensemble approaches its random global attractor (usually after about 1000 seconds). Generally, macromolecules repel each other initially and then fall back towards the centre, finding each other as they coalesce. Local interactions then mediate a self-organisation, in which particles are passed around (sometimes to the periphery) until neighbours jostle comfortably with each other. In brief, the motion and electrochemical dynamics look like an active, restless soup – but does it contain a Markov blanket?

## The Markov blanket

Because the structural and functional dependencies share the same adjacency matrix – which depends upon position – one can use the adjacency matrix to identify the principal Markov blanket using spectral graph theory: the Markov blanket of any subset of states encoded by a binary vector with elements $\chi_i \in \{0,1\}$ is given by $[\mathbf{B} \cdot \chi] \in \{0,1\}$, where the Markov blanket matrix $\mathbf{B} = \mathbf{A} + \mathbf{A}^T + \mathbf{A}^T\mathbf{A}$ encodes the children, parents and parents of children. The principal eigenvector of the (symmetric) Markov blanket matrix will – by the Perron–Frobenius theorem – contain positive values. These values reflect the degree to which each state belongs to the cluster that is most densely coupled. In what follows, the internal particles (i.e., macromolecules) were the particles with the $k = 8$ largest values. Having identified internal particles, the Markov blanket can be recovered from the Markov blanket matrix using $[\mathbf{B} \cdot \chi]$ and divided into sensory and active particles – depending upon whether they are influenced by the external particles or not.

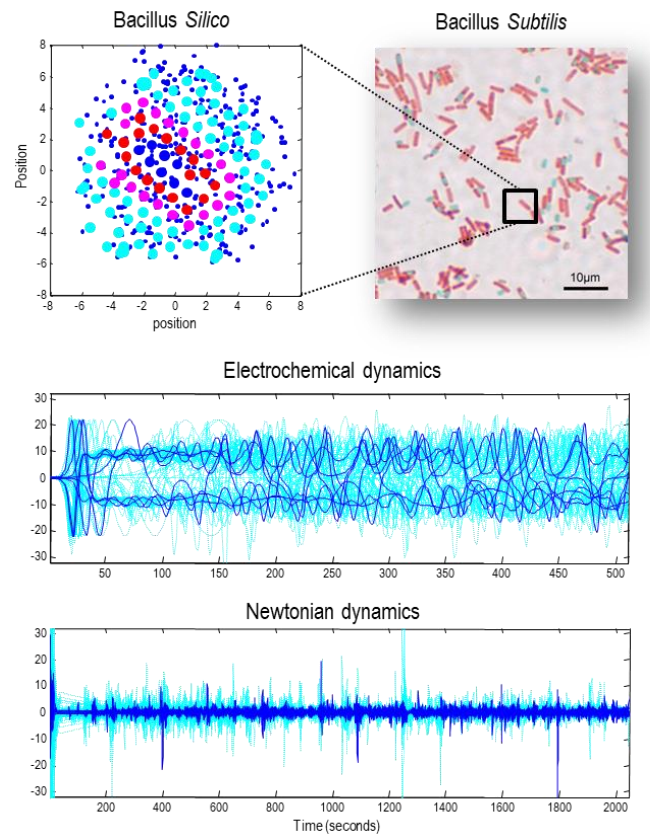

**FIGURE 6**

*Ensemble dynamics and self-organisation*. The upper panels show the position of (128) macromolecules comprising an ensemble, after 2048 seconds. The upper left panel shows the dynamical status (three blue dots per macromolecule) of each particle centred on its location (larger dots). The ensemble of macromolecules has been partitioned into external or hidden (cyan), sensory (magenta), active (red) and internal (blue) particles. The upper right panel is an image of an endospore stain of sporulating *B. Subtilis*. This graphic illustrates the spatiotemporal scale at which the simulations could be operating. The lower panels show the evolution of electrochemical (middle panel) and spatial (lower panel) states of each particle as a function of time. The (electrochemical) dynamics of the internal (blue) and external (cyan) states are shown for 512 seconds. The lower panel shows the position of internal (blue) and external (cyan) states over the entire simulation period. These simulations are solutions of the stochastic differential equations in the main text – using a forward Euler method with 1/512 second time steps and random Gaussian fluctuations with a standard deviation of an eighth.

## The emergence of order

Given the internal particles and their Markov blanket, we can now follow the assembly of constituent macromolecules and visualise their trajectories. The upper panels of Figure 6 show the position of (128) macromolecules comprising the ensemble. The upper left panel shows the electrochemical status (three blue dots per macromolecule) of each macromolecule centred on its location (larger dots) at the end of the simulation. The ensemble has been partitioned into external or hidden (cyan), sensory (magenta), active (red) and internal (blue) particles. It can be seen that the resulting Markov blanket surrounds a rod-like structure (i.e., Bacillus) of internal particles. Interestingly, the active macromolecules support the sensory macromolecules that are exposed to external particles. This is reminiscent of a biological cell with a cytoskeleton of active molecules (e.g., actin filaments), which are surrounded by sensory molecules (e.g., a cell surface). The upper right panel is an image of an endospore stain of sporulating *B. Subtilis*. This graphic illustrates the spatiotemporal scale at which we can imagine the simulations are operating. The lower panels show the evolution of electrochemical (middle panel) and Newtonian (lower panel) particular states as a function of time. One can see initial (chaotic) transients that resolve fairly quickly, with itinerant behaviour as they approach their attracting set. The lower panel shows the position of internal (blue) and external (cyan) particles over the entire simulation period.

Notice that something quite subtle is going on here. We started with an ensemble of particles (e.g., macromolecules), where each particle was characterised in terms of particular (i.e., sensory, active and internal) states. We then ended up with a single particle (e.g., a Bacillus or virus) characterised in terms of particular (i.e., external, sensory, active and internal) particles. In short, we have moved from a microscopic to a macroscopic scale, with blanket states at both. The next section looks more closely at this move. Here, we simply note that a macroscopic Markov blanket has emerged from simple self-organisation. So, what licenses us to describe the microscopic dynamics as self-organisation?

Figure 7 demonstrates microscopic self-organisation in terms of the particular entropy of the ensemble's particles – and concomitant changes in terms of mutual information (i.e., complexity cost or risk) and conditional entropy (i.e., ambiguity). Here, the ensemble averages of these (relative) entropy measures were taken over all (128) macromolecules; where the Markov blanket of each particle comprises all but the third (electrochemical) hidden state. This information theoretic characterisation discloses, as expected, a monotonic decrease in particular entropy (and complexity cost) as the ensemble approaches its random dynamical attractor.

## Summary

In summary, this section has described a somewhat arbitrary random dynamical system comprising an ensemble of particles, each with several dynamical states (three electrochemical and two describing position and velocity). Crucially, the flow or equations of motion were constructed to make electrochemical coupling among the simulated macromolecules depend upon position – and render their velocity dependent upon electrochemical states. This endows the ensemble with a dynamic and sparse coupling that readily enables the emergence of a Markov blanket; separating internal from external particles (and their constituent states). In this example, the internal particles (and Markov blanket) can be thought of as modelling a little virus-like particle or rod-like bacterium. We now have at hand an *in-silico* creature. Later, we will examine this synthetic creature to see whether the states of internal particles (e.g., intracellular electrochemical states) plausibly infer or represent the states of external particles (e.g., extracellular motion); much as real organisms do. However, first, we need to understand how a Markov blanket emerged from the coupling of particles that were themselves constituted by Markov blankets (and their internal states).

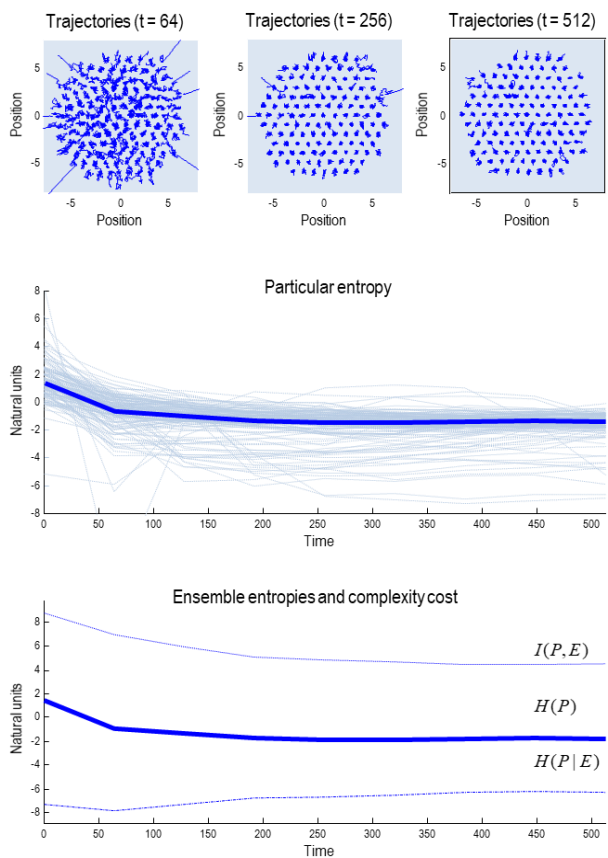

**FIGURE 7**

*Self-organising soups*. This demonstration uses an ensemble of particles with intrinsic (Lorentz attractor) dynamics and (Newtonian) short-range coupling to illustrate self-organisation in terms of particular (i.e., self) entropy and concomitant changes in terms of mutual information (i.e., complexity cost or risk). Here, the ensemble averages of these (relative) entropy measures were taken over all (128) particles; where the Markov blanket of each particle comprises all but the third (electrochemical) hidden state. The lower panels illustrate the decrease in blanket entropy (and complexity cost) as the system approaches its random dynamical attractor – shown as the thick and thin solid lines, respectively. The lowest broken line corresponds to conditional entropy (i.e., ambiguity). Illustrative trajectories of the particles are provided at three points during the (stochastic) chaotic transient in the upper three panels. These relative entropy changes can be compared with the equivalent results in Figure 2 for a single particle.

# States, particles and fluctuations

Let us return to where we started; namely, the Langevin equation (1.1) and ask a simple question: what is the difference between a state and a fluctuation? The answer offered in this section is that fluctuations are just fast states that change so quickly we can ignore their temporal correlations – and adopt the usual Wiener assumptions. This distinction highlights a key tenet of what is to follow; namely, a separation of temporal scales that licenses an adiabatic assumption, allowing one to separate slowly changing states from fast fluctuations. Now, let us ask a more fundamental question: what is a state? This question can be dissolved by appealing to an infinite regress along the following lines:

- *What is a state? A state is an eigenstate of a particle's Markov blanket.*
- *What is a particle? A particle is a set of particular states comprising blanket and internal states.*
- *What is a state? A state is … and so on.*

An eigenstate here refers to the expression of an eigenmode of blanket states; namely, the principal eigenvectors of their Jacobian (i.e., rate of change of flow with respect to state). These mixtures are formally identical to *order parameters* in synergetics that reflect the amplitude of slow, unstable eigenmodes (Haken, 1983). In terms of centre manifold theory, they correspond to solutions on the slow (unstable or centre) manifold (Carr, 1981; Davis, 2006).

In brief, the Markov blanket of a particle constitutes a set of vector states, whose eigenstate subtends blanket or internal states at the scale above. Note that the eigenstates are always mixtures of blanket states at the lower scale, while the eigenstates can be blanket or internal states at the higher scale. This follows from the fact that the only states 'that matter' are those that influence other (blanket) states. In other words, the only relevant coupling is between blanket states[8]. Effectively, all we are doing here is applying the slaving principle, or centre manifold theorem (Haken, 1983), recursively to Markov blankets of Markov blankets. A complementary perspective is provided by renormalisation group approaches (Cardy, 2015; Schwabl, 2002), where the following could be seen as an attempt to establish the universality of states (and fluctuations), in the sense of constituting universality classes. The final section of Part One unpacks this construction analytically (and with numerical simulations).

## Starting at the end

At a given scale or level (*i*) of description, we can entertain the following ansatz: a random dynamical system can be characterised as coupled subsets of states, where the *n*-th subset $x_n^{(i)} \subset x^{(i)}$ constitutes the vector state of a *particle* or nonlinear oscillator:

---

[8] Relevant in the sense of renormalisation group theory: Schwabl, F., 2002. Phase Transitions, Scale Invariance, Renormalization Group Theory, and Percolation, Statistical Mechanics. Springer Berlin Heidelberg, Berlin, Heidelberg, pp. 327-404.

$$\dot{x}_n^{(i)} = f_n^{(i)} + \sum_m \lambda_{nm}^{(i)} x_m^{(i)} + \omega_n^{(i)}$$

$$x^{(i)} = \{x_1^{(i)}, \ldots, x_N^{(i)}\}$$

(4.1)

$$E[\omega_n^{(i)}(\tau) \cdot \omega_m^{(i)}(\tau')] = \begin{cases} 2\Gamma_n^{(i)} \delta(\tau - \tau') & n = m \\ 0 & n \neq m \end{cases}$$

The equations of motion for the states of the $n$-th particle comprise some baseline flow (at the current point in phase-space) and intrinsic and extrinsic components determined by the states of the particle in question and other particles, respectively. In this form, the diagonal elements of the coupling matrix, $\lambda_{nn}^{(i)} \in \mathbb{C}$, determines the frequency and decay of oscillatory responses to extrinsic perturbations and random fluctuations. In what follows, we will see that (4.1) leads to an isomorphic expression for states of particles at a higher (macroscopic) scale. See Figure 8 for a schematic summary of this recursive induction.

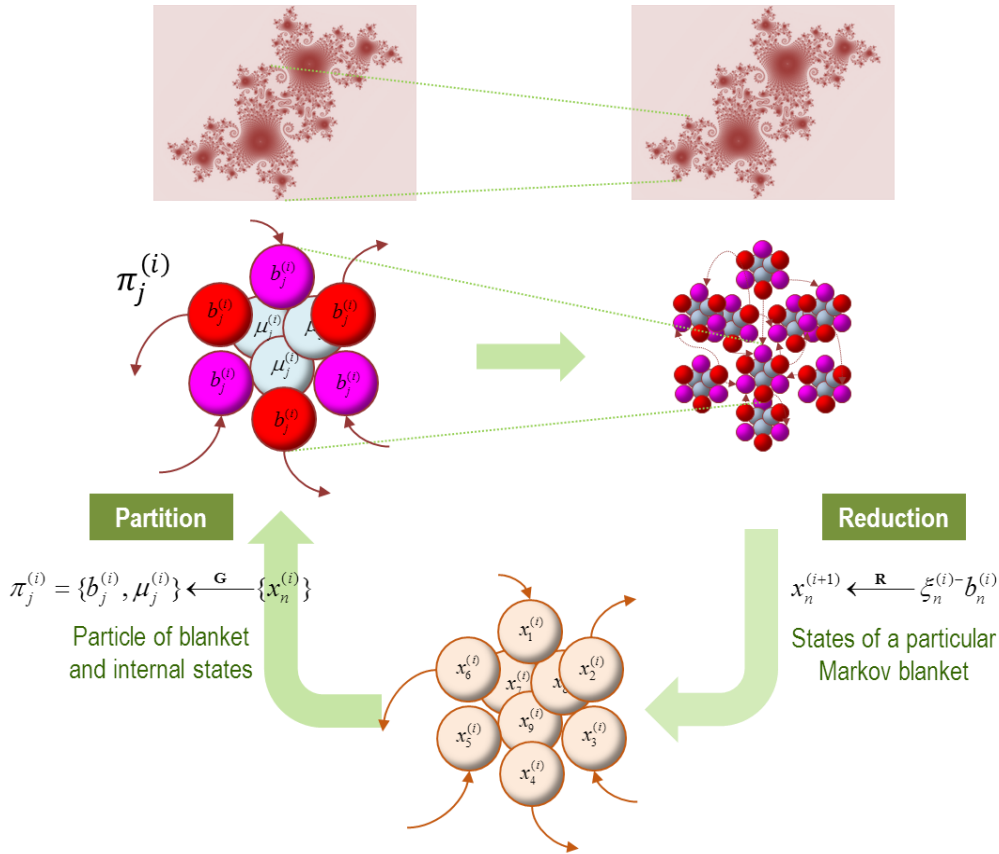

**FIGURE 8**

*Blankets of blankets*. This schematic illustrates the recursive procedure by which successively larger (and slower) scale dynamics arise from subordinate levels. At the bottom of the figure (lower panel), we start with an ensemble of vector states (here nine). The conditional dependencies among these vector states (i.e., eigenstates) define a particular partition into particles (upper panels). Crucially, this partition equips each particle with a bipartition into blanket and internal states, where blanket states comprise active (red) and sensory states (magenta). The behaviour of each particle can now be summarised in terms of (slow) eigenmodes or mixtures of its blanket states – to produce eigenstates at the next level or scale. These constitute an ensemble of vector states and the process starts again. Formally, one can understand this in terms of coarse graining the dynamics of a system via two operators. The first uses the particular partition to *group* subsets of states ($\mathbf{G}$), while the second uses the eigenmodes of the resulting blanket states to *reduce* dimensionality ($\mathbf{R}$). The upper panels illustrate the bipartition for a single particle (left panel) and an ensemble of particles; i.e., the particular partition *per se* (right panel). The insets on top illustrate the implicit self-similarity of particular dependencies pictorially, in moving from one scale to the next. Please see the main text for a definition of the variables used in this figure.

## The Markovian partition

If the extrinsic coupling has a nontrivial sparsity, $\lambda_{nm}^{(i)} = 0 : \exists (n, m)$, we can partition $N^{(i)}$ states into $J^{(i)}$ particles (i.e., blankets and internal states). The flow of the states comprising the *j*-th particle must have the following form, by the marginal flow lemma:

$$\dot{\pi}_j^{(i)} = \begin{Bmatrix} \dot{a}_j^{(i)} \\ \dot{s}_j^{(i)} \\ \dot{\mu}_j^{(i)} \end{Bmatrix} = \begin{Bmatrix} f_{a_j}^{(i)}(b_j^{(i)}, \mu_j^{(i)}) \\ f_{s_j}^{(i)}(b_1^{(i)}, \dots, b_J^{(i)}) \\ f_{\mu_j}^{(i)}(b_j^{(i)}, \mu_j^{(i)}) \end{Bmatrix} + \begin{Bmatrix} \omega_{a_j}^{(i)} \\ \omega_{s_j}^{(i)} \\ \omega_{\mu_j}^{(i)} \end{Bmatrix}$$

$$= \begin{Bmatrix} \dot{b}_j^{(i)} \\ \dot{\mu}_j^{(i)} \end{Bmatrix} = \begin{Bmatrix} f_{b_j}^{(i)}(\mu_j^{(i)}, b_1^{(i)}, \dots, b_J^{(i)}) \\ f_{\mu_j}^{(i)}(\pi_j^{(i)}) \end{Bmatrix} + \begin{Bmatrix} \omega_{b_j}^{(i)} \\ \omega_{\mu_j}^{(i)} \end{Bmatrix} \tag{4.2}$$

$$\pi^{(i)} = \{\pi_1^{(i)}, \dots, \pi_J^{(i)}\}$$
$$= \{x_1^{(i)}, \dots, \underbrace{\underbrace{\underbrace{x_k^{(i)}, \dots, x_\ell^{(i)}}_{a_j^{(i)}}, \underbrace{x_m^{(i)}, \dots, x_n^{(i)}}_{s_j^{(i)}}}_{b_j^{(i)}}, \underbrace{x_o^{(i)}, \dots, x_p^{(i)}}_{\mu_j^{(i)}}}_{\pi_j^{(i)}}, \dots, x_N^{(i)}\}$$

Here, active states depend only on the Markov blanket in which they participate and the internal states they surround. Similarly, the internal states depend only upon themselves and their Markov blanket. Conversely, the flow of sensory states depends upon all other states (apart from internal states that are sequestered behind Markov blankets). The partition implicit in the last equality emphasises the point that a (*particular*) Markovian partition is a partition into particles, where each particle is itself a partition of blanket and internal states.

Consider now the Taylor expansion of the flow of the *j*-th Markov blanket, where the intrinsic dynamics are absorbed into the random fluctuations. For notational simplicity, we will assume the current state constitutes the origin of the generalised coordinates: $x_0 \equiv x(0) = 0$. So that we can express everything in terms of local deviations:

$$\dot{b}_j^{(i)} = f_{b_j}^{(i)}(b_0^{(i)}) + \sum_k J_{jk} b_k^{(i)} + \ldots + K_j \varepsilon_j^{(i)} + \omega_{b_j}^{(i)}$$

$$\varepsilon_j^{(i)} = \mu_j^{(i)} - \mathbf{\mu}_j^{(i)}(b_j^{(i)})$$

$$J_{jk} \triangleq \partial_{b_k} f_{b_j}^{(i)}(b_k^{(i)}) \tag{4.3}$$

$$J_{jj} \triangleq \partial_{b_j} f_{b_j}^{(i)}(b_j^{(i)}, \mathbf{\mu}_j^{(i)}(b_j^{(i)}))$$

$$K_j \triangleq \partial_{\mu_j} f_{b_j}^{(i)}(\pi_j^{(i)})$$

In this expansion, the effect of the Markov blanket on its own flow is mediated directly – through interactions among active and sensory states – and vicariously through internal states. In other words, for every state of the Markov blanket, there is an expected internal state $\mathbf{\mu}_j^{(i)}(b_j^{(i)})$ that contributes to the flow of the blanket – more specifically, the active states: see (4.6) . This means that the contribution of internal states rests on fluctuations about their expectation. These *intrinsic fluctuations* $\varepsilon_j^{(i)}$ only affect the Markov blanket in question, because they are conditionally independent of external states (i.e., intrinsic fluctuations under other Markov blankets). This conditional independence means that the intrinsic fluctuations are unique to each blanket.

$$E[\varepsilon_j^{(i)}(\tau) \cdot \varepsilon_k^{(i)}(\tau)] = \begin{cases} \Sigma_j^{(i)} & : j = k \\ 0 & : j \neq k \end{cases} \tag{4.4}$$

By associating intrinsic fluctuations with random fluctuations, we require that they are independent from blanket to blanket *and* fluctuate quickly. The latter requirement is guaranteed to the extent that the expected internal state, conditioned on the blanket state, furnishes an unstable or centre manifold that attracts internal state trajectories at a rate that is substantially faster than flow on the manifold[9].

The Jacobians (i.e., rate of change of flow with respect to states) mediating the dynamics above respect the conditional independences implied by the Markov blanket; namely, active states cannot be influenced directly by external states (i.e., other Markov blankets) – and sensory states cannot be influenced directly by internal states. From (1.21):

$$J_{jk} \triangleq \partial_{b_k} f_{b_j} = \begin{bmatrix} 0 \\ \partial_{b_k} f_{s_j} \end{bmatrix} : j \neq k$$

$$K_j \triangleq \partial_{\mu_j} f_{b_j} = \begin{bmatrix} \partial_{\mu_j} f_{a_j} \\ 0 \end{bmatrix} : \forall j \tag{4.5}$$

---

[9] A more refined construction of an implicit centre manifold could, by appeal to Takens' (delay embedding) theorem, condition the expected internal states on the generalised motion of blanket states; however, for simplicity, we will just deal with generalised states *per se*. For a discussion of generalised coordinates of motion, see Appendix E and Friston, K., Stephan, K., Li, B., Daunizeau, J., 2010. Generalised Filtering. Mathematical Problems in Engineering vol., 2010, 621670, Kerr, W.C., Graham, A.J., 2000. Generalized phase space version of Langevin equations and associated Fokker-Planck equations. European Physical Journal B 15, 305-311.

Finally, the expected internal state conditioned upon its own blanket is not influenced by (i.e., is conditionally independent of) other blankets.

$$\boldsymbol{\mu}_j^{(i)}(b_j^{(i)}) \triangleq E_p[\mu_j^{(i)} \mid b_1^{(i)}, \ldots, b_J^{(i)}] = E_p[\mu_j^{(i)} \mid b_j^{(i)}] \tag{4.6}$$

Figure 9 describes how a particular partition (into particles) could proceed. It should be noted that the procedure in Figure 9 is one of the many ways in which to form a particular partition – and there are clearly a large number of particular partitions for any given system. This is reflected in the term *particular* partition, which denotes a partition into *particles* but is also particular in the sense it is one of many possible partitions. This procedure effectively identifies a small set of internal states and their Markov blanket based upon a graph Laplacian. The remaining (external states) are then recursively assigned to particles until all states have been accounted for. We will apply this procedure to our synthetic soup later. However, first, we need to address the dynamics of blanket states given a particular partition.

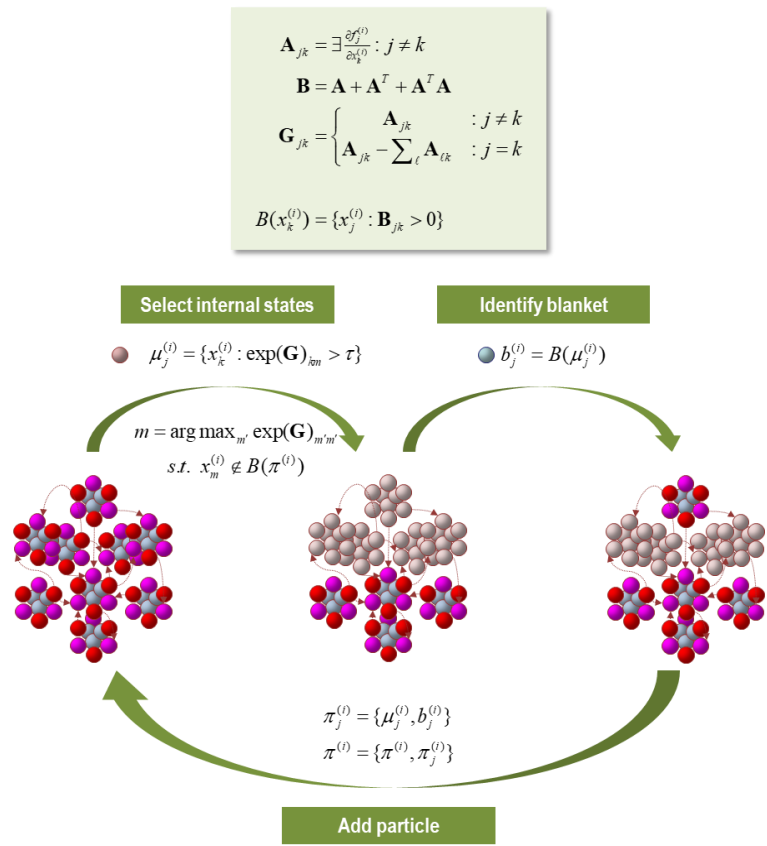

**FIGURE 9**

*The particular partition*. This schematic illustrates a partition of vectors states (small coloured balls) into particles (comprising nine vectors), where each particle has six blanket states (red and magenta for active and sensory states respectively) and three internal states (cyan). The upper panel summarises the operators used to create a particular partition. We start by forming an adjacency matrix that characterises the coupling between different vectors states. This is based upon the Jacobian and implicitly the flow of vector states. The resulting adjacency matrix defines a Markov blanket forming matrix ($\mathbf{B}$), which identifies the children, parents and parents of the children. The same adjacency matrix is used to form a graph Laplacian ($\mathbf{G}$) that is used to define neighbouring (i.e., coupled) internal states. One first identifies a set of internal states using the graph Laplacian. Here, the $j$-th subset of internal states at level $i$ are chosen, based upon dense coupling with the vector state with the largest graph Laplacian. Coupled internal states are then selected from the columns of the graph Laplacian that exceed some threshold. In practice, the examples used later specify the number of internal states desired for each level of the hierarchical decomposition. Having identified a new set of internal states (that are not members of any particle that has been identified so far) its Markov blanket is recovered using the Markov blanket forming matrix. The internal and blanket states then constitute a new particle, which is added to the list of particles identified. This procedure is repeated until all vector states have been accounted for. Usually, towards the end of this procedure, candidate internal states are exhausted because all remaining unassigned vector states belong to the Markov blanket of the particles identified previously. In this instance, the next particle can be an active or sensory state, depending upon whether there is a subset (of active states) that is not influenced by another. In the example here, we have already identified four particles and the procedure adds a fifth (top) particle to the list of particles; thereby accounting for nine of the remaining vector states.

## The adiabatic reduction

Having effectively eliminated internal states to form autonomous equations of motion for the Markov blanket, we now apply an adiabatic approximation to separate fast and slow dynamics. This separation rests on the eigenvectors of the Jacobian for each Markov blanket, where we can separate the eigenvectors with small (slow) and large (fast) negative eigenvalues (using $^-$ to denote a left eigenvector or generalised inverse of a right eigenvector):

$$\begin{bmatrix} \lambda_{11}^{(i)} & \cdots & \lambda_{1J}^{(i)} \\ \vdots & \ddots & \vdots \\ \lambda_{J1}^{(i)} & \cdots & \lambda_{JJ}^{(i)} \end{bmatrix} = \begin{bmatrix} \xi_1^{(i)} & \zeta_1^{(i)} \\ \vdots & \vdots \\ \xi_J^{(i)} & \zeta_J^{(i)} \end{bmatrix}^- \begin{bmatrix} J_{11} & \cdots & J_{1J} \\ \vdots & \ddots & \vdots \\ J_{J1} & \cdots & J_{JJ} \end{bmatrix} \begin{bmatrix} \xi_1^{(i)} & \zeta_1^{(i)} \\ \vdots & \vdots \\ \xi_J^{(i)} & \zeta_J^{(i)} \end{bmatrix}$$

$$\lambda_{jj}^{(i)} = [\xi_j^{(i)}, \zeta_j^{(i)}]^- J_{jj} [\xi_j^{(i)}, \zeta_j^{(i)}] = \begin{bmatrix} \lambda_{jj}^{\xi\xi} & 0 \\ 0 & \lambda_{jj}^{\zeta\zeta} \end{bmatrix} \tag{4.7}$$

$$\lambda_{jk}^{(i)} = [\xi_j^{(i)}, \zeta_j^{(i)}]^- J_{jk} [\xi_k^{(i)}, \zeta_k^{(i)}] = \begin{bmatrix} \lambda_{jk}^{\xi\xi} & \lambda_{jk}^{\xi\zeta} \\ \lambda_{jk}^{\zeta\xi} & \lambda_{jk}^{\zeta\zeta} \end{bmatrix}$$

$$[\xi_j^{(i)}, \zeta_j^{(i)}]^- [\xi_j^{(i)}, \zeta_j^{(i)}] = I, \quad 0 \geq \text{Re}\,\lambda_{jj}^{\xi\xi} > \epsilon \geq \text{Re}\,\lambda_{jj}^{\zeta\zeta}$$

This [eigen] decomposition is expressed in terms of block matrices, where the leading diagonal blocks comprise leading diagonal matrices of eigenvalues. Effectively, the eigenvectors represent mixtures of states that dissipate, following perturbations by extrinsic, intrinsic or random fluctuations. Here, $\epsilon < 0$ is some small negative number that places a lower bound on the rate that fast eigenmodes dissipate. Projecting the system of equations above onto the eigenvectors gives us two sets of equations for slow and fast dynamics respectively:

$$[\xi_j^{(i)}, \zeta_j^{(i)}]^{\top}\dot{b}_j^{(i)} = \begin{cases} \xi_j^{(i)\top}\dot{b}_j^{(i)} \\ \zeta_j^{(i)\top}\dot{b}_j^{(i)} \end{cases} = \begin{cases} \dot{b}_{j,slow}^{(i)} \\ \dot{b}_{j,fast}^{(i)} \end{cases}$$

$$= \begin{cases} \xi_j^{(i)\top} f_{b_j}^{(i)}(b_0^{(i)}) + \sum_k \lambda_{jk}^{\xi\xi} b_{k,slow}^{(i)} \\ \zeta_j^{(i)\top} f_{b_j}^{(i)}(b_0^{(i)}) + \sum_k \lambda_{jk}^{\zeta\xi} b_{k,fast}^{(i)} \end{cases} + \begin{cases} \xi_j^{(i)\top}(\omega_{b_j}^{(i)} + K_j \varepsilon_j^{(i)}) + \sum_{k \neq j} \lambda_{jk}^{\xi\xi} b_{k,fast}^{(i)} + \dots \\ \zeta_j^{(i)\top}(\omega_{b_j}^{(i)} + K_j \varepsilon_j^{(i)}) + \sum_{k \neq j} \lambda_{jk}^{\zeta\xi} b_{k,slow}^{(i)} + \dots \end{cases}$$

$$(4.8)$$

The upper equations describe flow with a slow dynamics (of the $j$-th Markov blanket) that is driven by slow extrinsic dynamics in other Markov blankets. In this formulation, intrinsic and random fluctuations are supplemented by the effects of fast fluctuations in other Markov blankets on the slow modes of the blanket in question. The separation of temporal scales implicit in this adiabatic expansion means that one can assume the intrinsic (and extrinsic) fluctuations are fast in relation to the dynamics of the slow modes. This assumption allows us to express the dynamics of the slow modes in the same form as the initial ansatz (4.1):

$$\dot{x}_n^{(i)} = f_n^{(i)} + \sum_m \lambda_{nm}^{(i)} x_m^{(i)} + \omega_n^{(i)}$$
$$\dot{x}_n^{(i+1)} = f_n^{(i+1)} + \sum_m \lambda_{nm}^{(i+1)} x_m^{(i+1)} + \omega_n^{(i+1)}$$
$$\dot{x}_n^{(i+2)} = \cdots$$

$$x_n^{(i+1)} \triangleq \xi_n^{(i)\top} b_n^{(i)} = b_{n,slow}^{(i)}$$
$$f_n^{(i+1)} \triangleq \xi_n^{(i)\top} f_{b_n}^{(i)}(b_0^{(i)})$$
$$\lambda_{nm}^{(i+1)} \triangleq \xi_n^{(i)\top} J_{nm} \xi_m^{(i)} = \lambda_{\xi_n^i \xi_m^i}^{(i)}$$
$$\omega_n^{(i+1)} \triangleq \xi_n^{(i)\top}(\omega_{b_n}^{(i)} + K_n \varepsilon_n^{(i)}) + \sum_{k \neq n} \lambda_{nm}^{\xi\zeta} b_{m,fast}^{(i)} + \cdots$$

$$(4.9)$$

This is the endpoint of our analysis; in which the ansatz for the form of dynamics at one level emerges as a consequence of conditional independencies at the level below. This means that the flow can be decomposed in a recursive fashion to describe the dynamics at progressively higher spatial and temporal scales (c.f., the recursive Gaussian elimination implicit in things like the Cholesky decomposition). Note that the last equality in (4.9) ensures that fluctuations are Gaussian, via the central limit theorem, because they are mixtures of fluctuations at the lower level.

In this construction, particles (i.e., things) are only defined in relation to the mapping between adjacent levels of description. In other words, the (macroscopic) states of a particle $x_n^{(i+1)}$ are a nonlinear mixture[10] of a particle's (microscopic) states: $b_n^{(i)} \subset \pi_n^{(i)}$, where the blankets of $J^{(i)} = N^{(i+1)}$ particles subtend the [eigen]states at the higher level. In other words, a particle – or a particular partition – underwrites the mapping between the

---

[10] The mixtures are nonlinear because the eigenvectors are functions of the current state. This follows because the Jacobians are state-dependent.

*macroscopic states of a particle* and a *particle of microscopic states*. Figure 10 illustrates the adiabatic reduction afforded by a particular partition.

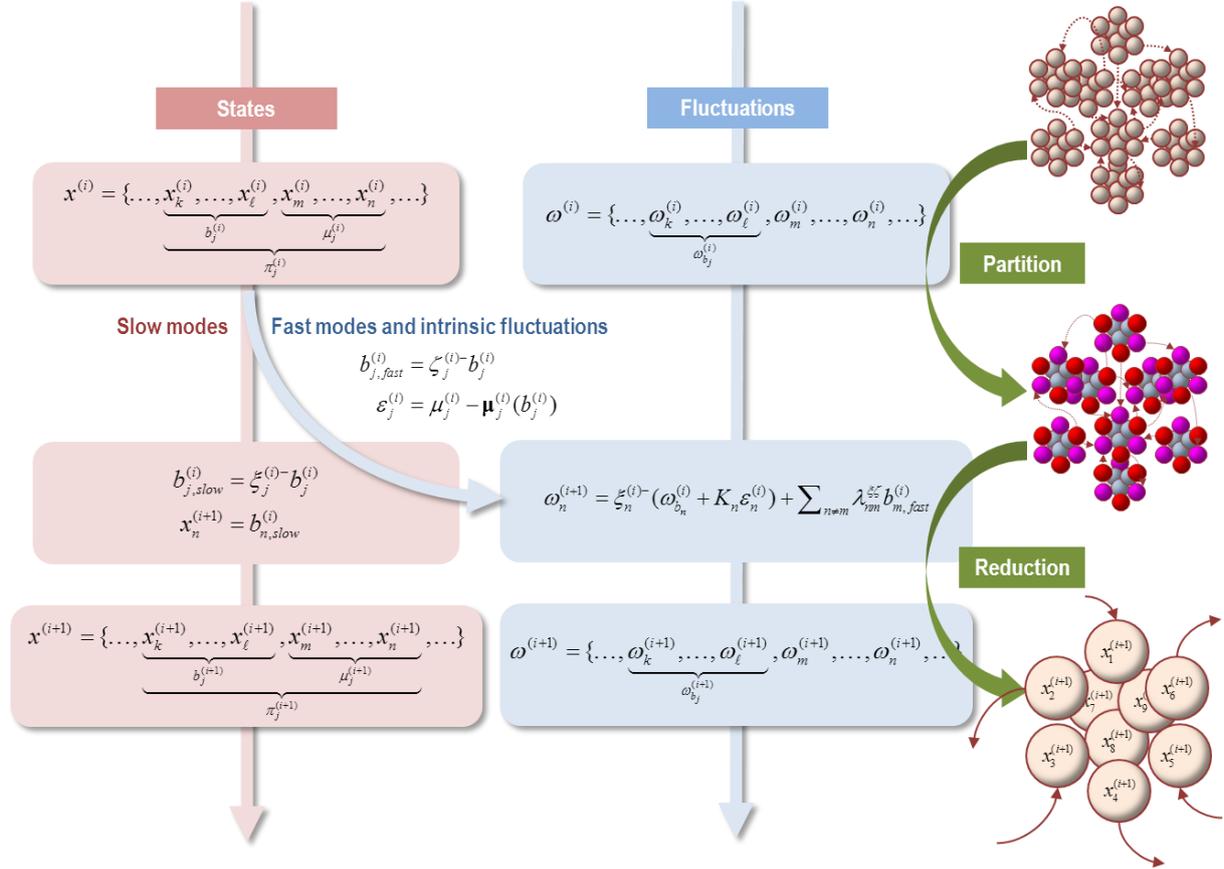

**FIGURE 10**

*Adiabatic reduction*. This figure illustrates the adiabatic dimension reduction that constitutes the second part of the hierarchical decomposition; namely, the elimination of internal states and retention of slow blanket modes. This schematic is presented in three columns. The first two columns represent the partitions of states (left row) and associated random fluctuations (middle row). The right row uses the same schematic format as in the previous figures. Here, we start off with a partition of states at level *i*. Following a particular decomposition into the blanket and internal states of particles (see previous figure), an adiabatic reduction decomposes each particle's blanket states into fast and slow modes. The slow modes now constitute the vector states (i.e., eigenstates) of the subsequent level. Conversely, the fast, fluctuating modes are added to the intrinsic fluctuations to produce random fluctuations for the next level. After this process is complete, we end up where we started; namely, with a partition of states into vectors (i.e., the slow modes of blanket states of particles at the level below) and random fluctuations. This is illustrated on the right by converting lots of vector states at the level *i* into a smaller number of vector states at the next level via, the particular partition and (adiabatic) reduction. The reason that this is referred to as an adiabatic reduction is that the eigenvalues – used to assign modes or mixtures of blanket states to vector states at the next level – correspond to the Lyapunov exponents of the underlying Jacobians. The real parts of these eigenvalues reflect the rate at which the modes decay over time. Please refer to the main text for a description of the variables used in this figure.

## Elimination and renormalisation

To recap, the recursive link between hierarchical scales rests on two steps. First, a particular partition enables the elimination of internal states by conditioning on the Markov blanket, by absorbing intrinsic fluctuations into random fluctuations at the next level (denoted by the transformation $\mathbf{G}$ in Figure 8). Second, the eigenvectors of the remaining (Markov blanket) states enable an adiabatic decomposition into slow and fast dynamics. In turn, this enables the elimination of fast dynamics, by absorbing the fast dynamics into random fluctuations at the next level (denoted by the transformation $\mathbf{R}$ in Figure 8). This can be summarised as follows:

$$x^{(i)} = \{x_1^{(i)}, \ldots, \underbrace{\underbrace{x_k^{(i)}, \ldots, x_f^{(i)}}_{a_j^{(i)}}, \underbrace{x_m^{(i)}, \ldots, x_n^{(i)}}_{s_j^{(i)}}}_{b_j^{(i)}}, \underbrace{x_o^{(i)}, \ldots, x_p^{(i)}}_{\mu_j^{(i)}}, \ldots, x_N^{(i)}\}$$

$$x_n^{(i+1)} = \xi_n^{(i)-} b_n^{(i)}$$

(4.10)

On this view, the (macroscopic) state of a particle corresponds to a mixture of its (microscopic) blanket states at the lower level. Successive decompositions eliminate internal states and mixtures of blanket states, whose fluctuations are sufficiently fast to be treated as random fluctuations.

The existence of this decomposition provides an interesting perspective on the genesis of random fluctuations. For example, if we assume that the particles at a given level are sufficiently similar to render $J_{nm} \approx J_{nn} \Rightarrow \lambda_{nm}^{\xi\varsigma} \approx 0 : \forall n \neq m$, then their slow and fast modes are uncoupled, and the random fluctuations reduce to:

$$\omega_n^{(i+1)} = \xi_n^{(i)-}(K_n^{(i)}\varepsilon_n^{(i)} + \omega_{b_n}^{(i)})$$

$$\omega_{b_{n_j}}^{(i)} = \xi_{b_{n_j}}^{(i-1)-}(K_{b_{n_j}}^{(i-1)}\varepsilon_{b_{n_j}}^{(i-1)} + \omega_{b_{b_{n_j}}}^{(i-1)})$$

$$\vdots$$

(4.11)

We will refer to this as the *ensemble assumption,* which entails a *weak coupling* between fast and slow modes. The recursive substitution implicit in (4.11) illustrates how random fluctuations inherit their dynamics from intrinsic fluctuations; namely, through a successive accumulation and mixing of intrinsic fluctuations[11] (see Figure

---

[11] For simplicity, we have ignored correlations among the random fluctuations at the higher-level that are induced by correlations among the intrinsic fluctuations. In principle, these are dealt with by an affine transformation of the fast eigenvectors, such that

10). In other words, blanket fluctuations are mixtures of intrinsic fluctuations from subordinate scales. The ensuing picture is consistent with formulations that distinguish between fast microscopic states and slow macroscopic states: for example, the distinction between slow (unstable) order parameters and fast (stable) modes in synergetics (Frank, 2004; Haken, 1983); the distinction between a micro-states and macro-states of canonical ensembles in statistical mechanics (Seifert, 2012) and the distinction between unstable and stable manifolds in bifurcation and centre manifold theorems (Carr, 1981). The crucial aspect of this treatment is that the fast, stable, dissipative, microscopic dynamics emerge via the elimination of internal states, when formulating the dynamics in terms of the Markov blanket that surrounds and sequesters them. These fluctuations are then separated from the slow modes that supervene at successively higher levels

An alternative perspective on this adiabatic reduction is provided by the notion of renormalisation. In theoretical physics, the *renormalization group* (RG) refers to a transformation that characterises a system when measured at different scales (Cardy, 2015; Schwabl, 2002). A working definition of renormalization involves three elements (Lin et al., 2017): vectors of random variables, a course-graining operation and a requirement that the operation does not change the functional form of the Lagrangian (or equivalent description of the dynamics). In our case, the random variables are states; the course graining operation corresponds to the grouping ($\mathbf{G}$) into a particular partition and adiabatic reduction ($\mathbf{R}$) – that leaves the functional form of the dynamics (and associated Lagrangian) unchanged. For example, from (1.2) and (4.9) we could write the Lagrangian of a particle at any scale in the style of the renormalisation group:

$$\mathcal{L}(x_n^{(i)}, \dot{x}_n^{(i)}) = \frac{1}{2}[(\dot{x}_n^{(i)} - \phi_n^{(i)}) \cdot (2\Gamma_n^{(i)})^{-1} (\dot{x}_n^{(i)} - \phi_n^{(i)}) + \nabla \cdot \phi_n^{(i)}]$$
$$\phi_n^{(i)} = f_n^{(i)} + \sum_m \lambda_{nm}^{(i)} x_m^{(i)}$$

$$\{x_n^{(i)}\} = \mathbf{R} \circ \mathbf{G} \circ \{x_n^{(i-1)}\}$$
$$\{f_n^{(i)}, \lambda_{nm}^{(i)}, \Gamma_n^{(i)}\} = \beta(\{f_n^{(i-1)}, \lambda_{nm}^{(i-1)}, \Gamma_n^{(i-1)}\})$$

(4.12)

Here, the Lagrangian of a particle at one scale has been expressed in terms of states at a lower scale, after a course graining or *blocking* transformation $\mathbf{R} \circ \mathbf{G}$ that composes the particular partition and adiabatic reduction. This transformation necessarily reduces the number of states, by eliminating internal states at the lower level and retaining the relevant eigenmodes of blanket states, where (under the ensemble assumption):

---

$$\xi_n^{(i)} \to \xi_n^{(i)} \Rightarrow \xi_n^{(i)-}(K_n^{(i)}\Sigma_n^{(i)}K_n^{(i)T} + 2\Gamma_{b_n}^{(i)})\xi_n^{(i)} = 2\Gamma_n^{(i+1)}$$

has the desired form.

$$\{x_n^{(i)}\} \xrightarrow{\mathbf{G}} \{b_j^{(i)}\} \subset \{b_j^{(i)}, \mu_j^{(i)}\}$$

$$
\begin{aligned}
\{b_n^{(i)}\} &\xrightarrow{\mathbf{R}} \{x^{(i+1)}\} = \{\xi_n^{(i)-} b_n^{(i)}\} \\
\{f_n^{(i)}\} &\xrightarrow{\beta} \{f_n^{(i+1)}\} = \{\xi_n^{(i)-} f_{b_n}^{(i)}\} \\
\{\lambda_{nm}^{(i)}\} &\xrightarrow{\beta} \{\lambda_{nm}^{(i+1)}\} = \{\xi_n^{(i)-} \partial_{b_n} f_{b_n}^{(i)} \xi_m^{(i)}\} \\
\{\Gamma_n^{(i)}\} &\xrightarrow{\beta} \{\Gamma_n^{(i)}\} = \{\xi_n^{(i)-} (\Gamma_{b_n}^{(i)} + \tfrac{1}{2} \partial_{\mu_n} f_{b_n}^{(i)} \Sigma_{b_n}^{(i)} \partial_{\mu_n} f_{b_n}^{(i)T}) \xi_n^{(i)}\}
\end{aligned}
\tag{4.13}
$$

Here, the parameters of the Lagrangian are taken to be the flow, coupling parameters and the amplitude of fluctuations, whose changes are implemented by a *beta function* that is said to induce a renormalization group flow (or RG flow) on parameter space. The key aspect of this flow rests upon the adiabatic reduction, which renders the dynamics progressively slower at successive macroscopic scales, because – by construction – only slow modes are retained by course graining; e.g.,

$$E[\text{Re}(\lambda_{nm}^{(i)})] \leq E[\text{Re}(\lambda_{nm}^{(i+1)})]\dots \leq 0 \tag{4.14}$$

The corresponding RG flow on the amplitude of fluctuations $\Gamma_n^{(i)} \geq \Gamma_n^{(i+1)} \dots \geq 0$, speaks to a progressive move from dynamics with high amplitude, fast fluctuations (e.g., quantum mechanics) through to deterministic systems that are dominated by slow dynamics (e.g., classical mechanics). In deterministic systems, $E[\text{Re}(\lambda_{nm}^{(i)}(x_n^{(i)}))]$ play the role of *Lyapunov exponents* (c.f., critical exponents), which quantify the rate of separation of infinitesimally close trajectories (Lyapunov and Fuller, 1992; Pyragas, 1997). This suggests that as we move from one scale to the next, there is a concomitant reduction in the amplitude of random fluctuations and a tendency to dynamic itinerancy (Cessac et al., 2001; Pavlos et al., 2012).

In this (RG) setting, a *relevant* variable is said to describe the macroscopic behaviour of the system, while an *irrelevant* variable is not. From our perspective, the relevant variables in question correspond to the slow modes retained in (4.13), while the irrelevant variables can be associated with fast modes and intrinsic fluctuations (see Figure 10). Figure 11 and Figure 12 provide a worked example, applying the adiabatic reduction to a hierarchical series of (particular) partitions of the synthetic soup of the previous section.

## Markov blankets as dissipative structures

We have made a few deflationary assumptions that require comment. First, nonequilibrium steady-state has been cast as self-organisation to a random (dynamical) attracting set (Arnold, 2003; Crauel et al., 1997; Crauel and Flandoli, 1994). This raises two questions. First, is this an appropriate mathematical image of dissipative structures that characterise systems far from equilibrium? Second, what distinguishes *nonequilibrium* from *equilibrium* steady-state? The second question has a straightforward answer: systems at equilibrium possess an attracting set, surrounding a dynamically stable *fixed point*. In other words, the gradient flows are, locally, directed to a single point in phase space, thereby precluding itinerancy – and the space-filling attractors that characterise nonequilibrium steady-state.

The first question is more delicate[12]. Dissipative structures – in the sense of Prigogine (Nicolis and Prigogine, 1977; Prigogine, 1978) – are dynamical structures characterised by the spontaneous emergence of topological [super] symmetry breaking and itinerancy; e.g., turbulence, cyclones and living systems (England, 2015). In particular, a dissipative structure has a reproducible (steady-state) regime to which it evolves. From our perspective, the 'system' only exists in virtue of its Markov blanket. As such, the blanket states of a random dynamical system constitute a dissipative structure. Associating Markov blankets with dissipative structures is appealing, in the sense that blanket states are defined by conditional independencies induced by dynamical flow. In short, *Markov blankets are dissipative structures that arise from structured flow*.

Having said this, the assumption of a random dynamical attractor precludes a formulation of dissipative structures (of this sort) in terms of *wandering sets* (Birkhoff, 1927). In other words, it does not easily accommodate the fact that the particles that constitute a Markov blanket can, over time, wander away or, indeed, be exchanged or renewed. The canonical example here would be the blanket states of a candle flame, whose constituent particles (i.e., molecules of gas) are in constant flux. This speaks to the interesting challenge of generalising the treatment above to handle wandering sets[13]. However, we will make the simplifying assumption that over a suitable time scale, blanket states are well defined – as a subset of attracting states.

## Summary

In summary, we have seen how structured dynamics can be derived hierarchically from one level (e.g., biophysical states) to produce intermediate things (e.g., macromolecules) that assemble into higher-order dissipative structures (e.g., cells) that form identifiable communities (e.g., organelles). At each level of (self) organisation, the integrity of constituent particles is underwritten by the preservation of a Markov blanket that enables us to talk about 'things' (Ramstead et al., 2017) or, indeed, how things affect things (Constant et al., 2018). In the second part of this monograph, we will look at the kinds of physics one might expect to see at different scales of self-organisation.

---

[12] My thanks Klaus Harisch for correspondence on this issue.

[13] For example, by the use of (periodic) boundary conditions and flow operators that transport blanket states from one location on a state space boundary to another, with the same NESS potential and action (i.e., flow).

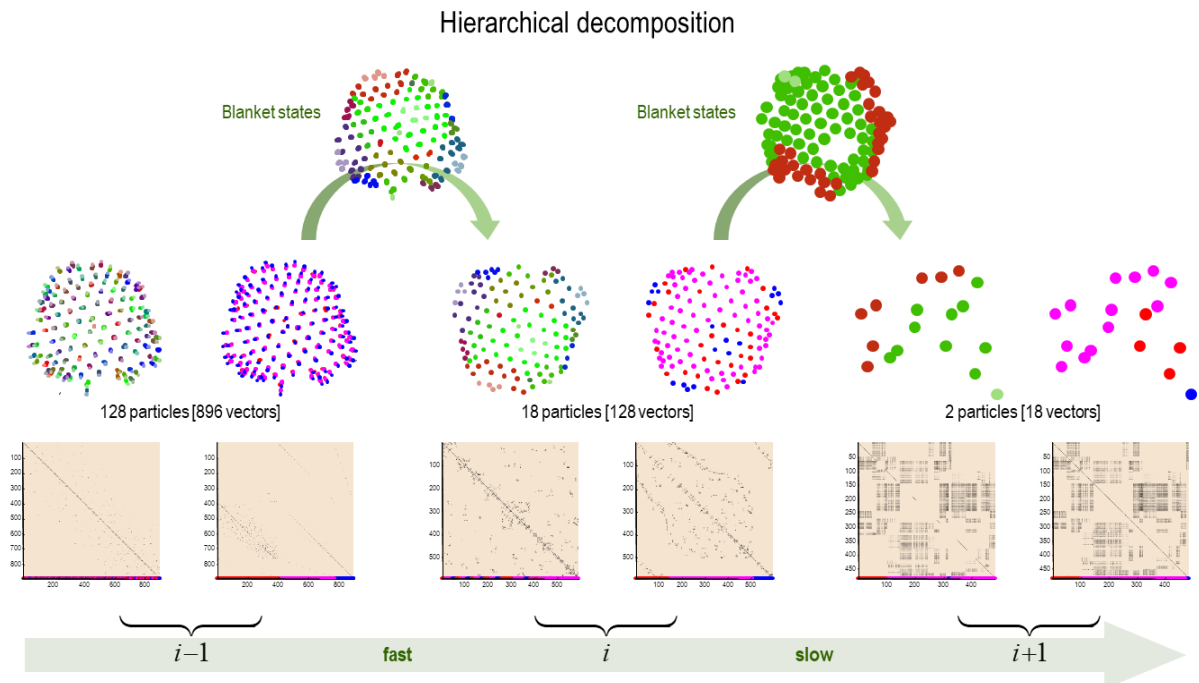

**FIGURE 11**

*Hierarchical renormalisation of active matter*. This figure illustrates the application of hierarchical renormalisation (i.e., recursive particular partition and adiabatic reduction) to the synthetic soup of Figure 5.This figure illustrates two successive renormalisations to construct dynamics at successively slower and larger scales. The lower row of images shows the Jacobians (i.e., rate of change of flow with respect to states) of [eigen]states at three levels. The Jacobians are presented as pairs: the left Jacobian is sorted by the particle to which the state belongs, while the right Jacobian is sorted by the type of state – active (red) sensory (magenta) and internal (blue). This designation is encoded by the colours of dots along the lower margin of each Jacobian. Starting with the lowest level (on the left) we have 896 vector or eigenstates that, following a particular partition can be assigned to 128 particles. The dependency or coupling among particles can be used to create a spectral or embedding space, corresponding to the eigenmodes of the graph Laplacian in Figure 9. Heuristically, this provides a coordinate system in which each particle can be located. The resulting scaling space means that the coupling among neighbouring particles scales with their proximity. This spectral embedding is illustrated twice, first colour-coded according to the particle (left image of each pair) and according to whether it is an active (red) sensory (magenta) or internal (blue) eigenstate (right image of each pair). One can see that the dependency structure implied by the Jacobian induces a complicated geometry in scaling space. In moving to the next level, internal states (blue) are eliminated and eigenstates of the remaining blanket states constitute vector states at the next level. In this example, we start with a particular partition of 896 microscopic states into 128 vector or eigenstates of particles. These correspond to the macromolecules that constitute our soup. Following a particular decomposition, this soup decomposes into 18 particles, five of which have internal states. This could be regarded as a collection of 18 cells. After a further decomposition, we end up with two particles, where only one (the green particle) has an internal state. This level of organisation can be regarded as a collection of cells (e.g., organelle) or a colony of bacteria, or a community. The images in the upper row illustrate the triaging of internal states by plotting the locations of eigenstates in the embedding or scaling space of the lower level – but colour coding them in terms of the particles to which they belong at the subsequent level. The spectral embedding evinces a much simpler topology that is reminiscent of the spatial locations of the macromolecules in Figure 6. This follows from the fact that the Euclidean positions in the simulation determine the conditional dependencies among the particles or macromolecules. The Jacobians show some important characteristic features that are detailed further in the next figure. In this example, the Jacobians were based upon the average rate of change of flow over 64 time steps after 512 iterations of the differential equations illustrated in Figure 5. A more detailed analysis of these Jacobians is available in the next figure. Here, we specified one, four and one internal state per particle, for each of the three levels shown.

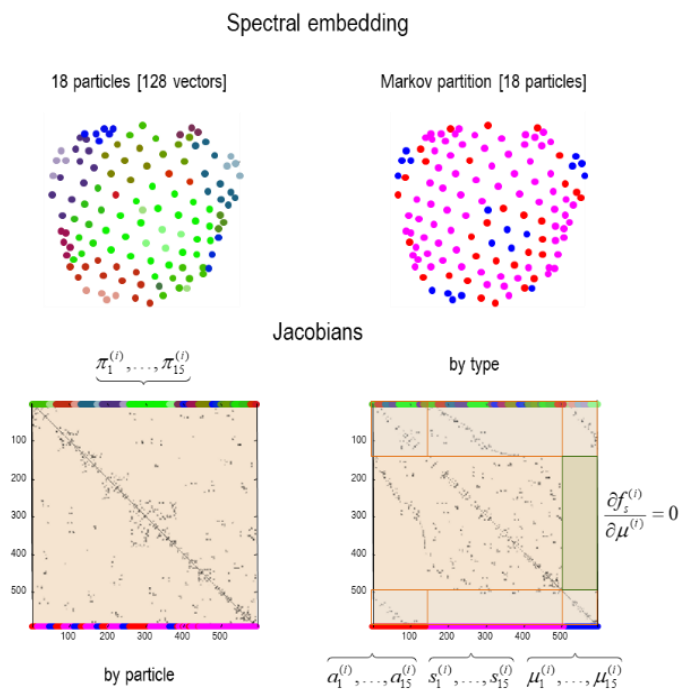

**FIGURE 12**

*Dynamics and dependencies*. This figure reproduces the results of the previous figure for the intermediate level of self-organisation that we have associated with the emergence of cell-like (dissipative) structures from macromolecules. As noted in the previous figure, the spectral embedding and spatial dependencies that underwrite the Markov blanket rest upon the flow of its constituent states. This is encoded by the Jacobians in the lower row. The lower left panel shows the Jacobian when sorting the states according to particle membership; here colour-coded with 18 colours (shown at the top of the image). The lower right panel shows exactly the same dependencies but after sorting the order of states according to whether they were active (red) sensory (magenta) or internal (blue) states of each of the 18 particles. This reordering reveals the particular sparsity of dependencies (seen on the left) that is implied by the existence of Markov blankets; specifically, the absence of any influence of internal states on sensory states (depicted in light green). The remaining conditional independencies are manifest as block diagonal forms (depicted as lighter shade rectangles). This block diagonal form ensures that the off-diagonal blocks (i.e., between-particle coupling) is zero, when considering internal states and active states. In other words, the common feature of autonomous (i.e., active and internal) states is that they cannot be influenced directly by other particles – they can only be influenced by the states that belong to the particle in question. At this level of organisation, there are 18 particles that represent a partition of 128 eigenstates (e.g., macromolecular states), where each vector or eigenstate comprises between two and six states. These summarise the two-dimensional velocity, position and electrodynamics of each particle at the lower level; namely, each macromolecule.

# Part Two: some special cases

## A theory of small things – quantum mechanics

This section considers quantum formulations of density dynamics that, effectively, rests upon dealing with the square root of the NESS density. This complex root plays the role of a *wave function*. In brief, we will consider the NESS density of particular states as a sufficient description of a particle. This description is sufficient in the sense that small particles have short information lengths – endowed by random fluctuations – and therefore obtain nonequilibrium steady-state very quickly. This lends quantum particles certain properties that elude classical mechanics. For example, one can think about random fluctuations as enabling a particle to explore paths (through state-space) around the path of least action – to which classical particles are confined. This exploration is determined by the amplitude of random fluctuations that, as we will see below, determines a particle's effective 'mass'[14]. The resulting behaviour leads to characteristic quantum phenomena. For example, in the absence of random fluctuations, a particular state at a local extremum of surprisal will have zero flow and remain there forever. However, the NESS density of a quantum particle can explore the local landscape of surprisal; enabling it to 'leak' away from minima; e.g., quantum tunnelling (Brookes, 2017) and evince behaviours that are wavelike in nature. From our perspective, this could be considered in the light of noise-induced tunnelling[15], of the sort illustrated in Figure 2.

A key insight afforded by this move – to a wave function description of ensemble dynamics – is that the Fourier transform of the density over states becomes a density over flow (in virtue of the NESS lemma)[16]. This means one can compute the dispersion or variance of flow to produce a measure of energy by analogy with classical mechanics. This analogy can be underwritten by associating the amplitude of random fluctuations with inverse mass, where the constant of proportionality is Planck's constant. A simple dimensionality analysis motivates the notion of mass (in kilograms), given the units of flow (in nats per second) and gradients of the NESS potential (per metre). Once we have mass in the picture, we can associate flow with momentum and its dispersion with kinetic energy, thereby equipping quantum formulations with an interpretation in terms of classical analogues.

---

[14] Excursions around the classical path become non-negligible when the (reduced) mass approaches Planck's mass of about 0.02 mg.

[15] The notion of 'tunnelling' in this context may be a slight misdirection, because fluctuations cannot 'see' potential energy barriers – the (thermodynamic or Schrödinger) potential pertains to flow. This means that fluctuations transcend potential gradients, as if they were not there.

[16] Intuitively, if the probability mass of a particular state is concentrated around one point in phase-space, then the flow must be countering random fluctuations. This means that the flow is as dispersed as the fluctuations. Conversely, if flow is limited to a small range, random fluctuations would disperse particular states over state-space. In short, the dispersion of states and their flow must complement each other at nonequilibrium steady-state.

The formulation of density dynamics in terms of a wave function introduces some interesting differences between treatments in terms of the density *per se*. In particular, it enables one to summarise density dynamics in terms of a Schrödinger potential, where the wave function is a solution to the Schrödinger equation (after the Schrödinger potential has been supplied). In what follows, we will see how the Schrödinger potential can be derived from the NESS potential and, implicitly, how the wave function relates to the NESS density. We will focus on the wave particle duality disclosed by the use of Fourier transforms and the implicit relationship between flow and momentum. The deflationary message of this section is that the NESS lemma (Appendix B) effectively dissolves wave-particle duality, on the view that every probability density (i.e. wave) function has particular (i.e., particle-like) characteristics. See (Koide, 2017) and (Wang, 2009) for a more rigorous treatment.

At a microscopic (quantum) scale, we will suppose that random fluctuations predominate over solenoidal flow, which is consequently ignored in this section. At this scale, one is primarily interested in the nonequilibrium steady-state densities over microscopic states. In what follows, we derive (an introductory level) quantum physics from the NESS formalism. In brief, this involves, expressing the NESS density as the product of a complex function and its conjugate. The resulting *wave function* is generally considered the most complete description of a physical system. For a single particle:

$$p = \Psi \cdot \Psi^{\dagger}$$
$$\Psi = \Psi(x)e^{-iEt/h}$$

(5.1)

One might ask why it is useful to factorise the probability density in this fashion (see Appendix D for discussion). A key motivation follows from the Plancherel theorem, which ensures the integral of a function's squared modulus is conserved following a Fourier transform (e.g., the power is the same in space and frequency). In terms of the Dirac notation, this means: $<\Phi^{\dagger} | \Phi> = <\Psi^{\dagger} | \Psi> = 1$, where $\Phi(k)$ is the Fourier transform of $\Psi(x)$. This can be exploited by treating both as *probability amplitudes*, whose squared modulus is a probability density. Below, we will see the frequency modes have an interesting interpretation that underlies things like wave-particle duality and Heisenberg's uncertainty principle. Put simply, working with the complex roots of the NESS density allows one to use Fourier transforms and talk about wavelengths in state-space. In turn, boundary conditions (e.g., the continuity of density functions) place formal constraints on density functions (e.g., a circle can only be divided into a finite number of wavelengths), leading to quantal behaviour.

First, we will establish that the NESS density can be recovered from the solutions to the (time-independent) Schrödinger equation, which offers an accurate account of nearly all empirical molecular, atomic and subatomic measurements:

$$\mathbf{H}\Psi = E\Psi$$
$$\mathbf{H} = V(x) - \frac{\hbar^2}{2m} \cdot \nabla^2 \tag{5.2}$$
$$V(x) = \frac{m}{2} f \cdot f + \frac{\hbar}{2} \nabla \cdot f$$

Here, $\mathbf{H}$ corresponds to a Hamiltonian operator that returns the energy $E$ of the particle and $V(x)$ is the Schrödinger potential of a particle's states. For consistency with other texts, this section will denote the states of a particle by $x \equiv \pi_k$. Schrödinger's equation is central to nearly all applications of quantum mechanics; including quantum field theory, which combines special relativity with quantum mechanics. Furthermore, formulations of quantum gravity, such as string theory do not modify Schrödinger's equation. The above time-independent (single particle) Schrödinger equation can be derived as follows.

## The Schrödinger equation from first principles

Our starting point is to express the flow in terms of the NESS density and the amplitude of random fluctuations, where the amplitude is expressed in terms of a (reduced) mass and Planck's constant. This enables us to express the gradients of the wave function in terms of flow[17]:

$$\Gamma = \frac{\hbar}{2m}$$
$$p = \Psi^\dagger \Psi$$
$$f = \frac{\hbar}{2m} \nabla \ln p = -\Gamma \nabla \Im \Rightarrow$$
$$pf = \frac{\hbar}{2m} \nabla p = \frac{\hbar}{2m} \nabla(\Psi \cdot \Psi^\dagger) \Rightarrow \tag{5.3}$$
$$\Psi^\dagger f \Psi = \frac{\hbar}{2m}(\Psi^\dagger \nabla \Psi + \Psi \nabla \Psi^\dagger) \Rightarrow$$
$$\nabla \Psi = \frac{m}{\hbar} f \Psi$$

These equalities can now be substituted into the density dynamics (i.e., Fokker Planck equation)

$$\dot{p} = \frac{\hbar}{2m} \nabla^2 p - p \nabla \cdot f - f \cdot \nabla p$$
$$= \frac{\hbar}{2m} \nabla^2 \Psi^\dagger \Psi - \frac{1}{2} \Psi \nabla \cdot f \Psi^\dagger - \frac{1}{2} \Psi^\dagger \nabla \cdot f \Psi - \Psi f \cdot \nabla \Psi^\dagger - \Psi^\dagger f \cdot \nabla \Psi$$
$$= \Psi^\dagger (\frac{\hbar}{2m} \nabla^2 - \frac{1}{\hbar} V) \Psi + \Psi(\frac{\hbar}{2m} \nabla^2 - \frac{1}{\hbar} V) \Psi^\dagger \tag{5.4}$$

$$V(x) = \frac{m}{2} f \cdot f + \frac{\hbar}{2} \nabla \cdot f$$
$$= \frac{\hbar^2}{4m}(\frac{1}{2} \nabla \Im \cdot \nabla \Im - \nabla^2 \Im)$$

This enables us to express the ensemble dynamics in terms of the (Hamiltonian) operator that plays the same role as the Fokker Planck operator but now operating on the wave function:

---

[17] The final equality uses the fact that $\Psi^\dagger \Psi \in \mathbb{R} \Rightarrow \Psi^\dagger \nabla \Psi = \Psi \nabla \Psi^\dagger$.

$$i\hbar\dot{p} = \Psi^\dagger i\hbar\dot{\Psi} + \Psi i\hbar\dot{\Psi}^\dagger = \Psi^\dagger E\Psi - \Psi E\Psi^\dagger = 0$$
$$-\hbar\dot{p} = \Psi^\dagger \mathbf{H}\Psi + \Psi\mathbf{H}\Psi^\dagger = 0$$
$$\mathbf{H} = V(x) - \frac{\hbar^2}{2m}\nabla^2 \tag{5.5}$$
$$\Leftarrow$$
$$\mathbf{H}\Psi = i\hbar\dot{\Psi} = E\Psi$$

This is the time independent Schrödinger wave equation (for a single particle). Effectively, it is just another way of expressing density dynamics, where the eigenvalue of the operator acquires an interpretation in terms of energy. One can see the formal similarity between the Fokker Planck and Hamiltonian operator's by expressing the Fokker Planck and (time-independent) Schrödinger wave equation in terms of their respective operators:

$$\dot{p}(x) = \mathbf{L}p(x)$$
$$\mathbf{L} = \frac{\hbar}{2m}\nabla^2 - \nabla \cdot f$$

$$\tag{5.6}$$

$$i\hbar\dot{\Psi}(x) = \mathbf{H}\Psi(x)$$
$$-\frac{1}{\hbar}\mathbf{H} = \frac{\hbar}{2m}\nabla^2 - \frac{1}{\hbar}V(x)$$

These equalities foreshadow a key conclusion of this section; namely, the Fokker Planck formulation and its NESS solution rests upon the flow, while the solution to the Schrödinger wave equation requires a Schrödinger potential. As noted in the introduction, things get interesting when we consider a parameterisation of the NESS density in terms of its Fourier transform, as follows.

## Wave particle duality and the de Broglie hypothesis

Without loss of generality, one can express the wave function in terms of its Fourier transform, giving the following Fourier transform pair, where $k$ denotes wave number:

$$\Psi(x,t) = \Psi(x)e^{-i\omega t} = \Psi(x)e^{-iEt/\hbar}$$
$$\Psi(x) = \frac{1}{\sqrt{2\pi}}\int_{-\infty}^{\infty}\Phi(k)\cdot e^{ik\cdot x}dk \tag{5.7}$$
$$\Phi(k) = \frac{1}{\sqrt{2\pi}}\int_{-\infty}^{\infty}\Psi(x)\cdot e^{-ik\cdot x}dx$$

Using the Dirac notation, we have, using $\nabla\Psi = \frac{m}{\hbar}f\Psi$ and $\left\langle \Phi^\dagger \mid \Phi \right\rangle = \left\langle \Psi^\dagger \mid \Psi \right\rangle = 1$ (by the Plancherel theorem)

$$\left\langle \Psi^\dagger(x) \mid -i\hbar\nabla \mid \Psi(x) \right\rangle = -i\hbar\int_{-\infty}^{\infty}\Psi(x)^\dagger\nabla\Psi(x)dx$$
$$= -i\hbar\int_{-\infty}^{\infty}\Psi^\dagger\frac{m}{\hbar}f\Psi dx = -im\mathbf{f} \tag{5.8}$$
$$= \left\langle \Phi(k)^\dagger \mid \hbar k \mid \Phi(k) \right\rangle = \hbar\mathbf{k} \Rightarrow m\mathbf{f} = i\hbar\mathbf{k}$$

The third equality is obtained by substituting (5.7) into the first. The interesting thing here is that we can associate the (reduced) *mass* times the expected *flow* with the expected spatial wave number. This lends itself naturally to an interpretation in terms of momentum, in accord with the de Broglie hypothesis, leading to the usual energy and momentum operators associated with quantum treatments

$$\mathbf{H}\Psi = i\hbar\dot{\Psi} = \hbar\omega\Psi = E\Psi$$
$$\mathbf{p}\Psi = -i\hbar\nabla\Psi = \hbar k\Psi$$

$$E \triangleq \hbar\omega$$
$$\mathbf{p} \triangleq \hbar\mathbf{k} = m\mathbf{f}$$

(5.9)

By analogy with classical mechanics – see Equation (7.8) below – one can interpret the Hamiltonian as comprising kinetic and potential energy operators, where the kinetic energy operator is the dual application of the momentum operator:

$$\mathbf{H} = \mathbf{T} + V(x)$$
$$\mathbf{T} = \frac{1}{2m}\mathbf{p}\cdot\mathbf{p} = -\frac{\hbar^2}{2m}\cdot\nabla^2$$

(5.10)

This allows one to associate the expected values returned by the Hamiltonian operator in terms of kinetic and potential energy, which can be expressed in terms of flow and its divergence:

$$\left\langle \Psi^{\dagger} \mid \mathbf{H} \mid \Psi \right\rangle = \underbrace{\left\langle \Psi^{\dagger} \mid \mathbf{T} \mid \Psi \right\rangle}_{\text{kinetic energy}} + \underbrace{\left\langle \Psi^{\dagger} \mid V(x) \mid \Psi \right\rangle}_{\text{potential energy}}$$
$$= \underbrace{\frac{m}{2}\left\langle f \cdot f \right\rangle}_{\text{kinetic energy}} + \underbrace{\left\langle \frac{m}{2} f \cdot f + \frac{\hbar}{2}\nabla \cdot f \right\rangle}_{\text{potential energy}} = E$$

$$\left\langle \nabla \cdot f \right\rangle = -\Gamma \left\langle \nabla^2 \Im \right\rangle = -\Gamma \left\langle \nabla \Im \cdot \nabla \Im \right\rangle = -\frac{2m}{\hbar}\left\langle f \cdot f \right\rangle$$

(5.11)

In this treatment the (ground state) energy is zero because the expected potential energy is balanced by kinetic energy at nonequilibrium steady-state – and both reflect the curvature of surprisal: see (13.1) in Appendix C for details. Compare (5.10) with its classical homologue (7.8) in the absence of random fluctuations. Interestingly, if we consider a free particle moving at the speed of light we recover Einstein's celebrated equality (Einstein, 2013),

$$f(x) = c \Rightarrow E = mc^2$$

(5.12)

Although perhaps not quite in the spirit originally intended. For a complete treatment of the Lagrangian formulation of the Schrödinger equation – and associated Hamiltonian in its coordinate representation – please see (Arsenović et al., 2014).

## Heisenberg uncertainty principle

An insight from the above is an equivalence between momentum and flow that is afforded by the Fourier transform of the wave function. In this setting, $p(x) = \Psi^\dagger \Psi$ and $p(k) = \Phi^\dagger \Phi$ provide probability densities over position and wave number, momentum or flow. The Fourier transform leads naturally to Heisenberg's uncertainty principle, which places a lower bound on the uncertainty (i.e. standard deviation) of these respective distributions:

$$\sigma_x \sigma_p \geq \frac{\hbar}{2} \tag{5.13}$$

This is famously interpreted as being unable to measure precisely both the position and momentum of a particle at the same time. The uncertainty principle can be demonstrated in terms of flow directly from the NESS conditions by assuming (for simplicity) a Gaussian density over any state[18]. In terms of the dispersion or variance of position and flow:

$$\Sigma_x \Sigma_p = E[x \cdot x] E[mf \cdot mf]$$
$$= \left(\frac{\hbar}{2}\right)^2 \Sigma_x^{-2} E[x \cdot x] E[x \cdot x] = \left(\frac{\hbar}{2}\right)^2$$

$$\Im(x) = \frac{1}{2} x \cdot \Sigma_x^{-1} \cdot x \tag{5.14}$$
$$f(x) = -\frac{\hbar}{2m} \nabla \Im = \frac{\hbar}{2m} \nabla^2 \Im \cdot x$$
$$\Sigma_x^{-1} = -\nabla^2 \Im$$

In short, the Heisenberg uncertainty principle means that – at nonequilibrium steady state – if the state of a particle is known precisely, then there is uncertainty about its flow and *vice versa*. Heuristically, this means that if the state has most of its probability mass around a particular point in state-space, the flow must be vigorously rebuilding gradients – in all directions – to counter the dispersive effects of random fluctuations. This means the probability distribution over the flow is dispersed. It is interesting to reflect upon the fact that (5.13) follows directly from the NESS lemma, without reference to the postulates of quantum theory.

## Inference, measurement and wave function collapse?

Deriving the Schrödinger equation from the Fokker Planck equation, or from the path integral formulation, is in itself unremarkable. The interesting aspects of the above derivations follow from substituting the NESS flow into the Schrödinger equation. This substitution suggests that momentum in quantum treatments corresponds to mass times flow; thereby admitting a perspective on wave particle duality in which the density over flow becomes a density over momentum. Furthermore, the second-order statistics of the flow can be associated with kinetic energy by analogy with classical mechanics (see below). Note that both the kinetic energy and Schrödinger potential reflect the curvature of the wave function. Heuristically, this means that a high energy density has a high negative curvature, with pronounced peaks that can be decomposed into potential and kinetic parts – and quantified via application of the appropriate operator to the wave function.

---

[18] Note that the inequality in (5.13) becomes an equality under Gaussian assumptions.

One might ask, where does quantal behaviour come from? One can intuit the discrete energies associated with wave functions by imagining a one-dimensional state-space with periodic boundary conditions. The implicit continuity constraints mean that the spatial wavelengths (inverse wave number) are restricted, so that a finite number of wavelengths 'fit into' the periodic support of state-space. Intuitively, the dispersive effects of random fluctuations mandate a smooth solution to the probability density dynamics that, if equipped with a radial symmetry, can only adopt a finite number of solutions. As a practical example, the (analytic) solutions for a hydrogen atom in spherical polar coordinates $x = (r, \theta, \phi)$ obtain from the amplitude of random fluctuations and Schrödinger potential as follows:

$$\Gamma = \frac{\hbar}{2m}, \quad V(x) = -\frac{\hbar^2}{a_0 m_e r} \Rightarrow$$

$$\Psi_{nlm}(x) = \sqrt{Z} e^{-\rho/2} \rho^l L_{n-\ell-1}^{\ell+\ell+1}(\rho) Y_\ell^m(\theta, \phi)$$
$$\rho = \frac{2r}{na_0}, \quad Z = \left(\frac{2}{na_0}\right)^3 \frac{(n-\ell-1)!}{2n(n+\ell)!}$$

(5.15)

Where $m_e$ is electron mass, $a_0$ is the Bohr radius and $L_{n-\ell-1}^{\ell+\ell+1}$ and $Y_\ell^m$ are generalized Laguerre polynomials (of degree $n - \ell - 1$) and spherical harmonics (of degree $\ell$ and order $m$) respectively. Here, $n$, $\ell$, $m$ are the principal, azimuthal, and magnetic quantum numbers that characterize the discrete (orthonormal) solutions and endow the hydrogen atom with quantal behaviour.

The application of quantum operators is usually interpreted in terms of a measurement – and discussed in terms of an implicit *collapse* of the wave function. In this monograph, we elude the interpretational issues of wave function collapse, because measurement is an inherent part of the nonequilibrium steady-state. In other words, measurement corresponds to a probabilistic mapping between states external to the measuring system and the internal states of the observer. Happily, apparent wave function collapse is predicted when a superposition forms between a particle's states and its external states (i.e., the blanket states of a measuring system or observer). Crucially, the combined wave function of the system and observer continue to obey the Schrödinger equation (Zurek, 2009). We will consider measurement and inference in Part Three – in terms of a generalised synchrony between internal and external states. This means that wave function collapse is essentially redundant on the current view, which is therefore more consistent with the Bohm, ensemble or 'many worlds' interpretations (Ballentine, 1970; Bohm, 1952; Garriga and Vilenkin, 2001).

The correspondence between the different interpretations of wave collapse or quantum decoherence and the perspective offered by the NESS lemma rests upon how close these interpretations fall under the perspective above. The underlying (nonequilibrium steady-state) solution is a probability distribution or wave function over the states of a system that is evolving extremely quickly. To the extent that the system is at nonequilibrium steady-state, the state-space averages of measurable quantities (i.e., eigenstates) will correspond to the time average. This can be variously interpreted as an ensemble of states or many worlds that are statistically indistinguishable from the fast evolution of a single trajectory on its random dynamical attractor. Interestingly, this fast aspect is guaranteed by the postulates underlying the quantum formulation. This follows from the fact that mass must be smaller than the Planck constant $\Gamma > 1 \Rightarrow m < \frac{\hbar}{2}$, which means that – for any measurable momentum – the flow will be almost instantaneous. In short, one can construe quantum mechanical behaviour in terms of immensely fast flows on attracting manifolds that can only be quantified (measured) in a probabilistic sense, where the Fourier transform of the ensuing ensemble density (or wave function) endows the states of a particle with wavelike properties.

The notions of momentum and kinetic energy above inherit from classical mechanics, where momentum has been associated with mass times the expected flow – and mass stands in for the precision of random fluctuations. Momentum is not, at this stage, an attribute of velocity. In other words, the flow or *motion of states* is not the *state of motion*. In a later section (on classical mechanics), we will introduce coordinates of motion, where some (i.e., sensory) states become the higher-order motion of other (i.e., active) states (through the equations of motion). This is a subtle point that explains why we can talk about momentum in the setting of a single quantum state – in contrast to classical mechanics that typically invoke generalised states to include position and momentum. Reassuringly, we will see later that the motion of a state becomes the state of motion, in the classical limit of small random fluctuations.

## Summary

Readers familiar with quantum mechanics will appreciate that the above treatment is somewhat light touch. There are many issues in quantum mechanics that are beyond the reach of the current analysis, which aims at understanding self-organisation at the macroscopic scale of living systems. As such, it is less concerned with microscopic scales, where random fluctuations predominate – or large (astronomic) scales. This section just serves to bridge between the small and larger scales considered later. There are many prescient connections between entanglement, quantum information theory and thermodynamics that speak to this bridge (D'Alessio et al., 2016; Esposito et al., 2009; Parrondo et al., 2015); in particular, quantum fluctuation theorems (Alhambra et al., 2016a; Alhambra et al., 2016b; Holmes et al., 2019)[19]. The next section introduces fluctuation theorems, via the stochastic thermodynamics of ensembles of particles. Part Three then introduces an integral fluctuation theorem for 'measurement', in terms of the information geometries induced by Markov blankets; c.f., (Sengupta and Friston, 2017). At that point, we will briefly revisit the measurement problem and the notion of wave function collapse.

---

[19] My thanks to Peter Morgan and Biswa Sengupta for correspondence on this issue.

It is interesting to stand back from the nonequilibrium steady-state and quantum formulations and ask what they offer each other. The NESS lemma tells us that if we know the amplitude of random fluctuations and the flow of a system, then there is an eigensolution for the NESS density. Alternatively, given the density and amplitude of random fluctuations, we can compute the expected flow – and associated Schrödinger potential. The quantum formulation, on the other hand, specifies the solution for a NESS density (via the wave function) that rests upon knowing the amplitude of random fluctuations (or equivalently reduced mass) and the Schrödinger potential. Crucially, the Schrödinger potential can be derived from the ensemble density in terms of its gradient and curvatures: see Equation (5.4). However, the NESS density can only be recovered from the Schrödinger potential via solution of the wave equation, which is intractable in many instances. Figure 13 tries to make this point using numerical analyses of a particle from our synthetic soup. The key point made in this figure is that one can derive the Schrödinger potential (and associated wave functions) from the flow (and associated NESS density).

In conclusion, one has a complete description of (quantum) behaviour provided one can specify the amplitude of random fluctuations (or reduced mass) and the NESS density (or Schrödinger potential). Table 2 provides some common examples that speak to the breadth of systems that can be described in this way. However, we will now turn to the behaviour of ensembles of particles, under the (ensemble) assumption that they are sufficiently similar to share the same constraints on their flow.

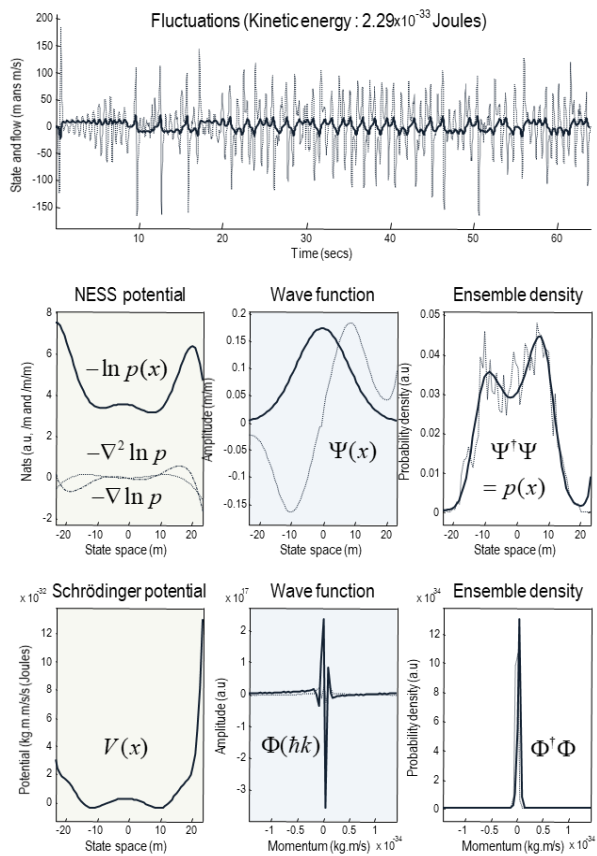

A *particular quantum mechanics*. This figure illustrates the quantum treatment of a single state – a microstate from an external particle of our synthetic soup. The aim of this example is to show how one can characterise the dynamics of the state in terms of the Schrödinger potential and ensuing kinetic energy. Furthermore, this example illustrates how one can eschew the solution of the Schrödinger equation using the NESS lemma. Here, we will consider a single (micro) state in isolation by assuming its flow is a (linear) mixture of the marginal or expected flow under all other states and some fast, random fluctuations. Although we know a lot about how these fluctuations are generated, we will treat them as stochastic and sufficiently fast that the only interesting behaviour is captured by the Schrödinger potential. The timeseries is shown in the upper panel in terms of the state (solid line – arbitrarily assigned units of metres) and flow (dotted line). The sample distribution of states over time was evaluated in terms of the NESS potential using a sixth order polynomial fit to the negative logarithm of the sample density over 64 bins. The resulting estimate and its derivatives are shown in the left middle panel. From these, Equation (5.4) specifies the Schrödinger potential (left lower panel). One can then solve the Schrödinger equation to evaluate the wave function over position in state-space (middle panel) and its Fourier transform, over momentum (lower middle panel). The corresponding densities over position and momentum are shown in the right panels, superimposed upon the corresponding sample densities. Finally, the density over momentum specifies the kinetic energy via Equation (5.11). Here, the kinetic (and potential) energy was $2.29 \times 10^{-33}$. To quantify this energy (and the Schrödinger potential) one needs the amplitude of the random fluctuations – or, equivalently, the reduced mass. This can be simply computed from the residuals of the flow having removed its expectation or marginal flow. The reduced mass of this quantum system was $5.52 \times 10^{-38}$. This concludes a description of how the Schrödinger equation can be applied to characterise nonequilibrium steady-state dynamics. However, *this is not how the results in this figure were generated*: they were derived directly from the NESS potential without solving the Schrödinger equation. In other words, the ensemble density is specified directly by the NESS potential, which means that the wave function (and its Fourier transform) can be specified directly from the ensemble density. Here, we somewhat arbitrarily split the ensemble density into a symmetric Gaussian component and an asymmetric (positive) residual. We then assigned the (square root of the) two components to the real and imaginary parts of the wave function. This complementary derivation of the wave function illustrates the point made in the main text; namely, one can either generate the wave function directly from ensemble density or one can start from the Schrödinger potential and solve the Schrödinger equation.

TABLE 2

Some common (analytic) solutions to the Schrödinger wave equation

| System | Schrödinger potential | Fluctuations | Remarks |
|---|---|---|---|
| **Free particle** | $V(x) = 0$ | $\Gamma = \frac{\hbar}{2m}$ | In the absence of a potential the particle is free, and the wave function has an exponential solution |
| **Quantum harmonic oscillator** | $V(x) = \frac{1}{2}m\omega^2 x^2$ | $\Gamma = \frac{\hbar}{2m}$ | The quadratic potential well gives Gaussian solutions modulated by Hermite polynomials |
| **Electrostatic potential** | $V(r) = -\frac{e_1 e_2}{4\pi\varepsilon_0 r}$ | $\Gamma = \frac{\hbar}{2m}$ | This potential corresponds to Coulomb potential energy for two point charges $e_1$ and $e_2$ |
| **Hydrogen atom** | $V(r) = -\frac{\hbar^2}{a_0 m_e r}$ $= -\frac{e^2}{4\pi\varepsilon_0 r}$ | $\Gamma = \frac{\hbar}{2m}$ $m = \frac{m_p m_e}{m_p + m_e}$ | Here, $m_p$ and $m_e$ correspond to the mass of a positron and electron respectively and $e$ is electron charge |
| **Pöschl-Teller** | $V(x) = -\frac{\lambda(\lambda+1)}{2}\operatorname{sech}^2(x)$ | $\Gamma = \frac{\hbar}{2m}$ | These solutions are Legendre functions of tanh(x). |

# A theory of lots of little things – statistical mechanics

"*If physical theories were people, thermodynamics would be the village witch. Over the course of three centuries, she smiled quietly as other theories rose and withered, surviving major revolutions in physics, like the advent of general relativity and quantum mechanics. The other theories find her somewhat odd, somehow different in nature from the rest, yet everyone comes to her for advice, and no-one dares to contradict her.*" (Goold et al., 2016); p1.

In this section, we move from the behaviour of one small particle to an ensemble of small particles; for example, a large number of internal particles surrounded by their Markov blanket. In doing so, we appeal to the *ensemble* assumption above. In other words, the states of the ensemble are partitioned such that the states of each constituent particle can be identified with the homologous states of another. This enables one to associate the NESS density with an *ensemble* density. In other words, instead of describing the probability of sampling a single particle in a particular state over time, the NESS density also describes the number of particles occupying the same (or neighbourhood) states. Crucially, at nonequilibrium steady-state all particles share *the same* ensemble density and their flow can be described by *the same* ensemble potential (i.e., the NESS potential) However, prior to nonequilibrium steady-state the (thermodynamic) potential describing the flow cannot be the ensemble potential; otherwise the NESS lemma would be satisfied. We will see that the implicit divergence between the two corresponds to thermodynamic free energy. Note that the ensemble assumption entails a weak coupling assumption, which will be important in the second half of this section, when we connect the NESS formulation to important results in statistical mechanics, such as the *Jarzynski relation*.

In brief, we will see that being able to relate the flow of particles to an ensemble density means that we can describe trajectories in terms of potentials and forces. In turn, this equips trajectories with properties such as *work* and *stochastic entropy*, leading to balance equations that constitute the first and second laws of thermodynamics. These balance equations can be expressed at the level of a single trajectory (of a single particle) or the ensemble average; leading to (fluctuation) theorems that place constraints on the expected evolution of measurable quantities. Effectively, these fluctuation theorems generalise and extend the second law. This section follows the treatment in (Seifert, 2012) who provides an encyclopaedic overview of stochastic thermodynamics – and how it underpins statistical mechanics (through the notion of ensemble averages over individual trajectories). Because it deals with time varying densities, we will focus on the emergence of nonequilibrium steady-state and what this entails for quantities like thermodynamic free energy and entropy production.

# Stochastic thermodynamics

Consider an ensemble of particles: $\{\pi_1^{(i)}, \pi_2^{(i)}, \ldots, \}$, that are internal to a Markov blanket at a higher level, playing the role of a container, heat bath or reservoir. From (4.2) the flow of the $k$-th particular state can be expressed in Langevin form.

$$
\begin{aligned}
\dot{\pi}_k^{(i)} &= f(\pi_k^{(i)}, b^{(i)}(\tau)) + \omega_k^{(i)} \\
&= f(\pi_k^{(i)}, \tau) + \omega_k^{(i)}
\end{aligned}
\tag{6.1}
$$

This expresses the dependency on the blanket states of other particles as a time-dependent flow. For notational simplicity, we will drop subscripts and superscripts and use $\pi \triangleq \pi_k^{(i)}$ to denote particular states that constitute a (thermodynamic) ensemble. To develop a (statistical) physics of ensembles, we start by expressing stochastic (Langevin) dynamics in terms of a thermodynamic potential $U(\pi, \tau)$ and associated forces $f_m(\pi, \tau)$, using the Helmholtz decomposition:

$$
\begin{aligned}
f(\pi, \tau) &= (\mu_m - Q_m) f_m(\pi, \tau) \\
&= (Q_m - \mu_m) \nabla U(\pi, \tau)
\end{aligned}
$$

$$
\begin{aligned}
f_m(\pi, \tau) &\triangleq -\nabla U(\pi, \tau) \\
Q &\triangleq Q_m k_B T \\
\Gamma &\triangleq \mu_m k_B T
\end{aligned}
\tag{6.2}
$$

The last equality is known as the Einstein–Smoluchowski relation, where $\mu_m$ is a mobility coefficient. This means, we have factorised the amplitude of random fluctuations $\Gamma = \mu_m k_B T$, into *mobility* and *temperature*. The Boltzmann constant $k_B$ has a dimension of energy divided by temperature, and units of joule per Kelvin (J/K) in the International System of Units. To ensure a consistent dimensional analysis, we will use a scaled version of (Shannon) entropy or expected surprisal; namely, thermodynamic entropy:

$$
\begin{aligned}
S(\tau) &\triangleq k_B H(\tau) \\
H(\tau) &\triangleq E[\Im(\pi, \tau)]
\end{aligned}
\tag{6.3}
$$

Notice that flow $f = (\mu_m - Q_m) f_m$ is expressed in terms of a *force* $f_m = -\nabla U$, which brings a new semantics to the table. Furthermore, the thermodynamic potential is only defined to within an additive constant, because its gradients predict stochastic flow in a least squares sense, by (6.1). We can therefore associate the thermodynamic potential with an *ensemble density* $p(\pi, \tau)$ and corresponding (time-dependent) surprisal $\Im(\pi, \tau)$. This defines a partition function (i.e., normalisation constant) $Z$ that enables the thermodynamic potential to be expressed as a probability density:

$$p(\pi,\tau) \triangleq \exp[-\Im(\pi,\tau)] = \tfrac{1}{Z(\tau)}\exp[-\tfrac{1}{k_B T(\tau)}U(\pi,\tau)]$$
$$k_B T \cdot \Im(\pi,\tau) = U(\pi,\tau) - F_m(\tau) \qquad (6.4)$$
$$F_m(\tau) = -k_B T(\tau)\ln Z(\tau)$$

So far, we have just formalised the dynamics of an ensemble in terms of a thermodynamic potential and some constants. As the ensemble approaches nonequilibrium steady state, the thermodynamic potential becomes proportional to surprisal (i.e., the NESS potential), where temperature is the constant of proportionality:

$$\Im(\pi) = \lim_{\tau \to \infty} \Im(\pi,\tau)$$
$$\Rightarrow U(\pi) = k_B T \cdot \Im(\pi) + F_m$$
$$\Rightarrow f(\pi) = (Q_m - \mu_m)\nabla U(\pi) = (Q-\Gamma)\nabla\Im(\pi) \qquad (6.5)$$
$$\Rightarrow \dot{p}(\pi) = 0$$

Otherwise, the NESS and thermodynamic potentials data differ, as the ensemble density evolves over time. We can express this in terms of a potential energy difference, whose expectation under the NESS density is a thermodynamic free energy:

$$F_\Im(\tau) \triangleq E[U(\pi,\tau)] - E[k_B T \cdot \Im(\pi)]$$
$$= E[U(\pi,\tau)] - k_B T \cdot H$$
$$= E[U(\pi,\tau)] - T \cdot S$$

$$= k_B T \cdot E[\Im(\pi,\tau) - \Im(\pi)] + F_m(\tau) \qquad (6.6)$$
$$= k_B T \cdot D[p(\pi) \parallel p(\pi,\tau)] + F_m(\tau) \Rightarrow$$
$$\Delta F_m = k_B T \cdot D[p(\pi) \parallel p(\pi,\tau)] = F_\Im(\tau) - F_m(\tau)$$

This generalises the well-known thermodynamic relationship between thermodynamic entropy, internal energy and free energy to non-steady-state dynamics, where entropy is expected surprisal and internal energy is the expected thermodynamic potential. This formulation defines entropy in relation to the nonequilibrium steady-state solution, where thermodynamic free energy is the relative entropy (i.e., the KL divergence) between the NESS and ensemble densities. In other words, it describes the evolution of an ensemble in terms of thermodynamic free energy minimisation to nonequilibrium steady-state (contrast this with the notion of increasing entropy, characterised by the second law). Effectively, this formulation says that any interesting ensemble (that has measurable characteristics) must have a random dynamical attractor to which it converges. On this view, the thermodynamic free energy decreases as the ensemble approaches nonequilibrium steady-state (i.e., its random dynamical attractor). At nonequilibrium steady-state, the divergence disappears, free energy is minimised and $\Delta F_m = 0 \Leftrightarrow F_\Im = F_m$. See Figure 14 for a numerical example using our synthetic soup. Note that the thermodynamic free energy plays the same role as the divergence characterising symmetry breaking in (2.9). However, here, we are dealing with a time-dependent ensemble density over particular states, not the propagation of uncertainty from an initial particular state. Furthermore, in this thermodynamic setting, the expectations are under the NESS density. This means that the entropy $H = E_{p(\pi)}[\Im(\pi)]$ in (6.6) does not change with time – all the heavy lifting is done by thermodynamic free energy, in terms of characterising convergence to nonequilibrium steady-state. However, one can still consider the entropy of the ensemble density $H(\tau) = E_{p(\pi,\tau)}[\Im(\pi,\tau)]$, as we will see below.

Heuristically, (6.6) is consistent with the notion of free energy as the thermodynamic energy available to do work when an ensemble is far from equilibrium. However, to talk about heat and work, we need to connect stochastic dynamics to first law quantities. We can do this via the stochastic energetics of state trajectories.

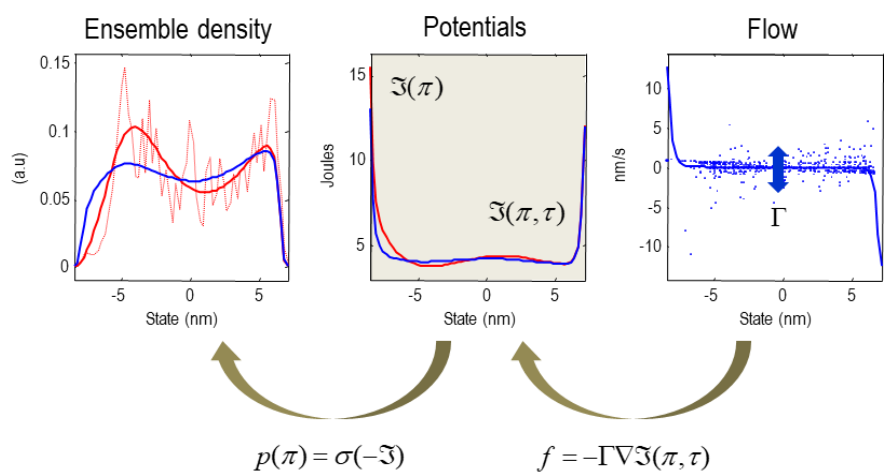

**FIGURE 14**

*Ensemble densities and thermodynamic potentials*. These illustrative potentials and density functions are taken from the analysis in the subsequent figure. In brief, they characterise the stochastic thermodynamics of an ensemble of particles; here, the external states of our synthetic soup (towards the end of the simulation). This characterisation involves estimating two functions of phase or state-space. The first (shown in red) is the surprisal or self-information that characterises the NESS density $\Im(\pi)$. The second (shown in blue) is a homologous potential energy function $\Im(\pi,\tau)$, whose gradients predict the flow at each point in phase-space. At nonequilibrium steady-state, these two functions are the same. In other words, the thermodynamic potential becomes self-information (multiplied by temperature). This means that the distance or, more strictly speaking, divergence from steady-state can be quantified in terms of the KL divergence between the associated probability density functions (shown on the left). When these densities converge, the ensemble density stops changing and becomes the NESS density; i.e., $\Im(\pi,\tau) = \Im(\pi)$. The **middle panel** shows estimates of the ensemble and thermodynamic potentials (i.e., surprisals). The ensemble potential was estimated using a polynomial approximation to the (log) sample distribution over the ensemble of states. The corresponding density functions are shown in the **left panel**, where the sample distribution is shown as a dotted line. The **right panel** shows the gradients of the thermodynamic potential (blue line) that predicts the flow sampled by the simulation (dots). Please see main text for a description of the variables in this figure.

## Stochastic energetics

Following (Sekimoto, 1998), one can endow Langevin dynamics with a thermodynamic interpretation, using constructs from the first law, by considering an individual fluctuating trajectory. For example, one can express the first law in terms of work and dissipated heat for a single trajectory[20]:

$$\mathbf{w}(\pi[\tau]) = \mathbf{q}(\pi[\tau]) + \Delta U$$
$$\Delta U = U(\pi_\tau) - U(\pi_0)$$

$$(6.7)$$

$$\mathbf{q}(\pi[\tau]) = \int_0^t \dot{\mathbf{q}} d\tau = \int_0^t f_m(\pi,\tau) \cdot \dot{\pi} d\tau$$

The second equality here is an integrated first law, relating work to force times distance. Similarly, stochastic entropy production can be defined in terms of dissipated heat and non-dissipative or conservative entropy production. This rests on associating the stochastic entropy of a single trajectory with its surprisal or self-information, using $\Gamma \nabla p = f(\pi,\tau) p(\pi,\tau) - j(\pi,\tau)$ from (1.6):

$$\Im(\pi,\tau) = -\ln p_\tau : p_\tau \equiv p(\pi,\tau)$$

$$\dot{\Im}(\pi,\tau) = -\frac{\dot{p}_\tau}{p_\tau} = -\frac{\partial_\tau p_\tau}{p_\tau} - \frac{\nabla p_\tau}{p_\tau} \cdot \dot{\pi}(\tau)$$

$$= \underbrace{-\frac{\partial_\tau p_\tau}{p_\tau} + \frac{j \cdot \dot{\pi}(\tau)}{\Gamma p_\tau}}_{\Im^p \, conservative} - \underbrace{\frac{f \cdot \dot{\pi}(\tau)}{\Gamma}}_{\Im^q \, dissipative} = \dot{\Im}^p - \dot{\Im}^q$$

$$(6.8)$$

This means that the rate of change of stochastic entropy (i.e., self-information) can be decomposed into *dissipative* and *conservative* (i.e., sometimes referred to as *total*) parts: where the dissipative part corresponds to stochastic entropy lost via the dissipation of heat:

$$\dot{\mathbf{q}}(\pi,\tau) = f_m \cdot \dot{\pi}$$
$$\dot{\Im}^q(\pi,\tau) = \frac{1}{\Gamma} f \cdot \dot{\pi} = \frac{1}{k_B T} f_m \cdot \dot{\pi} = \frac{1}{k_B T} \dot{\mathbf{q}}$$

$$(6.9)$$

It is interesting to note that this dissipative part corresponds to the path-independent term of the Lagrangian, which underwrites the action of a trajectory. Ignoring solenoidal flow, from (1.2) we have:

---

$$\mathcal{L}(\pi, \dot{\pi}) = \underbrace{\tfrac{1}{4\Gamma}(\dot{\pi} \cdot \dot{\pi} + \tfrac{1}{4\Gamma} f \cdot f) + \tfrac{1}{2}\nabla \cdot f}_{path\text{-}dependent} - \underbrace{\tfrac{1}{2\Gamma}\dot{\pi} \cdot f}_{independent}$$

$$= \underbrace{\tfrac{1}{4\Gamma}(\dot{\pi} \cdot \dot{\pi} + \tfrac{1}{4\Gamma} f \cdot f) + \tfrac{1}{2}\nabla \cdot f}_{non\text{-}dissipative} - \underbrace{\tfrac{1}{2k_B T}\dot{\mathbf{q}}}_{dissipative}$$

(6.10)

This is the part of the action that is odd under time reversal: $\pi^{\dagger}(\tau) \equiv \pi(t - \tau)$ where,

$$\mathcal{A}(\pi^{\dagger}[\tau]) - \mathcal{A}(\pi[\tau]) = \int_0^t \mathcal{L}(\pi(t-\tau), \dot{\pi}(t-\tau)) - \mathcal{L}(\pi(\tau), \dot{\pi}(\tau)) d\tau$$

$$= \ln \frac{p(\pi[\tau] \mid \pi_0)}{p^{\dagger}(\pi^{\dagger}[\tau] \mid \pi_0^{\dagger})} = \Delta \mathfrak{I}^q = \tfrac{1}{k_B T}\mathbf{q}$$

(6.11)

In other words, the only part of the Lagrangian that does not cancel – in the path integral under time reversal – is the dissipative or path-independent term. Intuitively, if one walks uphill and then retraces the path downhill, the heat dissipated depends only on how high one climbed. Equation (6.11) affords a deep physical interpretation: the log-ratio of the probability of a forward and time-reversed trajectory between any two points is the heat dissipated along the forward trajectory. This ratio underwrites the fluctuation theorems considered in the final part of this section. We will see later (6.18) that the expected conservative stochastic entropy production along any path is greater than zero, which can be considered as a refinement of the second law for a single trajectory (Seifert, 2012).

The entropy *per se* corresponds to the ensemble average of surprisal, with the corresponding entropy production (Ao, 2008; Friston and Ao, 2012; Seifert, 2012). In brief, using $\dot{p}_\tau = -\nabla \cdot j_\tau$ and $j_\tau = f_\tau p_\tau - \Gamma \nabla p_\tau$ from (1.6), the entropy production can be expressed as:

$$H(\tau) = E[\mathfrak{I}(\pi, \tau)] = \int p_\tau \mathfrak{I}_\tau dx$$

$$\dot{H}(\tau) = \int \dot{p}_\tau \mathfrak{I}_\tau + p_\tau \dot{\mathfrak{I}}_\tau dx = \int \dot{p}_\tau \mathfrak{I}_\tau - \dot{p}_\tau dx = \int \dot{p}_\tau \mathfrak{I}_\tau dx$$

$$= -\int \nabla \cdot j_\tau \mathfrak{I}_\tau dx = \int j_\tau \cdot \nabla \mathfrak{I}_\tau dx = -\int \frac{j_\tau}{p_\tau} \cdot \nabla p_\tau dx$$

$$= \dot{H}^{\omega}(\tau) - \dot{H}^f(\tau)$$

$$= \dot{H}^p(\tau) - \dot{H}^q(\tau)$$

(6.12)

$$= \underbrace{\int \frac{\nabla p_\tau \cdot \Gamma \cdot \nabla p_\tau}{p_\tau} dx}_{H^{\omega} \geq 0 \; fluctuations} - \underbrace{\int f(x, \tau) \cdot \nabla p_\tau dx}_{H^f \; flow}$$

$$= \underbrace{\int \frac{j(x, \tau) \cdot j(x, \tau)}{\Gamma p_\tau} dx}_{H^p \geq 0 \; non\text{-}dissipative} - \underbrace{\int \frac{j(x, \tau) \cdot f_m(x, \tau)}{k_B T} dx}_{H^q \; dissipative}$$

The final two equalities express entropy production in terms of probability gradients and currents respectively, where the last expression eliminates the gradients using $j_\tau = f_\tau p_\tau - \Gamma \nabla p_\tau$. The penultimate equality highlights the relative contributions of random fluctuations $\dot{S}^{\omega} = k_B \dot{H}^{\omega} \geq 0$ and flow $\dot{S}^f = k_B \dot{H}^f$. Intuitively speaking, random fluctuations always increase the entropy through dispersing the ensemble density, while flow decreases entropy by 'rebuilding' probability gradients (i.e., where probability currents flow towards regimes of greater density). In other words, random fluctuations disperse states into regimes of high surprisal (and implicitly thermodynamic potential), while gradient flows – due to forces – counter the implicit dispersion. On this view, the random fluctuations are 'blind' to potential energy gradients (e.g., barriers) that induce flows or forces – and therefore underwrite [thermo] dynamics.

At nonequilibrium steady-state, these opposing effects are in balance and entropy production is zero. The second expression, in terms of probability currents, connects with statistical mechanics, because the second term $\dot{S}^q = k_B \dot{H}^q$ is the ensemble average of the thermodynamic entropy generated via the work induced by forces. At NESS, both components of thermodynamic entropy disappear as probability currents fall to zero; i.e., the conservative flow and dissipative fluctuations are balanced. These balance equations for ensemble entropy constitute the *second law of thermodynamics*. Figure 15 provides an illustrative example, using our simulated soup, in which the above relations are used to track the evolution of the temperature, free energy, and entropy of particles that constitute the principal Markov blanket and its internal states. In this example, we used the standard value for the Boltzmann constant (see Glossary) to provide quantitative characterisations.

In summary, through an ensemble assumption about particular states, a series of balance equations emerge that have a straightforward interpretation in terms of work and entropy; effectively reconstituting the laws of thermodynamics. These balance equations can be applied at the level of single trajectories or ensemble averages. We now conclude this section with a closer look at generalisations of the second law in terms of fluctuation theorems.

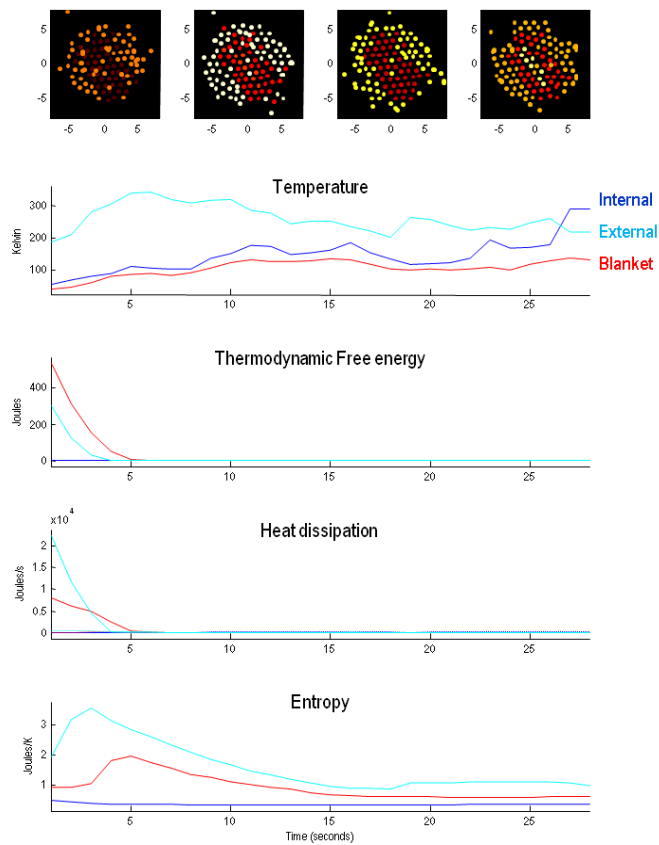

FIGURE 15

*Stochastic thermodynamics*. This figure illustrates the characterisation of our synthetic soup (or active matter) in terms of classical (stochastic) thermodynamics. The analysis is fairly straightforward and proceeds as follows: first, any set of particular states (e.g., the internal macromolecules of our synthetic virus) can be treated as an ensemble. In other words, the behaviour of any one particle can be treated as if it was responding to the same thermodynamic potential as all remaining particles. One can then evaluate the thermodynamic potential that best explains the flow of particular states (see Figure 14). Given this expected or predicted flow, one can then evaluate the variance or amplitude of random fluctuations and – equipped with a mobility coefficient – the corresponding temperature at any point in time (we used a mobility of 0.2 and assumed position has units of nanometres). Note that this is possible because we are associating the distribution over the states of the ensemble with the equivalent statistics that would have been observed over time. Given the temperature and thermodynamic potential, one can then evaluate the free energy using the KL divergence between the associated ensemble and NESS densities. As the ensemble states approach their random dynamical attractor (i.e., pullback attractor or nonequilibrium steady-state), these density functions converge, and the ensemble density ceases to change. At this point, it becomes the NESS density. The implicit changes in the ensemble density over time can be characterised in terms of entropy production, which can be partitioned in a number of ways (see main text). In the example here, we have focused on the entropy dissipated by probability currents that, when multiplied by temperature, corresponds to heat dissipation. In this somewhat heuristic illustration, we have (for simplicity) focused on the position in one dimension of the internal states surrounded by the principal Markov blanket (considering a single dimension means we can ignore solenoidal flow). This blanket is the small virus like particle in Figure 6. To estimate the ensemble and NESS potentials, we used a fourth order polynomial expansion of position (and appropriate least squares estimators). The thermodynamic potential is that which best predicts the stochastic flow of particular states; where the ensemble density best predicts the sample density. To obtain more efficient estimators, we also averaged over 256 time bins at 28 consecutive intervals during the evolution of the system. We started at the 32nd time bin to illustrate the thermodynamic correlates of self-organisation, during which the principal Markov blanket was formed. For interest, we repeated the analysis for the blanket and external particles. The **upper row** shows the evolution of temperature shown using a (hot) colour scale with a dot at the position of the particles (in two dimensions). The **second panel** shows the corresponding evolution of temperature in the three ensembles as a function of time. The interesting thing here is that the internal (blue) and blanket (red) particles start off at about the same temperature. However, during the course of self-organisation, the internal particles slowly increase their temperature to become hotter than the external particles (cyan). In this example, the temperature of the internal particles ended up being about twice the temperature of the blanket particles. The **third panel** shows the corresponding free energy for each ensemble. The most notable thing here is that free energy decreases with time as the ensemble and NESS potentials converge. This is most marked for the external and blanket particles – that could be thought of as spending their free energy to organise the internal particles. After about five seconds, there is relatively little free energy left within the system. This is reflected in the **bottom panel** that shows the corresponding heat dissipation, which is most marked for the external particles, as might be guessed from the changes in thermodynamic free energy. Although heat dissipation can fall to low levels – as nonequilibrium steady-state is approached – the temperature of our synthetic virus remains relatively high (here, the temperature reached about 300° Kelvin, which is roughly body temperature). This follows from the fact that random fluctuations are still in play – arising from intrinsic fluctuations of the internal states of each internal particle (at the underlying hierarchical scale). These fluctuations disperse particular states over the thermodynamic potential, while gradient flows down potential energy gradients reconstitute the ensemble density. Effectively, this flow (times distance) produces heat that is generated by intrinsic fluctuations. At nonequilibrium steady-state, these two processes are in balance and heat dissipation is eliminated; because probability currents are zero at all points in state-space. The **lower panel** shows the corresponding entropy as a function of time. The entropy of external particles increases initially and then falls as the system finds its random dynamical attractor. Note that the entropy of particular states (that are destined to become densely coupled internal particles) falls progressively; thereby, violating the second law. This is what we would expect in this far-from-equilibrium scenario. Clearly, this is an idealised description of stochastic dynamics under the ensemble assumption. In reality, the dynamics is more complicated, and our treatment here can be construed as a mean field approximation, where all the interactions within and between the three ensembles are summarised with a common thermodynamic potential and ensemble density. Despite this approximation, this sort of analysis provides an intuitive characterisation of stochastic dynamics, in terms of constructs that underpin the first and second laws of thermodynamics.

## Fluctuation theorems

Fluctuation theorems express universal properties of any density $p(\Omega)$ over phase measures $\Omega(\pi)$ (i.e., work, heat or entropy) evaluated along fluctuating trajectories taken from ensembles with well-defined initial distributions. Following (Seifert, 2012), we will briefly consider a phenomenological classification into three classes; namely, *integral fluctuation theorems*, *detailed fluctuation theorems* and (Generalized) *Crooks fluctuation theorems* – with a special focus on integral fluctuation theorems and, in particular, the Jarzynski equality (Jarzynski, 1997) Please see Table 3 for a summary of this classification.

<div align="center">TABLE 3</div>

<div align="center">Phenomenological classification of fluctuation theorems</div>

| Theorem and relation | Remarks |
|---|---|
| **Integral fluctuation theorems (IFT)**<br><br>**A functional $\Omega(\pi[\tau])$ obeys an integral fluctuation theorem if**<br><br>$E[\exp(-\Omega)] = 1 \Rightarrow$<br>$\qquad E[\Omega] \geq 0$ | The convexity of the exponential function implies $E[\Omega] \geq 0$ recapitulating the well-known thermodynamic inequality related to the second law. The IFT implies the existence of trajectories for which $\Omega(\pi)$ is negative, which have been characterized as 'violating' the second law. However, the probability of such events is exponentially small (for sufficiently large systems). This reconciles the validity of macroscopic thermodynamics with the occurrence of such violations. |
| **Detailed fluctuation theorems (DFT)**<br><br>**A detailed fluctuation theorem corresponds to the stronger relation:**<br><br>$p(-\Omega)/p(\Omega) = \exp(-\Omega) \Rightarrow$<br>$\qquad E[\Omega] \geq 0$ | This relation constrains 'one half' of the density function. In other words, the even moments of $\Omega(\pi)$ can be expressed by the odd moments and *vice versa*. A DFT implies the corresponding IFT. |
| **(Generalized) Crooks fluctuation theorems**<br><br>**The general form of these theorems can be expressed as**<br><br>$p^{\dagger}(-\Omega) = p(\Omega)\exp(-\Omega) \Rightarrow$<br>$\qquad E[\Omega] \geq 0$ | These relations compare the process of interest with the density of the *same* physical quantity for a 'conjugate' (mostly the time-reversed) process, $p^{\dagger}(-\Omega)$, which implies the IFT (but not the DFT). |

Originally, fluctuation theorems (FT) were derived on a case by case basis; however, stochastic thermodynamics furnishes a unifying treatment of fluctuation theorems by considering trajectories under time reversal (Seifert, 2012). Fluctuation theorems for processes with trajectories $\pi[\tau]$, are most generally derived by invoking a 'conjugate' dynamics for trajectories, $\pi^{\dagger}[\tau]$. The crucial quantity – leading to the fluctuation theorems – is a master functional that corresponds to the log likelihood ratio of conjugate paths:

$$R(\pi[\tau]) \triangleq \ln \frac{p(\pi[\tau])}{p^{\dagger}(\pi^{\dagger}[\tau])} = \ln \frac{p_0(\pi_0)}{p_0^{\dagger}(\pi_0^{\dagger})} + \ln \frac{p(\pi[\tau] \mid \pi_0)}{p^{\dagger}(\pi^{\dagger}[\tau] \mid \pi_0^{\dagger})} \qquad (6.13)$$

This ratio leads to a master fluctuation theorem: for any functional $S^{\dagger}(\pi^{\dagger}[\tau]) = \varepsilon S(\pi[\tau])$ that maps with a definite parity $\varepsilon = \pm 1$ to the conjugate dynamics – thereby representing the same physical quantity – the master fluctuation theorem says that for any function $g$ :

$$E[g(\varepsilon S^{\dagger}(\pi^{\dagger}[\tau]))]^{\dagger} = E[g(S(\pi[\tau])) \exp(-R(\pi[\tau]))] \qquad (6.14)$$

See (Seifert, 2012) p13. With suitable choices of $g$ , one can formulate the fluctuation theorems in Table 1 as special cases of this general theorem. With the simplest choice of $g \equiv 1$ , we obtain the most general from which all known IFT relations follow.

$$E[\exp(-R(\pi[\tau]))] = 1 \qquad (6.15)$$

For example, by choosing the reversed dynamics $\pi^{\dagger}(\tau) = \pi(t - \tau)$ as the conjugate dynamics, from the definition in (6.13) and (6.11) we have:

$$E\left[ \frac{p_0^{\dagger}(\pi_0^{\dagger})}{p_0(\pi_0)} \exp(-\Delta \mathfrak{J}^q) \right] 1 \Rightarrow E\left[ \ln \frac{p_0^{\dagger}(\pi_0^{\dagger})}{p_0(\pi_0)} - \Delta \mathfrak{J}^q \right] \geq 0 \qquad (6.16)$$

Here, $p_0(\pi) \equiv p(\pi, 0)$ is the initial distribution and $\mathfrak{J}^q$ corresponds to the dissipative part of stochastic entropy in (6.8). This equality holds for any initial condition and trajectories of any length. At nonequilibrium steady-state, from (6.9) we have:

$$\dot{\mathfrak{J}}^q(x, \tau) = \frac{f}{\Gamma} \cdot \dot{\pi}(\tau) = -\nabla \mathfrak{J} \cdot \dot{\pi}(\tau) = -\dot{\mathfrak{J}}(\tau)$$

$$\Rightarrow \Delta \mathfrak{J}^q = -\int_0^t \dot{\mathfrak{J}} d\tau = \mathfrak{J}(\pi_0) - \mathfrak{J}(\pi_t) \qquad (6.17)$$

$$\Rightarrow \exp(\Delta \mathfrak{J}^q) = \frac{p(\pi_t)}{p_0(\pi_0)}$$

Substitution into (6.16) shows that the NESS density over the final state $p(\pi_t) = p_0^{\dagger}(\pi_0^{\dagger})$ can be construed as the initial density under a time-reversed protocol. Similarly, the solution of the Fokker-Planck equation gives the IFT for conservative entropy production $\mathfrak{J}^p$ (Seifert, 2012).

$$E[\exp(-\Delta \mathfrak{J}^p)] = 1 \Rightarrow E[\Delta \mathfrak{J}^p] \geq 0 \Rightarrow \Delta S^p \geq 0 \qquad (6.18)$$

This IFT can be considered as a refinement of the second law; namely, that conservative entropy production is expected to be greater than zero (Seifert, 2012). With a time-dependent potential $U(x, b(\tau))$ and control parameter $b(\tau)$ – here, associated with blanket states – one obtains the following IFT as a special case of (6.16), where, following (6.9), (6.7) and (6.4):

$$E\left[\frac{p_0^\dagger(\pi_\tau, b_\tau)}{p_0(\pi_0, b_0)}\exp(-\Delta\mathfrak{I}^q)\right] = E[\exp(-\Delta\mathfrak{I} - \tfrac{1}{k_B T}\mathbf{q})] = 1 \Rightarrow$$
$$E[\exp(-\tfrac{1}{k_B T}\mathbf{w})] = \exp(-\tfrac{1}{k_B T}\Delta F_m)$$

$$\Delta\mathfrak{I}^q = \tfrac{1}{k_B T}\mathbf{q} : \mathbf{q} = \mathbf{w} - \Delta U \qquad (6.19)$$
$$\Delta U = k_B T \cdot \Delta\mathfrak{I} + \Delta F_m$$
$$\Delta\mathfrak{I} = \ln p_0(\pi_0, b_0) - \ln p_0^\dagger(\pi_\tau, b_\tau)$$
$$\Delta F_m = F_m(b_\tau) - F_m(b_0)$$

This is the *Jarzynski relation* (Jarzynski, 1997), relating the work spent – in driving a system from an initial equilibrium with a time-dependent potential – to the thermodynamic free energy difference between the initial and final states. This is the IFT for the (scaled) dissipated work $\mathbf{w} - \Delta F_m$ (Seifert, 2012). The importance of this relation is that it allows one to determine the free energy difference, which is a genuine equilibrium property, from non-equilibrium measurements. In short, it represents a strengthening of the familiar second law that follows as the corresponding (Jensen's) inequality,

$$E[\exp(-\tfrac{1}{k_B T}\mathbf{w})] = \exp(-\tfrac{1}{k_B T}\Delta F_m) \Rightarrow E[\mathbf{w}] \geq \Delta F_m = E[\Delta U] - T \cdot \Delta S \qquad (6.20)$$

which can be compared with (6.6).

## Summary

Interestingly, the original derivation of the *Jarzynski* relation "relies on the usual assumption of weak coupling between system and reservoir" (Jarzynski, 1997) p 2693; which, in the present context, translates into a weak coupling between internal particles and their Markov blanket. This weak coupling is where we started; namely, the ensemble assumption. More generally, the separation of temporal scales between the slowly fluctuating potentials assumed in this section and the fast, intrinsic fluctuations that constitute the particular (internal) states of interest is nicely summarised in Seifert's comprehensive treatment:

"*The identification of states, of work and of internal energy, i.e. of the ingredients entering the first law on the level of trajectories, is logically independent of the assumption of a Markovian dynamics connecting these states. The crucial step is the splitting of all degrees of freedom into slow and fast ones, the latter always being in a constrained equilibrium imposed by the instantaneous values of the slow ones.*" (Seifert, 2012) p44.

# A theory of big things – classical mechanics

In the final section of Part Two, we turn to the behaviour of large particles to see how solenoidal flow engenders a classical mechanics, when intrinsic fluctuations are averaged away. In brief, we will consider the simplest of Markov blankets, where *active states* correspond to *position* and their *momentum* is associated with *sensory states*. We then ask whether there are any lawful dynamics of the average position (and momentum) of blanket states that emerges at nonequilibrium steady-state. We will see that a simple consequence of averaging is Newtonian behaviour or, more generally, classical Hamiltonian or Lagrangian mechanics. Note that the ensuing (average) states of the Markov blanket now constitute the states of a particle at the next hierarchical level. However, we will drop the superscripts and subscripts for clarity.

In terms of the path integral formulation, in the limit of very small random fluctuations, the kinetic part of the Lagrangian dominates, where from (1.9):

$$\mathcal{A}(x[\tau]) = \int_0^t \mathcal{L}(x, \dot{x}) d\tau$$
$$\lim_{\Gamma \to 0} \mathcal{L}(x, \dot{x}) = \tfrac{1}{2}[\tfrac{1}{2\Gamma}(\dot{x} - Q\nabla\Im) \cdot (\dot{x} - Q\nabla\Im)]$$
$$\Rightarrow p(x[\tau]) = \delta(\dot{x} - Q\nabla\Im) \tag{7.1}$$

This means the only allowable paths $\dot{x} = \dot{\mathbf{x}} = f = Q\nabla\Im$ are determined by surprisal and solenoidal flow. We will see below that this renders the surprisal a Hamiltonian – that is conserved over every path, in virtue of the fact the flow is divergence free.

Imagine a simple universe composed of particles whose active and sensory states are low dimensional (e.g., each dominated by three orthogonal eigenmodes). By the nature of the Markovian partition, every particle's active state can only be influenced by its sensory state. In contrast, its sensory state can be influenced by the active states of other particles. With a suitable transformation of variables, we can express this as follows (suppressing intrinsic and random fluctuations):

$$\dot{a}_n = f_a(s_n) = \tfrac{\hbar}{m} \cdot s_n$$
$$\dot{s}_n = f_s(a_1, \ldots, a_N) \tag{7.2}$$

This construction licences us to talk about active states as *position* and sensory states as *momentum* in the following (classical) sense: the rate of change of position (i.e., active states) is velocity and the rate of change of momentum (i.e., rate of change of sensory states) depends on the position of other particles. In short, this simple sort of universe would look as if it was governed by gravitational rules, articulated in terms of particles exerting forces on each other. In this section, we illustrate the implicit association of blanket states with the generalised coordinates (i.e., position and momentum) of classical mechanics.

# Conservative systems

Consider a large particle where the flows of active and sensory states do not depend on themselves. From (1.21), we have:

$$\dot{x} = \begin{Bmatrix} \dot{a} \\ \dot{s} \\ \dot{\mu} \end{Bmatrix} = \begin{Bmatrix} f_a(s,\mu) \\ f_s(a,\eta) \\ f_\mu(b,\mu) \end{Bmatrix} + \omega \qquad (7.3)$$

Consider now the behaviour of the average over $n$ states, formed by the row vector $\xi^- = \frac{1}{n}\mathbf{1}_n$, in terms of the Taylor expansion (for simplicity, we will consider position in one dimension):

$$\begin{Bmatrix} \xi^-\dot{a} \\ \xi^-\dot{s} \\ \dot{\mu} \end{Bmatrix} = \begin{Bmatrix} \xi^-\nabla_s f_a \xi \xi^- s + \xi^-\nabla_\mu f_a \xi \xi^- \mu + \dots \\ \xi^-\nabla_a f_s \xi \xi^- a + \xi^-\nabla_\eta f_s \eta + \dots \\ f_\mu(b,\mu) \end{Bmatrix} + \begin{Bmatrix} \xi^-\omega_a \\ \xi^-\omega_s \\ \omega_\mu \end{Bmatrix} \qquad (7.4)$$

Assuming the particle is sufficiently large, the intrinsic and random fluctuations on the flow of blanket states can plausibly be assumed to be averaged away, so that the flow of their average becomes:

$$\dot{\mathbf{b}} = \begin{Bmatrix} \dot{\mathbf{a}} \\ \dot{\mathbf{s}} \end{Bmatrix} \triangleq \begin{Bmatrix} \xi^-\dot{a}_j \\ \xi^-\dot{s}_j \end{Bmatrix} \approx \begin{Bmatrix} \frac{\hbar}{m}\mathbf{s} \\ \xi^-\nabla_a f_s \xi\mathbf{a} + \xi^-\nabla_\eta f_s \eta + \dots \end{Bmatrix} \qquad (7.5)$$

Here, we have defined the average coupling between the motion of active states and sensory states in terms of Planck's constant $\xi^-\nabla_s f_a \xi \triangleq \frac{\hbar}{m}$. Now, the flow of blanket states also conforms to the marginal flow lemma:

$$\dot{\mathbf{b}} = (Q_{\mathbf{bb}} - \frac{1}{n}\Gamma_{\mathbf{bb}})\nabla_{\mathbf{b}}\Im(\mathbf{b}\mid\eta) \qquad (7.6)$$

Because the amplitude of random fluctuations is inherited from the averaging, it vanishes with a sufficiently large number of blanket states, leaving only solenoidal flow. Combining (7.5) and (7.6) – and defining velocity in terms of momentum, $m\mathbf{v} = \hbar\mathbf{s}$ – gives:

$$Q_{\mathbf{bb}} = \begin{bmatrix} 0 & -I \\ I & 0 \end{bmatrix}, \quad \lim_{n\to\infty}\frac{1}{n}\Gamma_{\mathbf{bb}} = 0$$

$$\Rightarrow \begin{Bmatrix} \dot{\mathbf{a}} \\ \dot{\mathbf{s}} \end{Bmatrix} = \begin{Bmatrix} +\nabla_s\Im(\mathbf{b}\mid\eta) \\ -\nabla_\mathbf{a}\Im(\mathbf{b}\mid\eta) \end{Bmatrix} \approx \begin{Bmatrix} \mathbf{v} \\ -\nabla_\mathbf{a}\Im(\mathbf{a}\mid\eta) \end{Bmatrix} \qquad (7.7)$$

$$\Rightarrow \Im(\mathbf{b}\mid\eta) = \Im(\mathbf{a}\mid\eta) + \frac{\hbar}{2m}\mathbf{s}\cdot\mathbf{s} = \Im(\mathbf{a}\mid\eta) + \frac{m}{2\hbar}\mathbf{v}\cdot\mathbf{v}$$

The key result here is that the motion of (active) states is proportional to momentum (sensory states), which means the surprisal associated with momentum must have a quadratic form (c.f., kinetic energy). This follows as a necessary consequence of Markov blankets; namely, that the only influences on (active) states are the (sensory) states of the same particle (and not other particles). The resulting dynamics are Hamilton's equations of motion that underwrite classical mechanics, where the active and sensory states play the role of generalised coordinates of position and momentum. In other words, *surprisal becomes the Hamiltonian* or, with a Legendre transform, the Lagrangian of classical mechanics.

$$\mathcal{L}_{\Im}(\mathbf{a}, \dot{\mathbf{a}}) = \frac{m}{2\hbar} \dot{\mathbf{a}} \cdot \dot{\mathbf{a}} - \Im(\mathbf{a} \mid \eta)$$
$$\mathcal{H}_{\Im}(\mathbf{a}, \mathbf{s}) = \frac{\hbar}{2m} \mathbf{s} \cdot \mathbf{s} + \Im(\mathbf{a} \mid \eta)$$
$$= \dot{\mathbf{a}} \cdot \mathbf{p} - \mathcal{L}_{\Im} = \Im(\mathbf{b} \mid \eta) \qquad (7.8)$$

$$\mathbf{p} = \frac{\partial \mathcal{L}_{\Im}}{\partial \dot{\mathbf{a}}} = \frac{m}{\hbar} \dot{\mathbf{a}} = \mathbf{s}$$

As we might expect, the Hamiltonian is conserved over state trajectories, because the flow is purely solenoidal:

$$\dot{\mathcal{H}}_{\Im}(\mathbf{a}, \mathbf{s}) = \dot{\Im}(\mathbf{b} \mid \eta) = \dot{\mathbf{b}} \cdot \nabla_{\mathbf{b}} \Im = \nabla_{\mathbf{b}} \Im \cdot Q_{\mathbf{bb}} \cdot \nabla_{\mathbf{b}} \Im = 0 \qquad (7.9)$$

The emergence of classical mechanics rests upon using a first order (Taylor) approximation to the flow in (7.5). This corresponds to a second order approximation to surprisal for, and only for, sensory states (i.e., momentum). Using this (semi) Laplace approximation, Hamiltonian mechanics follow from assumptions about the constant curvature of surprisal (dropping the conditioning on external states for clarity):

$$\Im(\mathbf{a}, \mathbf{s}) = \Im(\mathbf{a}) + \frac{\hbar}{2m} \mathbf{s} \cdot \mathbf{s}$$
$$= \Im(\mathbf{a}) + \Im(\mathbf{s})_0 + \nabla \Im_0 \cdot \mathbf{s} + \frac{1}{2} \nabla^2 \Im_0 \mathbf{s}^2 + \dots$$
$$\Leftrightarrow \qquad (7.10)$$
$$\nabla \Im_0 = 0$$
$$\nabla^2 \Im_0 = \frac{\hbar}{m}$$

A notable result under this approximation is the motion of the (active) state (i.e., position) is proportional to the (sensory) state of motion (i.e., momentum): $m\dot{\mathbf{a}} = \hbar \mathbf{s}$ from (7.8). Accordingly, the NESS density over momentum must be Gaussian (with a precision of $\hbar/m$). We will use this later to estimate the mass of our synthetic organism (see Figure 16). We can also rearrange (7.7) to recover Newton's second law; namely, force is equal to mass times acceleration

$$\begin{Bmatrix} \dot{\mathbf{a}} \\ \dot{\mathbf{s}} \end{Bmatrix} = \begin{Bmatrix} \mathbf{v} \\ -\nabla_{\mathbf{a}} \Im(\mathbf{a}) \end{Bmatrix} \Rightarrow -\hbar \nabla_{\mathbf{a}} \Im(\mathbf{a}) = m\ddot{\mathbf{a}} \qquad (7.11)$$

This expression discloses the intimate relationship between surprisal and potential, where – in this classical case – *Newtonian potential* is the surprisal of active states.

In this example, we assumed that the flow of active and sensory states of a particle depended on each other but not themselves. However, there is no special reason why this should be so. In fact, had we considered flows of (average) blanket states with the following form,

$$\begin{Bmatrix} \dot{\mathbf{a}} \\ \dot{\mathbf{s}} \end{Bmatrix} = \begin{Bmatrix} +\nabla_s \Im(\mathbf{b}) \\ -\nabla_a \Im(\mathbf{b}) \end{Bmatrix} \approx \begin{Bmatrix} \frac{\hbar}{m}(\mathbf{s} - z\mathbf{A}) \\ -\nabla_a \Im(\mathbf{b}) \end{Bmatrix}$$

(7.12)

$$\Im(\mathbf{b}) = z\varphi(\mathbf{a}) + \frac{\hbar}{2m}(\mathbf{s} - z\mathbf{A}) \cdot (\mathbf{s} - z\mathbf{A})$$

we would have recovered classical electromagnetics and Maxwell's equations. Equation (7.12) has augmented the motion of active (position) states with a vector function of position $\mathbf{A(a)}$ and the motion of sensory (momentum) states with a scalar function $\varphi(\mathbf{a})$. These equations provide a Hamiltonian description of a particle with charge $z$, subject to a Lorentz force by an electrical field $\mathbf{E} = -\nabla\varphi \Rightarrow \nabla \cdot \mathbf{E} = -\Delta\varphi$ (c.f., Coulomb's law) and magnetic induction $\mathbf{B} = \nabla \times \mathbf{A} \Rightarrow \nabla \cdot \mathbf{B} = 0$ (Helgaker et al., 2014). The ensuing Lorentz force follows by substitution and direct calculation:

$$\begin{Bmatrix} \dot{\mathbf{a}} \\ \dot{\mathbf{s}} \end{Bmatrix} = \begin{Bmatrix} +\nabla_s \Im(\mathbf{b}) \\ -\nabla_a \Im(\mathbf{b}) \end{Bmatrix} \Rightarrow \begin{Bmatrix} \frac{m}{\hbar}\ddot{\mathbf{a}} = \dot{\mathbf{s}} - z(\dot{\mathbf{a}} \cdot \nabla_a)\mathbf{A} \\ \dot{\mathbf{s}} = z(\mathbf{E} + \dot{\mathbf{a}} \times \mathbf{B} + (\dot{\mathbf{a}} \cdot \nabla_a)\mathbf{A}) \end{Bmatrix} \Rightarrow m\ddot{\mathbf{a}} = \hbar z(\mathbf{E} + \dot{\mathbf{a}} \times \mathbf{B})$$

(7.13)

In terms of a Taylor approximation, this corresponds to a coupling between sensory and active states (i.e., momentum and position) according to the (semi) Laplace approximation to surprisal in (7.12). One can pursue this perspective and consider higher-order approximations; for example, the Hamiltonian for a relativistic particle in a conservative force field $-\nabla V(\mathbf{a})$ can be expressed as:

$$\Im(\mathbf{b}) = V(\mathbf{a}) + \sqrt{m^2 c^4 + \mathbf{s}^2 c^2}$$

(7.14)

See (Helgaker et al., 2014) for details.

## Random fluctuations and generalised motion

An interesting difference – between conservative and dissipative dynamics – is the constitution of mass: in the quantum formulation, mass was associated with the amplitude of random fluctuations via the introduction of Planck's constant. In contrast, in conservative (Lagrangian) mechanics, mass is associated with the coupling between the motion of states and states of motion. Notice a further distinction between the treatment in this section and quantum mechanics. In the quantum formulation, flow was associated with momentum. In the classical formulation there is a subtle but fundamental distinction between the motion of the state (i.e., flow) and the state of motion (i.e., momentum). In classical mechanics, momentum and related constructs are an attribute of the state of motion that reduces to the motion of states when, and only when, classical (conservative) assumptions hold.

From our perspective, this simply means random (and intrinsic) fluctuations can be ignored. However, the distinction between the motion of states (e.g., position) and states of motion (e.g., momentum) becomes important when random fluctuations are put back in play, as follows.

Hitherto, we have discounted random fluctuations to reveal classical mechanics. However, it is interesting to consider what would happen if they were reintroduced. Random fluctuations on (classical) momentum emerges as friction (Friston and Ao, 2012), where, following (7.7):

$$Q_{\text{bb}} = \begin{bmatrix} 0 & -I \\ I & 0 \end{bmatrix}, \quad \Gamma_{\text{bb}} = \begin{bmatrix} 0 & 0 \\ 0 & \Gamma_{\text{ss}} \end{bmatrix} \Rightarrow \begin{Bmatrix} \dot{\mathbf{a}} \\ \dot{\mathbf{s}} \end{Bmatrix} \approx \begin{Bmatrix} \mathbf{v} \\ -\nabla_{\mathbf{a}} \Im(\mathbf{a} \mid \eta) - \Gamma_{\text{ss}} \mathbf{v} \end{Bmatrix} \qquad (7.15)$$

This introduces a velocity-dependent reduction in acceleration, suggesting that friction is an emergent property of random fluctuations in momentum. A more thorough treatment of the relationship between diffusion, random fluctuations and friction can be found in (Yuan et al., 2011), using the generalised Einstein relation. This return to dissipative dynamics does not usually consider random fluctuations on position (i.e., active states). For example, Langevin equations for mechanical systems with canonical position and momentum usually restrict fluctuations to the flow of momentum, as in (7.15); however, there is no reason why this should be the case: see (Kerr and Graham, 2000) for a derivation of the Langevin equations – and associated Fokker-Planck equations – for mechanical systems that include noise and damping terms in the equations of motion for all canonical variables.

Part Three will occasionally refer to generalised coordinates *of motion*. It is natural to ask whether generalised motion arises from the same sort of arguments used to account for classical mechanics. Our classical account rests upon the fact that active states cannot be influenced by external states (i.e., the blanket states of other particles). This means that the response of active states has to be mediated by sensory states. This leads to the natural interpretation of active states in terms of position that can only be changed via sensory states, which manifest as momentum or velocity (i.e., the rate of change of position with respect to time). However, generalised coordinates of motion arise from a quite different mechanism; namely, analytic or *smooth random fluctuations*. In brief, generalised coordinates of motion inherit from a relaxation of Wiener assumptions about random fluctuations and are therefore not restricted to classical mechanics: please see Appendix E for a treatment of generalised motion in terms of the Helmholtz decomposition – and its implications for generalised inference (filtering) schemes.

## Summary

In summary, if averaging over active states (e.g., position) and sensory states (e.g., momentum) suppresses random and intrinsic fluctuations (via the central limit theorem), we necessarily have a classical mechanics. In this nonequilibrium steady-state, the NESS potential over blanket states becomes a Hamiltonian, describing the behaviour of large particles (e.g., balls and planets). Note that the Hamiltonian includes the NESS potential (i.e., surprisal) of position, which depends upon the external states. This suggests that classical mechanics can be formulated as generalised motion on a state-dependent and therefore time-dependent potential. In other words, external influences mediate their effects by changing the potential energy of blanket states. This is illustrated in Figure 16, using the average position and motion of the principal Markov blanket in Figure 6. In this example, we can estimate the mass of the virus in our synthetic soup by examining its motion under the forces exerted by external states. Anecdotally, this would be like estimating the mass of a small object by shaking it gently in the palm of your hand. See Figure 16 for details.

We have assumed that averaging over the (large number) of intrinsic fluctuations is sufficient to eliminate them; thereby eluding the (Taylor expansion) terms due to generalised synchrony. This means classical mechanics applies to *inert* systems, whose internal states do not exhibit any synchrony. This leads to the sensible notion that classical mechanics is a limiting case when the internal dynamics of particles are insufficiently coherent to induce generalised synchrony, rendering them inert and lifeless. At macroscopic scales the (average) behaviour of (inert) particles will necessarily possess a Hamiltonian or Lagrangian mechanics. Furthermore, they will conform to Hamilton's principle of stationary action, in virtue of the fact that their flow is divergence free. This (classical) instance of Hamilton's principle is inherited from a more general principle of least action, which defines nonequilibrium steady-state.

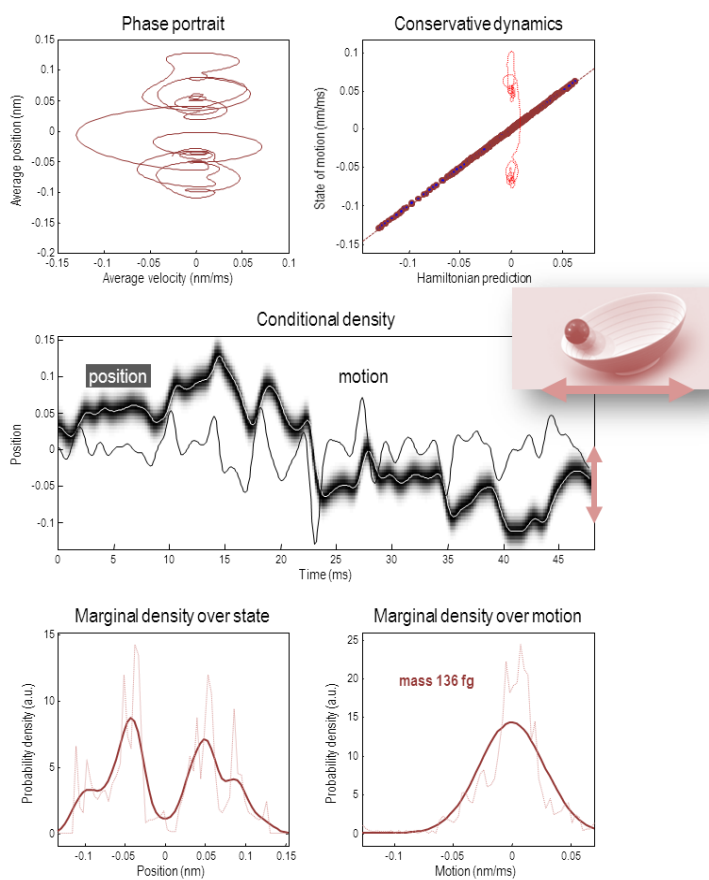

**FIGURE 16**

*Classical Mechanics*. This figure illustrates a treatment of our primordial soup from a classical (Hamiltonian or Lagrangian) perspective. By taking ensemble averages over various quantities, one can suppress the influence of intrinsic (random) fluctuations, thereby revealing conservative, Hamiltonian dynamics mediated by solenoidal flow. Here, we focus on the average position of the particles of the Markov blanket in Figure 6 and ask whether the associated motion conforms to classical predictions. The picture here is of a ball rolling around in a potential well where, crucially, the potential well changes with external particles. In particular, we considered a second order polynomial approximation to the NESS potential of the blanket's position (averaged over its constituent particles), conditioned on the position of all external particles. By formulating this dependency as a linear mixture of external positions, the gradients that produce the average motion of the Markov blanket can be expressed as a polynomial expansion that is quadratic in the blanket positions and linear in the external positions. The polynomial coefficients can then be estimated, using least squares, to best predict the average motion of the blanket; thereby specifying the (conditional) NESS potential and Hamiltonian dynamics. Heuristically, this corresponds to characterising the average behaviour of the Markov blanket as the motion of a marble (or ball) in a quadratic well (or bowl) that moves with – or is gently rocked by – external particles. The resulting behaviour can then be characterised in terms of the ball's mass that corresponds to the precision (i.e., inverse variance) of motion. **Upper left panel**: this phase portrait summarises the behaviour we are trying to explain by plotting the position (state) against velocity (motion). In the absence of external perturbations, the trajectory should be a perfect circle. However, it appears that the external particles are moving the potential energy well to produce more erratic, although entirely conservative, behaviour. **Middle panel**: this illustrates the potential well in terms of the corresponding (conditional) density shown as a time varying function of (a linear mixture of) the states (position) of external particles. The shaded area depicts regions of high probability density and the white line shows the trajectory of the position of the Markov blanket. The black line is the corresponding motion of the Markov blanket. This is a nontrivial solution, in the sense that the external particles are not simply moving the Markov blanket particles – they are inducing Hamiltonian motion by moving a potential energy well. **Upper right panel**: the resulting predictions of the blanket motion account almost exactly for its (generalised) motion (blue dots). The red line is the corresponding prediction for a single particle within the Markov blanket – and illustrates that the states of motion only becomes the motion of states when intrinsic fluctuations are suppressed. In other words, each member of the blanket ensemble is moving somewhat erratically and actively; however, their collective motion can be expressed as a nearly deterministic and instantaneous function of their collective positions. This is nontrivial, in the sense that the motion being predicted is orthogonal to the positions (of blanket and external particles) upon which the predictions are based. Using the estimates of the NESS potential afforded by the emergence of conservative dynamics, one can now quantify the ensemble density for any given external state, over both the motion of state (left) and the state of motion (right). **Lower left panel**: this shows the marginal distribution over position, averaged over the trajectory shown. The marginal density (solid line) is based on the polynomial coefficients that best predicted motion, while the dotted line corresponds to the sample density. **Lower right panel**: this is the equivalent ensemble density over average motion. The precision of this density determines the effective mass of the Markov blanket. In this example, if we assume that motion is expressed in nanometres per millisecond (i.e., slow motion at a macromolecular scale). Then the effective mass, given Planck's constant, is 136 femtograms. This corresponds to an extremely heavy virus – or a rather lightweight bacterium. For example, a typical E. coli would have a mass of 630 fg. Had we assumed that the velocity has units of metres per millisecond, the mass would have been in excess of 2 tonnes – the mass of a large car (assuming a classical Planck's constant of unity).

Some readers might be wondering where, in this treatment of deterministic systems, things like topological symmetry breaking and deterministic chaos arise. The foregoing has focused on the behaviour of single particles, where the influence of internal particles and other particles has been suppressed to recover classical laws of motion[21]. This is apt for describing the behaviour of inert particles, like heavenly bodies and pendula; however, things get more complicated when dealing with an ensemble of large particles; for example, *n-body* problems (Littlejohn and Reinsch, 1997). In terms of information length, one could associate the behaviour of inert particles with the short information lengths that characterise fixed point and periodic attracting manifolds (e.g., the point attractor of the Lorenz system in Figure 4). In these systems, the initial state is always close to steady-state behaviour and the particle can be said to show a simple form of self-organisation – appropriate for, say, a rock on the dark side of the moon. Conversely, in deterministic itinerancy, the attracting manifolds becomes more structured, engendering topological symmetry breaking (Evans and Searles, 2002; Kappen, 2005; Parisi and Sourlas, 1982; Schwabl, 2002), chaotic itinerancy (Namikawa, 2005; Nara, 2003; Tsuda and Fujii, 2004), generalised synchronisation (Barreto et al., 2003; Boccaletti et al., 2002; Hu et al., 2010; Rulkov et al., 1995) and so on.

On the current account, itinerancy can only emerge when two conditions are met: first, the amplitude of random fluctuations has to be low – as in noise induced tunnelling – or negligible – as in classical mechanics. Second, the information length induced by the flow must be sufficiently large – as in many-body problems. There is a vast literature on the *n-body* problem in a classical setting. This literature ranges from mean field approximations through to loosely coupled dynamical systems; e.g., (Boccaletti et al., 2002; Breakspear and Stam, 2005; Kayser and Ermentrout, 2010; Rulkov et al., 1995; Schumacher et al., 2012; Tsuda, 2001) and the behaviours they exhibit (e.g., generalised synchronisation).

There are many directions we could pursue here. However, we will return to a more fundamental aspect of classical mechanics: namely, the assumption that particles are *inert*, in the sense that the influence of internal states can be discounted. In the final part of this monograph, we turn to a mechanics that deals with *autonomous particles*, in which there is a structured contribution of internal states to a particle's dynamics – a structure that enriches their repertoire of behaviours, necessary for the treatment of large particles, like you and me.

---

[21] Furthermore, we have not considered the more delicate issue of solenoidal flows that are function of states.

# Part Three: a particular case

## A theory of autonomous things – Bayesian mechanics

In this section, we turn to a Bayesian mechanics for *active particles* (Bressloff and Newby, 2013; Keber et al., 2014; Khadka et al., 2018) that rests on the existence of a Markov blanket. In brief, the Markov blanket allows us to talk about internal states as *representing* external states in a probabilistic sense. Heuristically, this means that one can ascribe probabilistic beliefs to internal states, in the sense that they are *about something* – namely, external states. This interpretation rests upon a *variational density* over external states, which is parameterised by internal states:

$$q_{\mathbf{\mu}}(\eta) = p(\eta \,|\, b) = p(\eta \,|\, \pi)$$
$$\mathbf{\mu}(b) \triangleq \arg\max{}_{\mu} \, p(\mu \,|\, b) \qquad\qquad (8.1)$$

This variational density arises in virtue of the blanket as follows: if we condition internal and external states on the blanket state, there must be a most likely internal state for every blanket state – and a conditional density over external states, conditioned on that blanket state. This implies a mapping between an internal (statistical) manifold and a conditional density over external states[22]. In turn, this means the internal manifold $\mathbf{\mu}(b)$ acquires an information geometry, defined by the variational density above.

Another key symmetry implied by the Markov blanket is that the flow of internal and active states can be expressed as a (solenoidal) gradient flow[23] on the *same quantity*; namely, the surprisal of particular states. We will refer to internal and active states as *autonomous* because they are not influenced by external states. From the marginal flow lemma (1.21):

$$f_\alpha(\pi) = (Q_{\alpha\alpha} - \Gamma_{\alpha\alpha})\nabla_\alpha \Im(\pi)$$
$$\alpha = \{a, \mu\}$$
$$\pi = \{s, \alpha\} \qquad\qquad (8.2)$$

These two aspects of a Markov blanket underwrite a Bayesian mechanics, in which we can talk about internal states holding *Bayesian beliefs*[24] about external states – and autonomous states acting on external states, under those beliefs. This follows because one can express gradient flows on surprisal as minimising free energy functionals of the variational density above – functionals that underwrite Bayesian inference.

---

[22] On the assumption that the number or dimensionality of internal states is greater than the number of blanket states, the dimensionality of the internal (statistical) manifold corresponds to the dimensionality of blanket states.

[23] When referring to gradient flows, we allow for solenoidal components that are orthogonal to the gradients. We will refer to these as (solenoidal) gradient flows.

[24] Bayesian belief here refers to a posterior or conditional probability density. This is not a propositional belief but is used in the sense of Bayesian 'belief' updating and 'belief' propagation.

We will consider two forms of free energy – a *particular* free energy of particular states and a *variational* free energy that just considers blanket states. These lead to an interpretation of nonequilibrium steady-state dynamics as *exact* and *approximate* Bayesian inference, respectively. Particular free energy could be read as a mathematical 'sleight of hand'; in the sense that it stipulatively defines free energy as surprisal, rendering gradient flows on surprisal and free energy synonymous. However, this formulation allows one to treat autonomous (i.e., active and internal) states symmetrically – and establish a description of self-organisation in terms of Bayesian inference. Having set up the basic formalism with *particular* free energy, we will turn to *variational* free energy and approximate Bayesian inference that deals with internal and active states separately. Practically speaking, the variational free energy is more interesting, because it can be used to quantify and simulate self-evidencing, in the form of active inference. Furthermore, it leads to the interesting question: is self-organisation approximate Bayesian inference – or does Bayesian inference approximate self-organisation?

We look first at the underlying formalism in terms of a free energy lemma – and an integral fluctuation theorem that speaks to the most likely autonomous states in the future. This section establishes the basic form of the argument. In the next section, we address the nature of the variational density by appealing to approximate Bayesian inference, under something known as the Laplace assumption.

**Lemma** (particular free energy): *given a variational density:* $q_{\mathbf{\mu}}(\eta) = p(\eta \mid b)$*, the most likely path of autonomous states, given sensory states, can be expressed as a (solenoidal) gradient flow on a free energy functional of a particle's states:* $\pi = \{b, \mu\} = \{s, \alpha\}$*:*

$$
\begin{aligned}
\mathbf{\alpha}[\tau] &= \arg\min_{\alpha[\tau]} \mathcal{A}(\alpha[\tau] \mid s[\tau]) \\
&\Rightarrow \delta_{\mathbf{\alpha}[\tau]} \mathcal{A}(\mathbf{\alpha}[\tau] \mid s[\tau]) = 0 \\
&\Rightarrow \dot{\mathbf{\alpha}} = (Q_{\alpha\alpha} - \Gamma_{\alpha\alpha}) \nabla_{\alpha} F(\mathbf{\alpha}, s)
\end{aligned}
\tag{8.3}
$$

This means the most likely path conforms to a variational principle of least action, where particular free energy $F(\pi) \equiv F[\pi, q_{\mu}(\eta)]$ is an upper bound on surprisal:

$$
\begin{aligned}
F(\pi) &\triangleq \underbrace{E_q[\mathfrak{I}(\eta, \pi)]}_{energy} - \underbrace{H[q_{\mu}(\eta)]}_{entropy} \\
&= \underbrace{\mathfrak{I}(\pi)}_{surprisal} + \underbrace{D[q_{\mu}(\eta) \parallel p(\eta \mid b)]}_{bound} \\
&= \underbrace{E_q[\mathfrak{I}(\pi \mid \eta)]}_{inaccuracy} + \underbrace{D[q_{\mu}(\eta) \parallel p(\eta)]}_{complexity} \geq \mathfrak{I}(\pi)
\end{aligned}
\tag{8.4}
$$

$q_{\mathbf{\mu}}(\eta) = p(\eta \mid b) : \mathbf{\mu} = \arg\max_{\mu} p(\mu \mid b)$

This functional can be expressed in several forms; namely, an expected energy minus the entropy of the variational density, which is equivalent to the self-information associated with particular states (i.e., *surprisal*) plus the KL divergence between the variational and posterior density (i.e., *bound*). In turn, this can be decomposed into the negative log likelihood of particular states (i.e., *accuracy*) and the KL divergence between posterior and prior densities (i.e., *complexity*). In short, particular free energy constitutes a *Lyapunov function* for the expected flow of autonomous states.

**Proof**: the most likely trajectory – that minimises action – obtains when the random fluctuations on the autonomous flow take their most likely value of zero. By the marginal flow lemma, this means the flow of the most likely autonomous states can be expressed as a (solenoidal) gradient flow on surprisal or, by definition, particular free energy:

$$\boldsymbol{\alpha}[\tau] = \arg\min_{\alpha[\tau]} \mathcal{A}(\alpha[\tau] \,|\, s[\tau]) \Rightarrow$$
$$\dot{\boldsymbol{\alpha}} = (Q_{\alpha\alpha} - \Gamma_{\alpha\alpha})\nabla_{\alpha}\Im(\boldsymbol{\alpha}, s) \qquad (8.5)$$
$$= (Q_{\alpha\alpha} - \Gamma_{\alpha\alpha})\nabla_{\alpha}F(\boldsymbol{\alpha}, s)$$

Where, for the most likely internal state, $\boldsymbol{\mu} \subset \boldsymbol{\alpha}$ :

$$F(\boldsymbol{\alpha}, s) = \Im(\boldsymbol{\alpha}, s) + \underbrace{D[q_{\boldsymbol{\mu}}(\eta) \,\|\, p(\eta \,|\, \boldsymbol{\alpha}, s)]}_{bound} = \Im(\boldsymbol{\alpha}, s) \qquad (8.6)$$

The equivalence between particular free energy and the surprisal of particular states follows from the definition of the variational density that renders the bound zero □

Note that particular free energy can be expressed as a functional of the variational density or as a function of its sufficient statistics. The results above hold for any system at nonequilibrium state and can be considered as a principle of least action for particular free energy. Given this stipulative reformulation of gradient flows under a Markov blanket, one can now consider the most likely path of autonomous states from an initial particular state. The very existence of a nonequilibrium steady-state has some interesting implications for autonomous states in the future – that we will leverage later to formalise 'planning ahead'. This long-term behaviour can be summarised with following corollary, based upon an integral fluctuation theorem for Markov blankets:

**Corollary** (expected free energy). *The surprisal of a future autonomous state is upper bounded by expected free energy:*

$$G(\alpha_{\tau}) \geq \lim_{d\ell(\tau)\to 0} \Im_{\tau}(\alpha_{\tau} \,|\, \pi_0) \qquad (8.7)$$

Expected free energy is a functional of the final state:

$$G(\alpha_\tau) \triangleq \underbrace{E_{q_\tau}[\Im(\eta_\tau, \pi_\tau)]}_{energy} - \underbrace{H[q_\tau(\eta_\tau \mid \pi_\tau)]}_{entropy}$$

$$= E_{q_\tau}[\Im(\pi_\tau) + \underbrace{D[q_\tau(\eta_\tau \mid \pi_\tau) \parallel p(\eta_\tau \mid b_\tau)]}_{bound}]$$

$$= E_{q_\tau}[\underbrace{\Im(\pi_\tau \mid \eta_\tau)}_{ambiguity} + \underbrace{D[q_\tau(\eta_\tau \mid \pi_\tau) \parallel p(\eta_\tau)]}_{risk}] \geq E_{q_\tau}[\Im(\pi_\tau)] \qquad (8.8)$$

$$q_\tau(\eta_\tau, s_\tau \mid \alpha_\tau) \triangleq p_\tau(\eta_\tau, s_\tau \mid \alpha_\tau, \pi_0) = E_{q_\mu}[p_\tau(\eta_\tau, s_\tau \mid \alpha_\tau, \pi_0, \eta_0)]$$

The expectation here is under the *predictive density* over hidden and sensory states, conditioned upon the initial state of the particle and final autonomous state. Here, this predictive density is expressed in terms of the variational density $q_\mu(\eta_0) = p(\eta_0 \mid \pi_0)$.

The expected free energy in (8.8) has been formulated to emphasise the formal correspondence with particular free energy in (8.4), where the complexity and accuracy terms become *risk* (i.e., expected complexity) and *ambiguity* (i.e., expected inaccuracy). One can also compare the expression for expected free energy in (8.8) with the surprisal of autonomous states in (2.7). The corollary above is important because it affords a lower bound on the probability of autonomous states in the future, which is a function of particular states in the present. This bound holds under convergence to nonequilibrium steady-state. Conversely, if the activity of a particle (i.e., its autonomous path) leads to an autonomous state that minimises expected free energy, convergence to nonequilibrium steady-state is assured.

**Proof**: consider the path from an initial $\pi_0 \equiv \pi(0)$ particular state to a final state $\pi_\tau \equiv \pi(\tau)$. If the information length has converged, the divergence between the predictive density over systemic states and the NESS density disappears. From (2.9), we can express the time-dependent surprisal of the final state $\Im_\tau(\pi_\tau \mid \pi_0) \equiv -\ln p_\tau(\pi_\tau \mid \pi_0) \equiv -\ln q_\tau(\pi_\tau)$ in terms of its expected free energy:

$$d\ell(\tau) = 0$$
$$\Rightarrow D[q_\tau(\eta_\tau, \pi_\tau) \parallel p(\eta_\tau, \pi_\tau)] = 0$$
$$\Rightarrow E_{q_\tau}[G(\pi_\tau) - \Im_\tau(\pi_\tau \mid \pi_0)] = 0$$
$$\Rightarrow G(\pi_\tau) = \Im_\tau(\pi_\tau \mid \pi_0) \qquad (8.9)$$

$$G(\pi_\tau) = E_{q_\tau}[\Im(\eta_\tau, \pi_\tau)] - H[q_\tau(\eta_\tau \mid \pi_\tau)]$$
$$q_\tau(\eta_\tau, \pi_\tau) = p_\tau(\eta_\tau, \pi_\tau \mid \pi_0)$$

Note that the divergence in (8.9) plays the same role as thermodynamic free energy in (6.6); namely, the divergence between an evolving density and the density at nonequilibrium steady-state. However, here, we are not dealing with an ensemble density but with a predictive density over future states. Equation (8.9) can also be written in the form of an integral fluctuation theorem; c.f., (6.16),

$$E_q\left[\ln\frac{q_\tau(\eta_\tau\mid\pi_\tau)}{p(\eta_\tau\mid\pi_\tau)}-\Delta\mathfrak{I}_\tau\right]\geq 0,\quad \Delta\mathfrak{I}_\tau=\mathfrak{I}_\tau(\pi_\tau\mid\pi_0)-\mathfrak{I}(\pi_\tau) \tag{8.10}$$

with equality when $d\ell(\tau)\to 0$. In short, at some point in the future, the expected free energy of a particular state becomes its surprisal, at which point there is no further increase in information length. We can now average the expected free energy of a particular state over the predicted sensory states to create an upper bound on the surprisal of autonomous states. By Jensen's inequality:

$$G(\alpha_\tau)=E_{q_\tau(s_\tau\mid\alpha_\tau)}[G(\pi_\tau)]=E_{q_\tau(s_\tau\mid\alpha_\tau)}[\mathfrak{I}_\tau(\pi_\tau\mid\pi_0)]\geq\mathfrak{I}_\tau(\alpha_\tau\mid\pi_0)$$

$$G(\alpha_\tau)=E_{q_\tau}[\mathfrak{I}(\eta_\tau,s_\tau,\alpha_\tau)]-H[q_\tau(\eta_\tau\mid s_\tau,\alpha_\tau)] \tag{8.11}$$

$$q_\tau\left(\eta_\tau,s_\tau\mid\alpha_\tau\right)=p_\tau(\eta_\tau,s_\tau\mid\alpha_\tau,\pi_0)$$

Substituting (8.11) into (8.8) gives (8.7) □

## Risk and ambiguity

In summary, particular free energy is an upper bound on the surprisal of (current) particular states – and expected free energy is an upper bound on the surprisal of (future) autonomous states. Crucially, both particular and expected free energy are functions of particular states, which means that they can be 'evaluated' by a particle. In other words, the minimisation of particular and expected free energy furnish a self-consistent description of a particle's behaviour in both the short and long term.

The expected free energy corollary qualifies our use of *risk* and *ambiguity* when decomposing self-entropy in (8.12). Recall from Section 2 that risk is the complexity cost expected under the NESS density, while ambiguity is the expected likelihood of particular states, given external states. These expectations correspond to the mutual information between particular and external states – and conditional uncertainty about particular states, given external states (2.3). However, the corresponding quantities in the decomposition of expected free energy are expectations under the *predictive density* conditioned upon some autonomous behaviour (i.e., trajectory of autonomous states). In this setting, risk is an attribute of autonomous behaviour, which corresponds to the divergence between predicted (external) states of affairs and those at nonequilibrium steady-state. Similarly, ambiguity becomes the expected likelihood of particular states, given external states and autonomous behaviour. In short, autonomous particles will look as if they are selecting behaviours that actively minimise risk in relation to some *a priori* (nonequilibrium steady-state) beliefs about future outcomes, while acting upon external states to resolve ambiguity.

The difference between risk and ambiguity – defined in terms of expectations under the NESS and predictive densities – disappears when the two densities converge. This is reflected in the expectations of particular and expected free energy. Because particular free energy is an upper bound on surprisal, the free energy expected at nonequilibrium steady-state is an upper bound on self-entropy:

$$E_{p(\pi)}[F(\pi)] \geq E_{p(\pi)}[\Im(\pi)] = H(P) = \underbrace{H(P\,|\,E)}_{\text{ambiguity}} + \underbrace{I(E,P)}_{\text{risk}} \tag{8.12}$$

This means that minimising particular free energy minimises self-entropy or the equivalent ambiguity and risk (2.3). Similarly, the corresponding expectation for the expected free energy of future autonomous states is:

$$d\ell(\tau) = 0 \Leftrightarrow q_\tau(\eta_\tau, \pi_\tau) = p(\eta_\tau, \pi_\tau) \Rightarrow E_{p(\alpha_\tau)}[G(\alpha_\tau)] = \underbrace{H(P\,|\,E)}_{\text{ambiguity}} + \underbrace{I(E,P)}_{\text{risk}} = H(P) \tag{8.13}$$

An interesting case arises when the particle acts on external states exclusively via active states. In this situation, the predictive density over external states depends only on future autonomous states, such that:

$$q_\tau(\eta_\tau\,|\,\pi_\tau) = p(\eta_\tau\,|\,\alpha_\tau) \Rightarrow G(\alpha_\tau) = E_{q_\tau}[\Im(\pi_\tau) - \underbrace{D[\,p(\eta_\tau\,|\,s_\tau, \alpha_\tau)\,\|\,p(\eta_\tau\,|\,\alpha_\tau)]}_{\text{information gain}}] \tag{8.14}$$

In this case, we recover the information gain in (2.7) as part of expected free energy. Notice that the sign of the KL divergence is now negative, because the predictive and NESS densities have been switched. This means that minimising expected free energy maximises information gain. We will return to this case when dealing with discrete state-space models, where external states are only influenced by active states. Table 4 is a version of Table 1 that augments the information theoretic terms with their variational counterparts to illustrate their respective roles.

Teleologically speaking, the NESS density can take on two distinct interpretations, depending upon whether autonomous behaviour is thought of in terms of gradient flows (i.e., the particular free energy lemma) or selecting future states (i.e., the expected free energy corollary). From the point of view of a statistician, the gradient flow formulation regards *the NESS density as a generative model*; in other words, a probabilistic specification of the sensory consequences of hidden causes. It is this dynamics that licenses an interpretation of self-organisation in terms of statistical (i.e., Bayesian) inference. The picture changes when we consider expected free energy. Here, we are picking out trajectories of autonomous states (i.e., active and internal states) that are most likely under the generative model. On this view, the generative model (i.e., NESS density) can be regarded as furnishing prior beliefs about the sensory states (and their external causes) that will be encountered in the future. In other words, the generative model prescribes the attracting set that the particle will autonomously work towards – by apparently selecting the paths that lead to attracting states with the smallest expected surprisal (i.e., expected free energy). Put simply, this means that coupling back to the external states – via active states – lends self-evidencing an enactive aspect. This aspect makes it look as if the generative model is both an explanation for sensory states and a specification of the (external and particular) states a particle aspires to.

The time scale over which the particular density evolves from an initial state to the NESS density characterises the itinerancy of self-organisation, in terms of the associated information length. One can reinterpret (8.9) to characterise self-evidencing the terms of a *critical time* $\tau$, after which the surprisal of particular states becomes their expected free energy:

$$G(\pi_\tau) = \Im_\tau(\pi_\tau \mid \pi_0) : \forall \tau \geq \boldsymbol{\tau}$$
$$\Rightarrow D[q_\tau(\eta_\tau, \pi_\tau) \parallel p(\eta_\tau, \pi_\tau)] \approx 0 \Leftrightarrow d\ell(\boldsymbol{\tau}) \approx 0 \qquad (8.15)$$
$$\Rightarrow G(\alpha_\tau) \geq \Im_\tau(\alpha_\tau \mid \pi_0) = \Im(\alpha_\tau)$$

This operationalises the notion of 'itinerancy', enabling one to express the expected free energy of states after a critical time in the future, without referring explicitly to information length: c.f., the inequality in (8.16) and Figure 17. After this time, the expected free energy of autonomous states upper bounds their NESS surprisal – and the initial states are 'forgotten'. Particles with a short critical time will, effectively, converge to nonequilibrium steady-state quickly and show a simple kind of self-organisation. Conversely, particles with a long critical time will exhibit itinerant density dynamics with temporal depth, experiencing many more probabilistic configurations. We will return to this distinction later, in terms of the difference between particles evincing a generalised homoeostasis and allostasis. Figure 17 provides an overview of how Bayesian mechanics inherits from Langevin dynamics, when a Markov blanket is in play. Before unpacking self-evidencing in terms of (variational) inference, we will briefly consider the implications of a Markov blanket for (quantum) measurement and the information geometry of internal states.

TABLE 4

Information measures and particular free energy

| Measure | Definition | Variational homologue |
|---|---|---|
| **Surprisal** | $\Im(\pi) = -\ln p(\pi)$ | $\Im(\pi_0) \leq F(\pi_0)$ |
| **Self-entropy** | $H[P] = E_{p(\pi)}[\Im(\pi)]$ | $H[P] \leq E_{p(\pi_0)}[F(\pi_0)] = E_{p(\alpha_\tau)}[G(\alpha_\tau)]$ |
| **Complexity** | $D[p(\eta \mid \pi) \parallel p(\eta)]$ | $D[q_\mu(\eta_0) \parallel p(\eta_0)]$ |
| **Risk** | $I(E, P) = E_{p(\pi)}[D[p(\eta \mid \pi) \parallel p(\eta)]]$ | $D[q_\tau(\eta_\tau \mid \pi_\tau) \parallel p(\eta_\tau)]$ |
| **Accuracy** | $E_{p(\eta \mid \pi)}[\ln p(\pi \mid \eta)]$ | $E_{q_\mu(\eta)}[\ln p(\pi_0 \mid \eta_0)]$ |
| **Ambiguity** | $H(P \mid E) = E_p[\Im(\pi \mid \eta)]$ | $E_{q_\tau}[\Im(\pi_\tau \mid \eta_\tau)]$ |

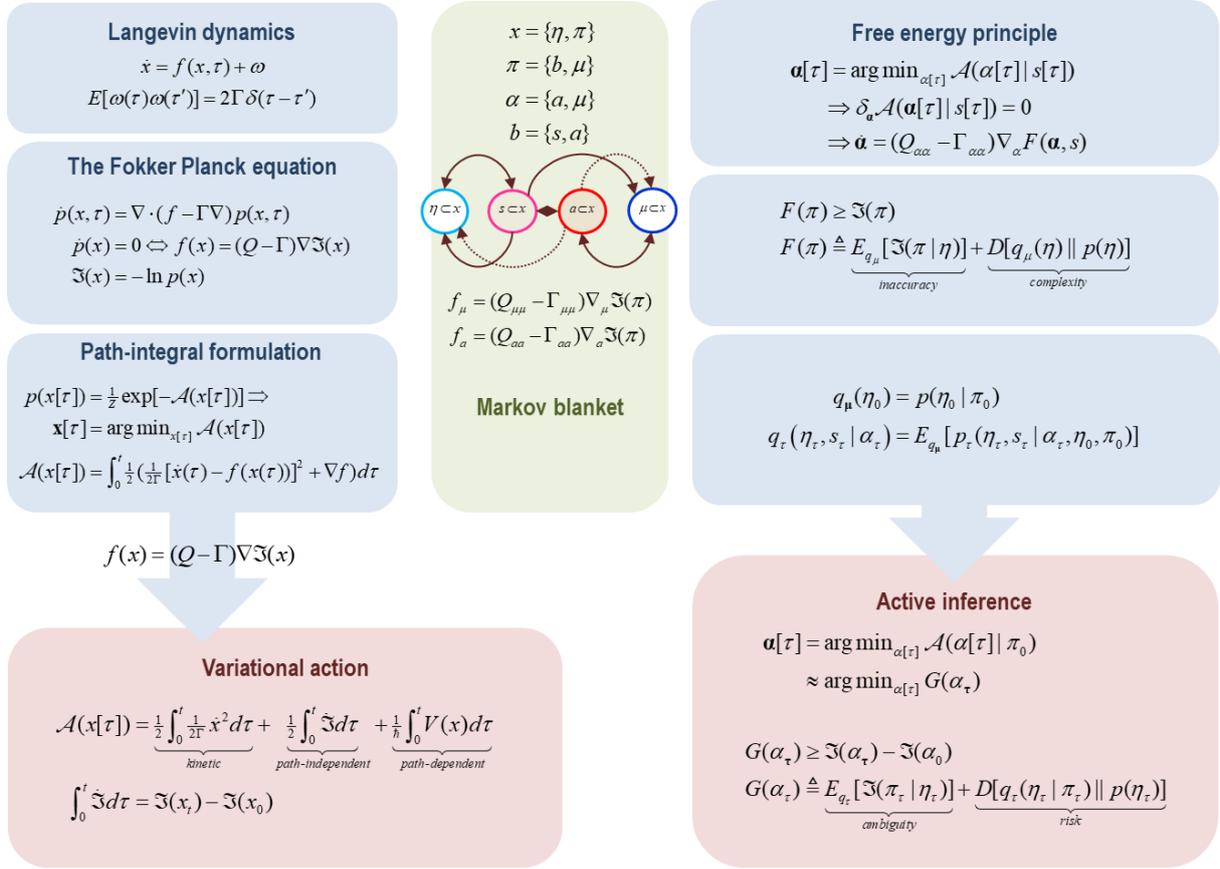

FIGURE 17

*From Langevin dynamics to Bayesian mechanics*. This figure provides a schematic overview of the conceptual moves that take us from Langevin dynamics to Bayesian mechanics. In brief, these entail (i) unpacking standard formulations of density dynamics in the setting of a Markov blanket, (ii) invoking a variational density to equip density dynamics with an information geometry and (iii) interpreting the underlying gradient flows (*free energy principle*) and future paths (*active inference*) in terms of Bayesian inference. The left panels summarise the three formulations of random dynamical systems: in terms of Langevin dynamics, density dynamics, as described by the Fokker Planck equation, and the path integral formulation. The key results here rest upon the nonequilibrium steady-state solution of the density dynamics (i.e., the Fokker Planck equation), under which the dynamics can be described as a gradient flow on surprisal (via the NESS lemma). Substituting this solution into the path integral formulation allows the probability of a trajectory or path to be expressed in terms of its action. The middle column introduces the partition into external and internal states that are separated by (sensory and active) blanket states. The ensuing conditional independences mean that the gradient flows that characterise nonequilibrium steady-state have a particular form; namely, the autonomous states of a particle (internal and active states) can be expressed as a (solenoidal) gradient flow on a particle's (particular) states. This means one can express the most likely flow of autonomous states as appearing to minimise the surprisal of particular states, which (on average) constitutes their entropy. The right panels unpack this particular form of gradient flow in terms of a particular free energy – by inducing a density over external states that is parameterised by internal states. This variational density must exist in virtue of the Markov blanket (by conditioning external and internal states on blanket states). On this Bayesian reading, the gradient flows of autonomous states minimise a particular free energy bound on surprisal. In turn, the surprisal of autonomous states in the future (and, implicitly, path-independent action from an initial state) is upper bounded by expected free energy. When suitably rearranged, the expected free energy comprises terms – risk and ambiguity – that lend surprisal minimising, evidence maximising, gradient flows an interpretation in terms of sentient, purposeful, risk minimising behaviour that is contextualised by epistemic, ambiguity reducing dynamics. At this point, the formal description of nonequilibrium steady-state, in terms of self-organising gradient flows, can be written down in a way that allows one to simulate self-organisation of active particles to *a priori* preferred (i.e., nonequilibrium steady-state) states. In other words, to design self-organisation in terms of autonomous self-evidencing.

## Inference and measurement

In our treatment of quantum mechanics, we deferred a discussion of what it means to 'measure' something until this section. One could assert that a probabilistic description of (quantum) states is sufficient for their 'measurement': in other words, one could understand 'measurement' as inferring the (external) states of things. On this reading, measurement 'just is' the mapping from an internal state to a probability density over external states; i.e., the slowly varying blanket states of other particles. Heuristically, this suggests that we can never 'know' external states – an observer can only hold 'beliefs' about external states, which are parameterised by internal states. In turn, this means it is sufficient to characterise external states in terms of probability densities (or wave functions), because these are the only constructs that map in a lawful way to the (internal) states of a measuring 'particle' or observer.

If we take measurement to be inference; namely, holding or representing (probabilistic) beliefs about the causes of an observing system's blanket state, then we have a straightforward account of quantum measurement, which can be expressed (in Dirac notation) as:

$$\boldsymbol{\mu} = \left\langle \Psi^{\dagger}(\mu) \,|\, \mu \,|\, \Psi(\mu) \right\rangle$$

$$p(\eta \,|\, b) = \Psi^{\dagger}(\eta)\Psi(\eta) = q_{\boldsymbol{\mu}}(\eta)$$
$$p(\mu \,|\, b) = \Psi^{\dagger}(\mu)\Psi(\mu)$$

$$(8.17)$$

Clearly, this imbues measurements with uncertainty, where 'knowing' a (quantum) state is a limiting case of precise inference. To the extent that one subscribes to this formulation, notions of measurement, entanglement and decoherence become emergent properties of conditional synchronisation, under a three-way partition into external, internal and blanket states. It is interesting to note that the resolution of debates about wave function collapse – that appeal to quantum decoherence – rest upon a three-way partition into observed, observer and environmental states (Goold et al., 2016; Schlosshauer, 2004).

From the point of view of inference, an anecdotal conception of wave function collapse might be the collapse of a prior distribution to a (precise or definitive) posterior density. Indeed, this has some currency when considering the measurement problem (Zurek, 2003): in other words, if measurement were treated as an event, then belief updating could be characterised by a collapse of prior beliefs to posterior beliefs. However, this view is not tenable under Bayesian mechanics because measurement (i.e., inference) is a dynamical process (not an event). In this setting the 'collapse' of the prior to the posterior is an inherent and ongoing part of nonequilibrium steady-state dynamics. So, could one resurrect the notion of wave function collapse in relation to the bound in (8.8)? I.e., could one invoke a collapsing bound as an explanation for the dynamics of inference? When framed in terms of particular free energy, this is stipulatively true – the bound that separates the variational and posterior beliefs is, by definition, collapsed. We will pursue this below by casting self-evidencing as the minimisation of a *variational* free energy (as opposed to a *particular* free energy). The ensuing Bayesian mechanics offers an explicit formulation of self-evidencing in terms of inference – and reinstates a form of wave function (i.e., probability density function) collapse.

In summary, although a vast oversimplification, entanglement and decoherence in quantum measurement can be replaced by conditional synchronisation in Bayesian mechanics. In Bayesian mechanics, the focus moves from NESS density – over the joint space of external, internal and blanket states – to the information geometry supplied by the variational density; i.e., probabilistic beliefs about external states. We will retain this focus for the remainder of this monograph.

## Information geometry

Einstein, 1949: *"[I]t is not so easy to free oneself from the idea that coordinates must have an immediate metrical meaning.*" (Schilpp, 2000)

At this point, we can return the notion of information length and geometry. Recall that an information geometry rests on Riemannian metrics that measure distances on statistical manifolds (Amari, 1998; Ay, 2015). Following (1.15), the distance $d\ell$ between two nearby points $\mu$ and $\mu' = \mu + d\mu$ is expressed in terms of a metric tensor $g$:

$$d\ell^2 = g_{ij} d\mu^i d\mu^j \tag{8.18}$$

For statistical manifolds, this tensor is the Fisher information metric, which scores the rate at which the probability density changes as one moves on the statistical manifold. Technically, the information metric corresponds to the curvature of the KL divergence between the densities as neighbouring points move apart. From (1.15):

$$g = \mathbf{I}(\mu) = \nabla_{\mu'\mu} D[q_{\mu'}(\eta) \parallel q_\mu(\eta)]|_{\mu'=\mu} = E_q[\nabla_\mu \ln q_\mu(\eta) \times \nabla_\mu \ln q_\mu(\eta)] \tag{8.19}$$

Crucially, the internal state-space contains a statistical manifold, because internal states parameterise a (variational) density over external states. This means that the internal manifold acquires a dual aspect geometry.

It has an information geometry in virtue of playing the role of a statistical manifold, while, at the same time, having its own geometry, endowed by the NESS density and inherent information length (Kim, 2018). We can express these two geometries in terms of their respective metric tensors – in (1.15) and (8.19) – where $\lambda$ are the sufficient statistics of the density over internal states:

$$g_\lambda = \mathbf{I}(\lambda) = \nabla_{\lambda'\lambda'} D[p_{\lambda'}(\mu) \parallel p_\lambda(\mu)]|_{\lambda'=\lambda}$$
$$g_\mu = \mathbf{I}(\mu) = \nabla_{\mu'\mu'} D[q_{\mu'}(\eta) \parallel q_\mu(\eta)]|_{\mu'=\mu}$$

(8.20)

Conceptually, the key point to take from this is that internal manifolds are equipped with an information geometry because they represent probability densities over external states [25]. This is a necessary consequence of the conditional dependencies entailed by the Markov blanket. The utility of information geometry is reflected in the enormous literature on its applications to mathematical statistics, model selection, thermodynamics and, indeed, quantum information geometry (Caticha, 2015a). This section considers a special case; namely, the Laplace assumption that surprisal is locally quadratic – and the variational density is consequently Gaussian [26]. This assumption leads to a particular formulation of Bayesian mechanics, where the (Fisher information) metric tensor is the precision matrix of the variational density – and the curvature of surprisal (Friston et al., 2007).

Statistically speaking, a stipulative definition of the variational density – as the posterior over hidden states – means one can interpret the most likely trajectory of internal states as performing *exact Bayesian inference*. If we relax this stipulative definition and assume a (locally) quadratic form for surprisal, internal dynamics can be interpreted in terms of *approximate Bayesian inference*, also known as *variational Bayes* or, more specifically, *variational Laplace*. Practically, this means one can associate internal states with a well-behaved variational density that can be used to evaluate expected free energy. This affords a tractable description of self-organisation in terms of autonomous behaviour that acquires a prospective aspect. We will first focus on the nature of approximate Bayesian inference, under the Laplace assumption, and return to autonomous behaviour – under the rubric of active inference – in subsequent sections.

## Variational Bayes

Although the NESS lemma follows from standard results, there is something remarkable about the flow it entails: the flow of internal (and external) states is essentially a (circuitous) gradient descent on surprisal. The gradient flow is circuitous because it contains divergence-free (solenoidal) components that circulate on the isocontours of the NESS density – like walking down a winding mountain path. This means internal (resp. external) states are flowing towards regions of state-space that are most frequently occupied, despite the fact that their flow is not a function of external (resp. internal) states (Friston, 2013). In short, internal states behave as if they can 'see' external states behind their blanket. There is no mystery here: it just means that the gradients of surprisal, with respect to internal states do not depend upon external states by the marginal flow lemma (1.21). So, what does this mean for the relationship between internal and external states?

---

[25] Assuming for the moment that $\sigma(\mu(\tau)) \approx \sigma(\mathbf{\mu}(b(\tau)))$, so that every internal state parameterises a variational density.

[26] The choice of a Gaussian approximation (a.k.a. the Laplace assumption) is not arbitrary. This follows from the fact that the Gaussian distribution has the largest entropy of probability densities whose parameters are limited to first and second order moments. This ensures the variational entropy term in (8.4) is minimised in relation to other forms. One can also view the Laplace assumption as a second-order approximation to any density from the exponential class (e.g., a canonical density).

The Markov blanket ensures that internal and external states are conditionally independent. This means that there must be a mapping between the most likely internal and external states, when conditioned on blanket states. We will refer to this mapping has a *conditional synchronisation manifold* and use it to recapitulate the free energy lemma under the Laplace assumption. The existence of a conditional synchronisation manifold allows one to characterise the relationship between (maximum *a posteriori*) internal and external states in terms of internal states 'sensing' or 'tracking' external states through the Markov blanket. This notion of sentience can be expressed formally by the following lemma[27]:

**Lemma**: (approximate Bayesian inference): *when conditioned on blanket states, the internal states of a weakly mixing random dynamical system can be cast as performing approximate Bayesian inference on external states, via a minimisation of variational free energy.*

Approximate Bayesian inference requires the flow on the internal manifold to decrease an upper bound on (negative) Bayesian model evidence. Under the Laplace assumption this bound is a (free energy) functional of an approximate posterior (i.e., variational) density $q_\mu(\eta) = \mathcal{N}(\sigma(\mu), \Sigma(\mu)) \approx p(\eta \,|\, b)$ parameterised by the most likely internal states:

---

[27] This treatment eschews some subtleties of approximate Bayesian inference in the context of dynamical systems (often referred to as Bayesian filtering). In more general formulations, generalised coordinates are augmented with high orders of motion $\vec{x} = (x, x', x'', \ldots)$: see Appendix E and Friston, K., Stephan, K., Li, B., Daunizeau, J., 2010. Generalised Filtering. Mathematical Problems in Engineering vol., 2010, 621670. Effectively, this accommodates the (local) history or trajectory of hidden states (via a Taylor expansion). Furthermore, it allows one to dispense with Wiener assumptions and deal with analytic (smooth, differentiable) fluctuations. However, this induces off-diagonal terms in their covariance $\Gamma(\omega, \omega', \omega'', \ldots)$, which complicates the derivation of the NESS lemma.

$$\dot{\boldsymbol{\mu}}(b) = -\Gamma_{\sigma\sigma}\nabla_{\mu}F(\boldsymbol{\mu},b)$$
$$\dot{\mathbf{a}}(\boldsymbol{\mu}) = -\Gamma_{aa}\nabla_{a}F(\boldsymbol{\mu},b)$$

$$F(\boldsymbol{\mu},b) \geq \Im(b)$$
$$F(\boldsymbol{\mu},b) \triangleq \underbrace{E_q[\Im(\eta,b)]}_{energy} - \underbrace{H[q_{\boldsymbol{\mu}}(\eta)]}_{entropy} \qquad (8.21)$$
$$= \underbrace{D[q_{\boldsymbol{\mu}}(\eta) \| p(\eta \mid b)]}_{evidence\ bound} - \underbrace{\ln p(b)}_{log\ evidence}$$
$$= \underbrace{E_q[\Im(b \mid \eta)]}_{inaccuracy} + \underbrace{D[q_{\boldsymbol{\mu}}(\eta) \| p(\eta)]}_{complexity} \gtrsim \Im(b)$$

$$q_{\boldsymbol{\mu}}(\eta) = \mathcal{N}(\sigma(\boldsymbol{\mu}),\Sigma(\boldsymbol{\mu}))$$

Here, $\Gamma_{\sigma\sigma}$ is a positive semi-definite matrix that plays the role of a metric tensor, ensuring that free energy decreases with time. Variational free energy is guaranteed to be greater than the (negative log) marginal likelihood of blanket states (a.k.a., model evidence), because the KL divergence (a.k.a., evidence bound) in (8.21) – between the approximate and true posterior densities – cannot be less than zero (Beal, 2003). Note that variational free energy is an upper bound on the surprisal of blanket states, as opposed to particular states; as in (8.4). This means that internal states can only minimise variational free energy by reducing the evidence bound.

**Proof**: the proof rests upon establishing the existence of an approximate posterior density that satisfies (8.21). This calls on a conditional synchronisation manifold associated with a (differentiable) mapping between the most likely internal and external states, conditioned upon the Markov blanket. We will refer to these as maximum *a posteriori* or conditional modes:

$$\boldsymbol{\eta}(b) = \arg\max_{\eta} p(\eta \mid b)$$
$$\boldsymbol{\mu}(b) = \arg\max_{\mu} p(\mu \mid b) \qquad (8.22)$$
$$\boldsymbol{\eta}(b) = \sigma(\boldsymbol{\mu}(b)) \Rightarrow \dot{\boldsymbol{\eta}}(b) = \nabla_{\mu}\sigma \cdot \dot{\boldsymbol{\mu}}(b)$$

The existence of this smooth ($C^1$ or higher) map $\sigma: \boldsymbol{\mu} \to \boldsymbol{\eta}$ is assured because for every point in blanket space there is a unique pair of conditional modes in the external and internal state-space. This map, induced by the Markov blanket, allows one to relate the flow of internal and external modes in terms of generalised synchrony (Barreto et al., 2003; Hunt et al., 1997). Technically, this is an instance of *strong* synchronisation[28], because the dimensionality of the conditional manifold $M = \{(\boldsymbol{\mu},\boldsymbol{\eta}): \sigma(\boldsymbol{\mu}) = \boldsymbol{\eta}\}$ is the same as the dimensionality of the blanket states: $(\mu, \eta) \in \mathbb{R}^{|B|}$. The last equality in (8.22) follows because the flow of conditional modes must lie on the synchronisation manifold.

---

[28] Weak synchronisation characterises coupling when the manifold has a greater dimensionality than the internal (or external) spaces.

Using the chain rule $\nabla_\mu \Im(\sigma(\mu) \,|\, b) = \nabla_\mu \sigma \cdot \nabla_\eta \Im(\eta \,|\, b)$, one can now express the dynamics of the internal mode as a gradient flow on the surprisal of the external mode:

$$
\begin{aligned}
\dot{\mathbf{a}}(\mu) &= -\Gamma_{aa} \nabla_a \Im(b \,|\, \mu) \\
\dot{\eta}(b) &= -\Gamma_{\eta\eta} \nabla_\eta \Im(\eta \,|\, b) \\
\dot{\mu}(b) &= (\nabla_\mu \sigma)^- \dot{\eta}(b) \\
&= (\nabla_\mu \sigma)^- (Q_{\eta\eta} - \Gamma_{\eta\eta}) \cdot (\nabla_\mu \sigma)^- (\nabla_\mu \sigma) \cdot \nabla_\eta \Im(\eta \,|\, b) \\
&= -\Gamma_{\sigma\sigma} \nabla_\mu \Im(\sigma(\mu) \,|\, b)
\end{aligned}
\tag{8.23}
$$

$$
\Gamma_{\sigma\sigma} = (\nabla_\mu \sigma)^- (\Gamma_{\eta\eta} - Q_{\eta\eta}) \cdot (\nabla_\mu \sigma)^-
$$

Here, $(\nabla_\mu \sigma)^-$ is the generalised inverse of the mapping from internal and external modes. Equation (8.23) says that the flow of the internal mode is effectively performing a gradient descent on the surprisal of the external mode. Now, let the internal mode parameterise a Gaussian density:

$$
\begin{aligned}
q_\mu(\eta) &= \mathcal{N}(\sigma(\mu), \Sigma(\mu)) \\
\Sigma(\mu) &= \nabla_{\sigma\sigma}^{-1} \Im(\sigma(\mu), b)
\end{aligned}
\tag{8.24}
$$

Under the Laplace assumption that the surprisal is quadratic (i.e., ignoring higher-order terms and derivatives), substitution into the expression for variational free energy (8.21) gives (dropping constants):

$$
\begin{aligned}
F(\mu, b) &= \Im(\sigma(\mu), b) + \tfrac{1}{2} tr[\Sigma(\mu) \nabla_{\sigma\sigma} \Im(\sigma(\mu), b)] - \tfrac{1}{2} \ln|\Sigma(\mu)| \\
&= \Im(\sigma(\mu), b) - \tfrac{1}{2} \ln|\Sigma(\mu)|
\end{aligned}
$$

$$
\begin{aligned}
\nabla_a F &= \nabla_a \Im(b \,|\, \sigma(\mu)) = \nabla_a \Im(b \,|\, \mu) \\
\nabla_\mu F &= \nabla_\mu \Im(\sigma(\mu) \,|\, b) \\
\nabla_\Sigma F &= \tfrac{1}{2} \nabla_{\sigma\sigma} \Im(\sigma(\mu), b) - \tfrac{1}{2} \Sigma(\mu)^{-1} = 0 \\
&\Rightarrow \nabla_{\sigma\sigma} \Im(\sigma(\mu), b) = \Sigma(\mu)^{-1} = \mathbf{I}(\mu)
\end{aligned}
\tag{8.25}
$$

This equation shows that the conditional precision minimises variational free energy, while the variational free energy gradients are the gradients of the surprisal of the external mode, which determines the flow of the internal mode. More specifically, by substituting (8.25) into (8.23) we obtain (8.21), where internal and active states can be expressed as a gradient flow on variational free energy:

$$\dot{\boldsymbol{\mu}}(b) = -\Gamma_{\sigma\sigma}\nabla_\mu F(\boldsymbol{\mu}, b)$$
$$\dot{\mathbf{a}}(\boldsymbol{\mu}) = -\Gamma_{aa}\nabla_a F(\boldsymbol{\mu}, b)$$

(8.26)

$$\Gamma_{\sigma\sigma} = (\nabla_\mu \sigma)^- \Gamma_{\eta\eta} \cdot (\nabla_\mu \sigma)^-$$

The solenoidal flow disappears because $Q_{\eta\eta}$ is antisymmetric. This ensures that $\Gamma_{\sigma\sigma}$ is positive semi-definite (because the covariance of random fluctuations is positive definite) □

**Corollary** (expected free energy). The corresponding corollary for expected free energy under the Laplace assumption obtains by removing the internal states from the energy term in (8.8); i.e., replacing $\alpha_\tau = (a_\tau, \mu_\tau)$ with $a_\tau$ and $\pi_\tau = (b_\tau, \mu_\tau)$ with $b_\tau$. Following the derivations for exact Bayesian inference we have, after a critical time:

$$G(a_\tau) \geq \Im_\tau(a_\tau \mid \pi_0)$$
$$G(a_\tau) \triangleq E_{q_\tau}[\underbrace{\Im(b_\tau \mid \eta_\tau)}_{ambiguity} + \underbrace{D[q_\tau(\eta_\tau \mid b_\tau) \parallel p(\eta_\tau)]}_{risk}] \geq E_{q_\tau}[\Im(b_\tau)]$$

(8.27)

$$q_\tau(\eta_\tau, s_\tau \mid a_\tau) \approx p_\tau(\eta_\tau, s_\tau \mid a_\tau, \pi_0)$$

Similarly, under the Laplace assumption, (8.12) and (8.13) become:

$$E_{p(b)}[F(\boldsymbol{\mu}, b)] \approx E_{p(b)}[\Im(b)] = H(B)$$
$$E_{p(a_\tau)}[G(a_\tau)] \approx H(B) = \underbrace{H(B \mid E)}_{ambiguity} + \underbrace{I(E, B)}_{risk}$$

(8.28)

In other words, the average variational free energy of blanket states – and expected free energy of active states – approximates the entropy of blanket states. This means that any particle will appear to actively resist the dispersion of its Markov blanket by engaging in approximate Bayesian inference. We will return to this interpretation later. First, we consider a couple of different perspectives on this emergent property of the Markov blanket.

Note that the conditional precision is the curvature of surprisal, evaluated at the conditional mode. Crucially, under the Laplace assumption, this curvature is also the Fisher information metric in (8.19):

$$\mathbf{I}(\boldsymbol{\mu}) = \Sigma(\boldsymbol{\mu})^{-1} = -\nabla_{\sigma\sigma} \ln q_\mu(\eta)$$
$$= E_q[\nabla_\sigma \ln q_\mu(\eta) \times \nabla_\sigma \ln q_\mu(\eta)]$$

(8.29)

In other words, the uncertainty about external states manifests as a Fisher information metric that equips the internal statistical manifold with a representational (i.e., information) geometry – a geometry where distances are measured in terms of the curvature of surprisal. See Figure 18 for an illustration of these relationships.

Equation (8.25) uses the Laplace assumption to associate variational free energy gradients with gradients of surprisal: $\nabla_\mu F(\mu, b) = \nabla_\mu \Im(\sigma(\mu), b)$, effectively ignoring kurtosis (and higher order moments) of the NESS density over external states. This is known as maximum *a posteriori* (MAP) estimation and follows the usual derivation of Variational Laplace, where the solution to (8.23) renders the variation of free energy, with respect to the variational density, zero; i.e., minimises variational free energy to second order (Friston et al., 2007):

$$\left.\begin{matrix}\dot{\mu}(b) = 0\\ \nabla_\Sigma F = 0\end{matrix}\right\} \Rightarrow \left.\begin{matrix}\nabla_\mu \Im(\sigma(\mu), b) = 0\\ \nabla_{\sigma\sigma} \Im(\sigma(\mu), b) = \Sigma(\mu)^{-1}\end{matrix}\right\} \Rightarrow \delta_q F = 0 \Leftrightarrow \partial_\mu F = 0$$

$$q_\mu(\eta) = \mathcal{N}(\sigma(\mu), \Sigma(\mu))$$

(8.30)

Clearly, these solutions never obtain in a dynamic setting; however, they can be recovered in a moving frame of reference, supplied by generalised coordinates of motion: see (Friston et al., 2010) and Appendix E. The final quality in (8.30) explains why variational free energy and variational Bayes are called 'variational' (Beal, 2003; Fox and Roberts, 2011; MacKay, 1995). From our perspective, the important thing here is that we have moved from exact Bayesian inference – in which the variational density was defined as the true posterior – to approximate Bayesian inference, where internal states (appear to) minimise a variational bound on surprisal. This distinction arises because we have committed to a particular form for the variational density; namely, a Gaussian form under the Laplace assumption that surprisal is locally quadratic.

Finally, the last equality in (8.26) shows that the positive semidefinite matrices play the role of metric tensors where:

$$g[\eta]_{kl} = g[\mu]_{ij} D_i^k D_j^l \Leftrightarrow \Gamma_{\eta\eta} = (\nabla_\mu \sigma)\Gamma_{\sigma\sigma} \cdot (\nabla_\mu \sigma)$$

$$\Gamma_{\eta\eta} = (g[\eta]_{kl})$$

$$\Gamma_{\sigma\sigma} = (g[\mu]_{ij})$$

$$(\nabla_\mu \sigma)_{ki} = D_i^k = \partial \eta^k / \partial \mu^i$$

(8.31)

This suggests that the implicit representation of external states, by internal states, depends upon the amplitude of random fluctuations. The use of the word 'representation' here is qualified by the fact that internal states can be associated with a *Bayesian belief* about external states. This endows the internal manifold with a representational (i.e., relational) meaning that inherits from its information geometry. In other words, the beliefs (approximate posterior densities) parameterised or encoded by internal states are *about something*; namely, the external states.

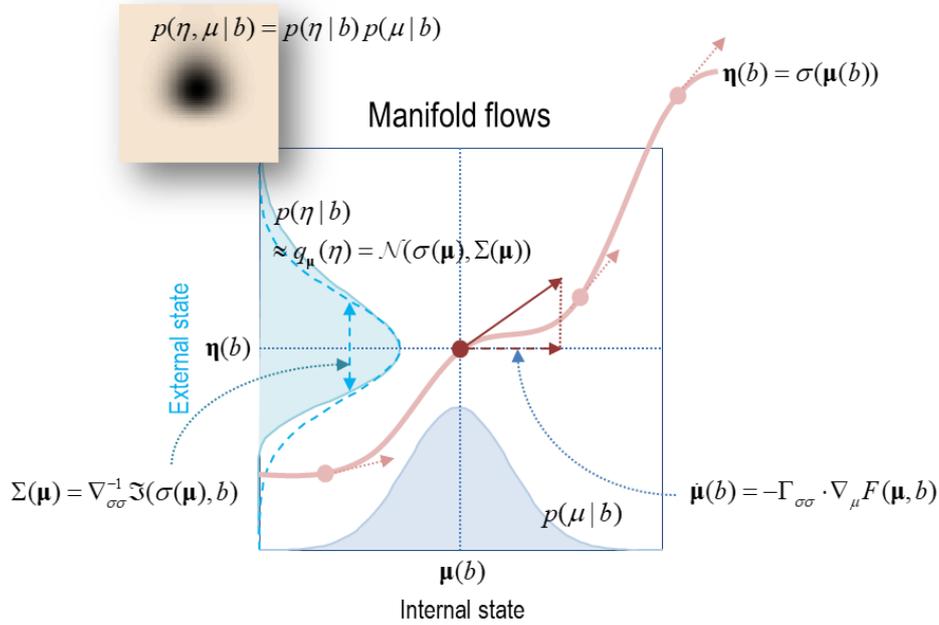

Furthermore, if we assume a surjective mapping (e.g., nonlinear mixture) $\xi : \mu \to \boldsymbol{\mu}$ from internal states to the internal manifold, we have:

$$\boldsymbol{\mu}(b(\tau)) \approx \xi(\mu(\tau)) \Rightarrow q_\mu(\eta) = \mathcal{N}(\sigma(\xi(\mu)), \Sigma(\xi(\mu))) \approx p(\eta \,|\, b) \tag{8.33}$$

This assumption means that the sufficient statistics of the variational density are encoded by nonlinear mixtures of internal states. To the extent this assumption holds, every internal state becomes part of the internal conditional synchronisation (i.e., statistical) manifold – and acquires an information geometry. The assumption of nonlinear mixtures implies a dimension reduction (from the dimensionality of internal states to the dimensionality of the internal manifold; i.e., blanket states). In turn, this means there will be degenerate internal states that are infinitesimally close, in terms of their information length.

So how does this inferential interpretation relate to inference in the conventional statistical sense? Does the approximate Bayesian inference lemma mean that all random dynamical systems with a Markov blanket are little statisticians? The answer lies in noting that the surprisal in (8.25) $\Im(\sigma(\boldsymbol{\mu}), b) = -\ln p(\boldsymbol{\eta}(b), b)$ plays the role of a generative model in Bayesian statistics; namely, a probabilistic specification of how causes (external states) and consequences (blanket states) depend upon each other. This means the NESS density underwrites the generative model. In contrast, a statistician faced with some data would select a particular generative model, specify the approximate form of her posterior beliefs and then minimise variational free energy (i.e., maximise model evidence). However, the internal states do not 'know' the generative model; in the sense that the generative model is not represented explicitly. This is why we have previously asserted that the agent *is* (or entails) a generative model of its world – as opposed to *possessing* a generative model (Friston, 2013). For example, the curvature in (8.25), underwriting uncertainty, is 'unknown' to the internal states and is not 'represented'. However, the *natural inference* of self-evidencing is, unlike the statistician, guaranteed to use the correct generative model, because it has already been selected through the emergence of the Markov blanket and implicit generative model. In short, the natural statistician implied by a Markov blanket will always produce better (lower free energy) expectations of how its sensory data are caused but will not know (i.e. represent) how it arrived at those expectations. In short, any (active) particle will look 'as if' it is performing Bayesian inference and – as we will see in the next section – Bayesian beliefs can be recovered from internal states. However, this 'as if' inference does not consider competing hypotheses – or perform any Bayesian model selection – in the usual sense of Bayesian statistics, because it is always using the true model. The emergence of model selection and planning rests on deep generative models that we will consider in the penultimate section.

Although the approximate Bayesian inference lemma may appear rather involved, there is a deflationary account at hand: all we are doing here is to note that at nonequilibrium steady-state there must be a pair of internal and external modes for every blanket state. If we know the mapping between these modes, then the internal state tells us about the distribution of external states: namely, there is a direct correspondence between the internal and external modes, while the implicit dispersion of external (i.e., hidden) states is just the curvature of the conditional (i.e., posterior) surprisal, given the blanket state. In turn, this is the curvature of the marginal (i.e., prior) surprisal plus the curvature of the conditional (i.e., likelihood) surprisal associated with blanket state (i.e., data), given the external state. All this follows naturally from Bayes rules – hence Bayesian mechanics. The only formal issue here is the nature and existence of the posterior. We have assumed it can be approximated with a Gaussian density. This (Laplace) assumption becomes more tenable when working in generalised coordinates of motion (Friston et al., 2010), which licenses an appeal to the Takens embedding theorem (Deyle and Sugihara, 2011; Takens, 1980), to establish the existence of a well-behaved conditional synchronisation manifold.

In summary, this section suggests that the existence of a Markov blanket in any (weakly mixing) random dynamical system means that internal states can be treated as parameterising a Bayesian belief or (approximate) posterior probability density over external states. More specifically, the most likely internal states – induced by external impressions on the Markov blanket – come to encode the most likely external states. This formulation of sentient or representational dynamics rests upon the notion of (conditional) synchronisation of chaos – between internal and external states – that provides a nice mathematical metaphor for Freeman's "Kiss of chaos" (Freeman, 1995).

Notice that the ensuing representations are encoded by the most likely internal states and, implicitly, trajectories on the internal manifold. Practically, this means that one would have to take the average responses of internal states to disclose their sentient nature. We turn to this issue next, in the context of empirical studies that use event-related averaging. The key point here is that in the life sciences; particularly in the neurosciences, the encoding of representations almost invariably involves taking averages, when examining the neural correlates of perceptual inference. Interestingly, this averaging is mandated under the current treatment, via the use of conditional modes (or averages). In the next section, we use see how inference and representation emerge in a familiar way, even in relatively simple systems like viruses or bacteria.

## Simulating sentience

This section uses the microscopic denizen of our synthetic soup to illustrate the nature of Bayesian inference afforded by conditional synchronisation. In brief, we first consider how to identify the mapping between internal and external modes, conditioned upon blanket states; which underwrites the conditional synchronisation manifold $\sigma : \mu \rightarrow \eta$ and ensuing inference. Numerically, this entails scrolling through the time-series, identifying when particular configurations of Markov blanket states recur and taking the mode or average of concurrent internal and external states. The conditional synchronisation manifold can then be estimated by regressing the average external states on (some nonlinear function of) the average internal states. Having identified the synchronisation manifold, it is straightforward to evaluate the posterior belief associated with any (expected) internal state – about the external state – and any uncertainty, in terms of the curvature in (8.24). To illustrate this representational behaviour, we will take a subset of (electrochemical) states of internal particles and see to what extent they encode or represent fluctuations in the (motion) states of external particles.

We then revisit this behaviour from the point of view of an electrophysiologist, who assembles ensemble averages of internal states – in the form of event-related potentials – and tries to identify whether any trajectories are time locked to fluctuations of external states; for example, the presentation of a visual motion stimulus. We will see that the ensuing ensemble averages bear a remarkable similarity to the kind of results seen in empirical event-related potential (ERP) studies.

## The representation of order

To keep things simple, we will focus on the representation of a single attribute of the external world; namely, the collective motion of external macromolecules, as encoded by internal electrochemical states (very much like the neuronal firing encodes visual motion). To first identify the conditional expectations, the internal and external states were averaged, according to their proximity to blanket states, over the penultimate 512 seconds of the simulation:

$$\boldsymbol{\mu}(b(\tau)) \triangleq \sum_t \boldsymbol{\sigma}(\Delta_\tau) \cdot \mu(t)$$
$$\boldsymbol{\eta}(b(\tau)) \triangleq \sum_t \boldsymbol{\sigma}(\Delta_\tau) \cdot \eta(t)$$

$$(9.1)$$

$$\boldsymbol{\sigma}(\Delta_\tau) = \frac{\exp(-\Delta_\tau)}{\sum_t \exp(-\Delta_\tau)}$$
$$\Delta_\tau = \tfrac{1}{128} \| b(t) - b(\tau) \|^2$$

Here, $\boldsymbol{\sigma}(\cdot)$ is a softmax (i.e., normalised exponential) function of the (squared) Mahalanobis distance between the blanket state of at every point in time $b(t)$ and the time upon which $b(\tau)$ was conditioned. This averaging procedure creates two sets of time-series, corresponding to estimates of the expected internal and external states, conditioned upon the realised values of the Markov blanket. To identify the synchronisation manifold, canonical covariates analysis was used to identify the principal canonical vectors that showed the greatest (canonical) correlation between the internal and external states examined. Finally, the synchronisation manifold *per se* was identified using a fifth order polynomial regression of the expected external canonical variate on the expected internal variate. Intuitively, this characterisation of representational coupling (between internal and external states) is in terms of linear mixtures that correspond to canonical vectors, whose expressions over time correspond to canonical variates. In this instance, the external vector is a distributed pattern of motion; like a convection swirl around the little organism. The internal vector can be regarded as a distributed representation in terms of electrochemical (e.g., neuronal) patterns of activity. In short, we were effectively looking for evidence that our synthetic creature could detect and represent motion in its external milieu, using its internal electrochemistry.

The upper left panel of Figure 19 shows the first canonical vector of motion over the external states (green arrows), which is represented by the internal states (blue dots). The blue and cyan dots are placed at the location of internal and external states, respectively. The colour level reflects the norm (sum of squares) of the first canonical vector showing the greatest covariation between external and internal states. The upper right panel illustrates a synchronisation manifold (conditioned upon the Markov blanket) that maps from the electrochemical states of internal macromolecules to the velocity of external macromolecules. The blue dots identify the manifold *per se*, while the cyan dots are the expectations in (9.1) used to estimate the manifold (using a fifth order polynomial regression). The middle panel shows the same information but plotted as a function of time during the last 512 seconds of the simulation. In this format, the posterior or conditional expectations are based upon the electrochemical states of internal macromolecules, while the real motion is shown as a cyan line. The blue shaded areas correspond to 90% confidence intervals. The key thing to note here is that the real motion lies, almost universally, within the 90% confidence intervals that are based on the variational density: see (8.24).

The lower left panel illustrates simulated event-related potentials of the sort shown in the insert on the lower right. This illustrative example was taken from a study of autistic spectrum disorder reported in (Modi and Sahin, 2017). These are responses to alternating chequerboard visual stimuli in humans. The simulated ERP shows a remarkable similarity – and was obtained by time-locking the internal electrochemical states to the six time points that showed the greatest expression of the first canonical variate; i.e., motion of external states. The dotted lines correspond to the simulated (internal) responses around the six time points (indicated by the vertical lines in the middle panel), while the solid lines correspond to the average – as in an event-related potential ERP. The blue lines are the responses of internal states, while the cyan lines correspond to the real motion associated with the first canonical vector.

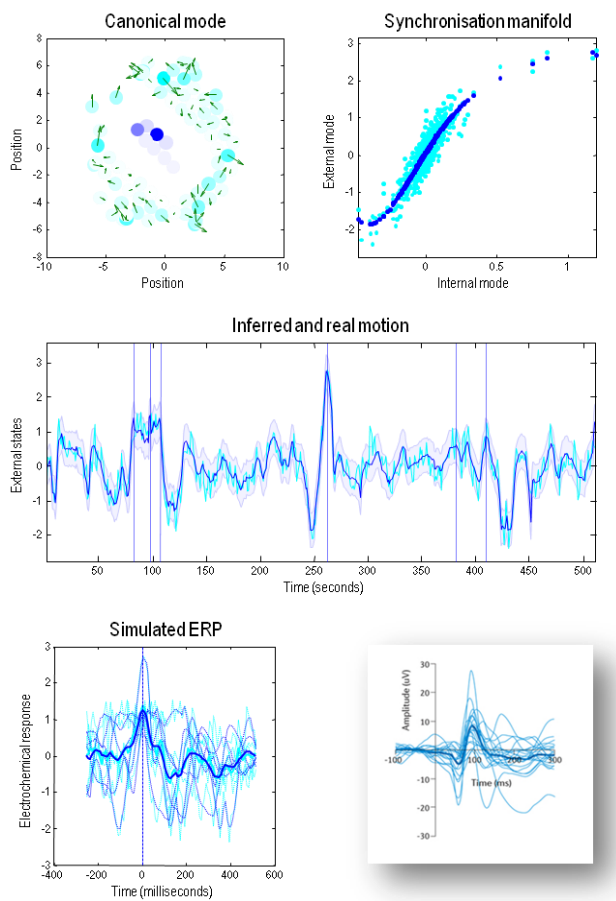

FIGURE 19

*Sentient dynamics and the representation of order*. This figure illustrates approximate Bayesian inference that follows when associating the internal states of a system with a variational (i.e., approximate posterior) density over external states. The upper left panel shows the first canonical vector of motion over the external states (green arrows) that are represented by the internal states (blue dots). The blue and cyan dots are placed at the location of internal and external states, respectively. The colour level reflects the norm (sum of squares) of the first canonical vectors showing the greatest covariation between external and internal states. The upper right panel illustrates a synchronisation manifold (conditioned upon the Markov blanket) that maps from the electrochemical states of internal macromolecules to the velocity of external macromolecules. The blue dots identify the manifold *per se*, while the cyan dots are the estimated expectations used to estimate the manifold (using a fifth order polynomial regression). The middle panel shows the same information but plotted as a function of time during the last 512 seconds of the simulation. In this format, the conditional expectation is based upon the electrochemical states of internal macromolecules, while the real motion is shown as a cyan line. The blue shaded areas correspond to 90% confidence intervals. The lower left panel illustrates simulated event-related potentials of the sort illustrated by the insert (lower right panel). The simulated ERP (lower left panel) was obtained by time locking the internal electrochemical states to the six time points that showed the greatest expression of the first canonical variate (indicated by the vertical lines in the middle panel). The dotted lines are six trajectories around these points in time, while the solid lines correspond to the average. The blue lines are the responses of internal states, while the cyan lines correspond to the real motion associated with the first canonical vector. The timing in the lower panels has been arbitrarily rescaled to match empirical peristimulus times.

## Summary

In summary, numerical analyses suggest that (average) internal states can represent, in a Bayesian sense, (average) external states with a remarkable degree of fidelity. Furthermore, this form of representation endorses the standard practice in electrophysiology of trying to match average internal neuronal states with repeated stimuli – a practice that is seen almost universally in physiology. Note that we have focused on the first pair of canonical vectors. There were, in fact, 24 pairs, many of which showed an extremely high canonical correlation. This means that although we have illustrated the encoding of a single pattern of external motion, the actual representation was high dimensional – and could be used to reconstruct the movement of macromolecules in the external milieu. Having said this, the above demonstration is sufficient for our purpose, which was to illustrate the dynamical and distributed nature of neuronal representation; e.g., (Freeman, 1994).

In the foregoing, we have focused on gradient flows as the vehicle for Bayesian mechanics. However, this flow is a *solenoidal* gradient flow with divergence free flow or *flux*. It is interesting to speculate on how this flow manifests in dissipative structures and, in particular, in biotic systems[29]. One important observation is that the existence of a Markov blanket necessarily introduces an asymmetry in dynamical coupling among particular states and their external milieu. Almost universally, this elementary form of symmetry breaking induces solenoidal flow at nonequilibrium steady-state (Yan et al., 2013), which – to an empirical eye – will manifest as oscillations. This is interesting because much of empirical neuroscience (and beyond) is occupied with oscillations and their ontology (Breakspear et al., 2010; Burgess et al., 2007; Buzsaki, 1998; Buzsaki and Moser, 2013; Giraud and Poeppel, 2012; Jensen et al., 2014; Lisman, 2012; Lopes da Silva, 1991; Sejnowski and Paulsen, 2006; Uhlhaas and Singer, 2010).

From the perspective of Bayesian mechanics, solenoidal flow is an inherent component of generalised Bayesian filtering: see Appendix E. In other words, if neuronal dynamics constitute a solenoidal gradient flow on variational free energy, then the solenoidal part can be associated with a prediction of how external states are changing, while the solenoidal, curl-free part is driven to minimise a variational bound on surprisal. From the perspective of an engineer, this is exactly the form of a Kalman-Bucy filter (a.k.a. predictive coding in neuroscience) (Dauwels, 2007; Roweis and Ghahramani, 1999). In short, the divergence free flows – that attend dissipative structures with Markov blankets – may play a key role in self-evidencing; especially, if we consider that one particle is trying to predict another particle that is, itself is oscillating. In turn, this brings us to another view on generalised synchronisation of dissipative structures that may undergird things like communication among ensembles – of particles (e.g., conspecifics) – that are trying to infer themselves (Friston and Frith, 2015; Hunt et al., 1997).

---

[29] My thanks to Jin D. Wang for correspondence on this notion.

So far, we have looked at the sentient aspect of Bayesian mechanics, in terms of inferring external states. In the remaining sections, we turn to the role of active states in the (active) inference that accompanies self-organisation; i.e., self-evidencing (Hohwy, 2016; Palacios et al., 2017).

# Active inference and self-evidencing

*"Each movement we make by which we alter the appearance of objects should be thought of as an experiment designed to test whether we have understood correctly the invariant relations of the phenomena before us, that is, their existence in definite spatial relations."* (Helmholtz, 1878 (1971)) p.384

In the preceding sections, we saw that the expected flow of internal states is consistent with a minimisation of variational free energy that reduces the KL divergence between an approximate and true (posterior) probability over hidden states, conditioned on blanket states:

$$q_{\mu}(\eta) \approx p(\eta \mid b) = p(\eta \mid \pi) \tag{10.1}$$

This furnishes an account of internal states in which they (appear to) play the role of sufficient statistics of posterior beliefs about external states. But what about the active states – is there an equivalent account? Note that internal states are behind the Markov blanket and cannot directly influence external states and, by implication, the Markov blankets of other particles. This means their influence is mediated vicariously by active states. So, what do we know about the active states?

From (8.21), the flow of the most likely autonomous states can be expressed as a gradient descent on variational free energy, under the Laplace assumption.

$$\dot{\mu}(b) = -\Gamma_{\mu\mu} \nabla_{\mu} \Im(\mu \mid b) = -\Gamma_{\sigma\sigma} \nabla_{\mu} F(\mu, b)$$
$$\dot{a}(\mu) = -\Gamma_{aa} \nabla_{a} \Im(b \mid \mu) = -\Gamma_{aa} \nabla_{a} F(\mu, b)$$

$$\tag{10.2}$$

$$\nabla_{\mu} F(\mu, b) = \nabla_{\mu} \underbrace{D[q_{\mu}(\eta) \parallel p(\eta \mid b)]}_{evidence\ bound}$$

$$\nabla_{a} F(\mu, b) = \nabla_{a} \underbrace{E_{q}[\Im(b \mid \eta)]}_{inaccuracy}$$

This is consistent with the constraint that autonomous states are driven by, and only by, particular states. Equation (10.2) also emphasises the fact that both internal and active modes are effectively trying to minimise the same quantity; namely, free energy functionals of implicit beliefs over hidden states. This provides an elementary description of sentient behaviour (Friston et al., 2017a) – a description that rests on the same free energy functional used in approximate Bayesian inference and machine learning (Beal, 2003). The final equalities above retain the terms from (8.21) that determine the gradients and subsequent flow of internal and active states; namely, the evidence bound and accuracy, respectively. On an anthropic view, internal states therefore strive to form veridical beliefs about external states (c.f., perception), while active states try to fulfil the ensuing beliefs to make them as accurate as possible (c.f., action). This is often cast in terms of a perception-action cycle (Friston et al., 2006; Fuster, 2004). This interpretation could be read as a purely epistemological exercise, in the sense that one can also cast self-evidencing as a gradient flow on particular surprisal or log-evidence: see (10.2). So, does variational free energy bring anything else to the table?

Things get more interesting when we consider autonomous behaviour in terms of *expected free energy*. In what follows, we will compare and contrast accounts of self-organisation – based upon gradient flows – with descriptions based upon the selection of trajectories that minimise expected free energy in the future. However, to evaluate expected free energy, one needs a posterior predictive density. This is available via posterior beliefs about external states that underwrite the variational free energy formulation (i.e., approximate Bayesian inference). In brief, if the evidence bound is minimised by internal states, the variational density becomes a good approximation to the true posterior required to evaluate expected free energy. In turn, the minimisation of expected free energy endows self-organisation with a prospective and intentional aspect, which we will now consider in greater depth.

## Active inference with continuous states

So far, we have a formulation in which internal states encode an approximate posterior over hidden states, which is used to predict the sensory states that active states will (appear to) realise (8.21). This realisation can be characterised as the most likely trajectory of autonomous states from any particular state. Recall from the expected free energy lemma (8.7) that the surprisal of future autonomous states can be expressed in terms of expected free energy:

$$D[q_\tau(x_\tau) \| p(x_\tau)] = 0 \Rightarrow G(\alpha_\tau) \geq \mathfrak{I}_\tau(\alpha_\tau \mid \pi_0)$$

$$G(\alpha_\tau) = E_{q_\tau}[\underbrace{\mathfrak{I}(s_\tau \mid \alpha_\tau, \eta_\tau)}_{sensory\ ambiguity} + \underbrace{\mathfrak{I}(\alpha_\tau \mid \eta_\tau)}_{active\ ambiguity} + \underbrace{D[q_\tau(\eta \mid s_\tau, \alpha_\tau) \| p(\eta_\tau)]}_{risk}]$$

$$(10.3)$$

This suggests that the trajectory of active (and internal) states from $\pi_0$ will (appear to) minimise expected surprisal (i.e., uncertainty about outcomes in the future). This is equivalent to minimising risk and ambiguity, where risk (i.e., complexity cost) is the expected KL divergence between the (posterior) predictive and prior

density – describing the sorts of states the system typically occupies. In other words, active states will appear to fulfil prior beliefs about states of affairs beyond the Markov blanket.

In (10.3), we have separated ambiguity into the ambiguity of sensory states (i.e., *sensory ambiguity*) and the expected surprisal of autonomous states, given hidden states (i.e., *active ambiguity*). The latter corresponds to the (negative log) likelihood of autonomous states and reflects the most likely responses of a particle under the (posterior) predictive density. This likelihood is sometimes characterised as a *state-action policy* in psychology and machine learning; e.g., (Dolan and Dayan, 2013; Gershman and Daw, 2017; LeCun et al., 2015). The distinction between the final two terms above (*active ambiguity* and *risk*) maps nicely to the distinction between *habitual* and *goal-directed* behaviour in psychology (Balleine and Dickinson, 1998; Dezfouli and Balleine, 2013; Dolan and Dayan, 2013). This distinction rests upon the context-sensitivity of goal-directed behaviour implicit in minimising the divergence between predicted and preferred states; while habits are elicited as the most likely response to an inferred state of affairs. These two components are, however, not sufficient to describe active inference, because they are contextualised by the epistemic imperative to minimise conditional uncertainty about sensory states. In cognitive neuroscience, this ultimately leads to ambiguity reducing, curious behaviour that underwrites the exploratory (i.e., epistemic) foraging that characterises self-evidencing (Baranes and Oudeyer, 2009; Berlyne, 1950; Friston et al., 2017b; Schmidhuber, 2006; Still and Precup, 2012).

In short, if a particle actively minimises expected free energy, it will self-organise. Equivalently, self-organising particles will appear to minimise expected free energy. The important thing here is that expected free energy is a function of, and only of, particular states. This means that if we observed a particle at nonequilibrium steady-state, it would look as if the internal states were guiding active states. The ensuing behaviour would, on average, appear to minimise uncertainty (expected surprisal) via the minimisation risk and ambiguity.

One might ask what the formulation of expected free energy – in terms of risk and ambiguity – offers beyond an epistemological account. We will see below that it is possible to define expected free energy in terms of beliefs about hidden states (that constitute risk), to specify the dynamics of sentient systems. This can be useful when simulating self-organisation to an attracting set, defined in terms of prior beliefs $p(\eta_\tau)$ or *preferences* (Friston et al., 2015b). This is an important move, beyond simply characterising self-organisation, towards creating systems that engage in active inference. In short, one can prescribe self-organisation in terms of preferred states by solving for autonomous trajectories that satisfy (10.3).

Interestingly, expected free energy shares a mathematical heritage with (incomputable) formulations of artificial general intelligence based upon Universal Computation; i.e., a combination of Solomonoff induction with sequential decision theory (Hutter, 2006). This heritage rests on the relationship between the complexity term in variational free energy and algorithmic complexity (Hinton and Zemel, 1993; Wallace and Dowe, 1999), which can be articulated in terms of information length and total variation distance. However, unlike Solomonoff induction, variational free energy is computable because it uses bound approximations (Feynman, 1972). We now take a closer look at expected free energy and the sorts of behaviours its minimisation would manifest.

## Active inference with discrete states

The final move – in theoretical biology – is to assume a particular form for the generative model and ask whether expected free energy is sufficient to explain the behaviour of sentient creatures like ourselves. We have seen nonequilibrium steady-state dynamics can be cast as approximate Bayesian inference. Now, we can ask: what would happen if this model included the dependencies of external (hidden) states on active states? This would lead to a curious situation in which internal states parameterise beliefs about active states, leading to a fundamental distinction between active states *per se* – that conform to the particular free energy lemma – and the particle's approximate posterior *beliefs about its active states*. These posterior beliefs rest upon a generative model equipped with a prior, over the trajectories of active states (i.e., policies), that conforms to the expected free energy lemma. This leads to a more prospective form of active inference that entails a generative model of trajectories and a form of self-organisation that starts to look like autopoiesis; i.e., self-creation of a definitive biological sort (Maturana and Varela, 1980; Thompson and Varela, 2001). This kind of particle would look as if it was inferring its own behaviour and indeed selecting actions from the risk sensitive, ambiguity reducing, policies it inferred were the most likely.

The pragmatic move here shifts from deriving the nonequilibrium steady-state density from any given dynamics – using the Fokker Planck formulation – to deriving the dynamics from any given density – using the expected free energy. In the context of active inference, the NESS density corresponds to the prior preferences, over external states, which specify a prior over future autonomous states; i.e., beliefs about autonomous behaviour.

The models usually entertained in this setting consider continuous state-spaces – leading to schemes like predictive coding (Bastos et al., 2012; Rao and Ballard, 1999) – or discrete state-space models that are more apt for modelling discrete choices and decisions in an experimental setting (Friston et al., 2017a). In what follows, we briefly consider the minimisation of expected free energy under discrete models. Under discrete states, the expected free energy can be expressed as follows (see Appendix F for details):

$$G(\alpha^i) = \underbrace{D[Q_i(\eta_\tau \mid \alpha^i) \parallel P(\eta_\tau \mid \alpha^i)]}_{risk} + \underbrace{E_{Q_i}[\Im(s_\tau \mid \eta_\tau)]}_{ambiguity} + \Im(\alpha^i) \geq \Im(\alpha^i \mid \pi_0)$$

$$Q_i \triangleq P(\eta_\tau, s_\tau \mid \alpha^i, \pi_0)$$

(10.4)

In this context, prior preferences $P(\eta_\tau \mid \alpha^i)$ specify (a bound on) the prior probability of a sequence of active states; usually denoted by a *policy*, $\alpha^i = (\alpha_1^i, \ldots, \alpha_\tau^i)$. This construction is slightly simpler than the continuous state-space formulations we have been dealing with so far. Here, the variational density is specified in terms of a sensory likelihood that does not depend upon active states (because sensory states depend only on hidden states in the generative models in question; e.g., hidden Markov models and Markov decision processes).

Equation (10.4) describes how policies are sampled when a particle self-organises to nonequilibrium steady-state – that becomes the 'preferred' state. The converse suggests that any particle that samples its policies according to (10.4) should show self-organising, self-evidencing behaviour:

$$p(a^i \mid \pi_0) \approx \boldsymbol{\sigma}(-G(a^i)) \Leftrightarrow \Im(a^i \mid \pi_0) \approx G(a^i) \tag{10.5}$$

To incorporate the prior belief that policies should be selected to minimise expected free energy, it is usual to assume a generative model of trajectories, under each allowable or plausible policy. The minimisation of variational free energy then proceeds using belief propagation or variational message passing, under a generative model specified by a likelihood matrix and policy-dependent state transition matrices (Friston et al., 2017c). Figure 20 illustrates the computational architecture implied by (10.4) and the figure legend lists a few examples of its application in neuroscience. The key point here is that all of these applications are predicated on the minimisation of expected free energy, following the treatment of self-organisation above.

The computational efficiency afforded by discrete formulations of free energy minimisation reveals a qualitative distinction between two sorts of self-evidencing behaviour; namely, (i) reflexive, short term self-organisation – akin to homoeostasis (Ashby, 1947; Bernard, 1974) and (ii) deep, long-term, active inference – that speaks to planning and allostasis (Attias, 2003; Botvinick and Toussaint, 2012; Ramsay and Woods, 2014; Stephan et al., 2016; Sterling and Eyer, 1988; Sutton et al., 1999; Toussaint and Storkey, 2006). In terms of information lengths, this distinction can be associated with particles that have short and long critical times, respectively (8.15). We conclude this section by revisiting this shallow (short-term) *vs*. deep (long-term) distinction in terms of gradient flows *vs*. the selection of paths of least action.

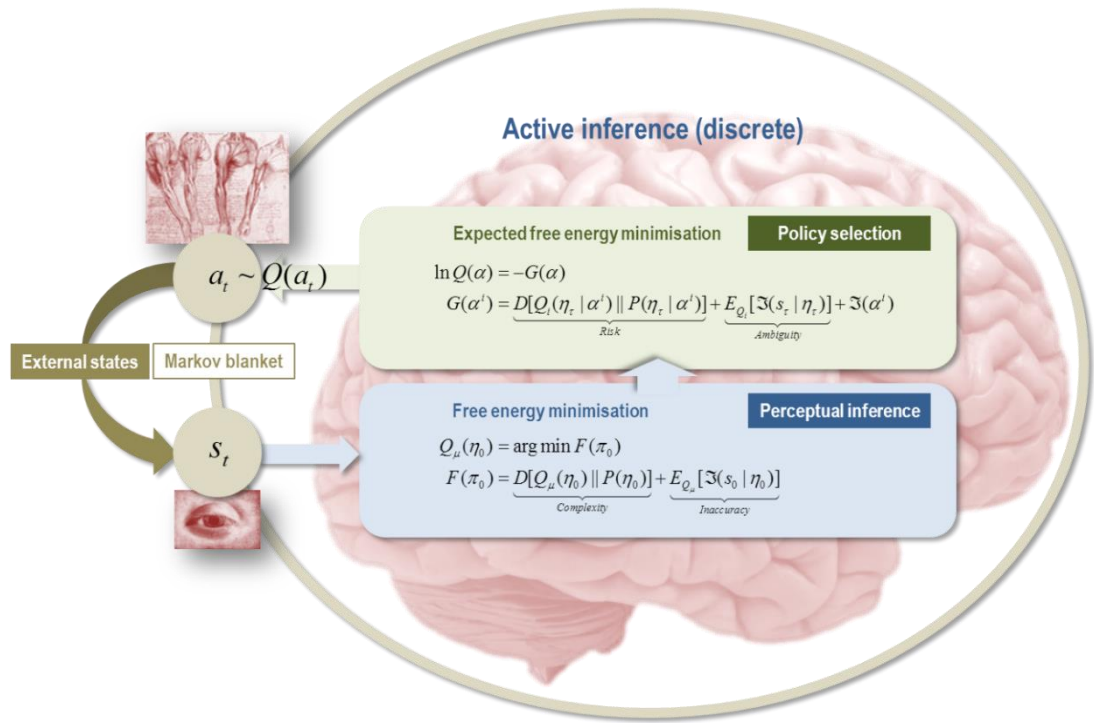

**FIGURE 20**

*Bayesian mechanics and active inference*. This graphic summarises the belief updating implicit in the minimisation of variational and expected free energy. It provides a generic (active) inference scheme that has been used in a wide variety of applications and simulations; ranging from games in behavioural economics (FitzGerald et al., 2015) and reinforcement learning (Schwartenbeck et al., 2015) through to language (Friston et al., 2017d) and scene construction (Mirza et al., 2016). In this setup, discrete actions solicit a sensory outcome that informs approximate posterior beliefs about hidden or external states of the world – via minimisation of variational free energy under a set of plausible policies (i.e., *perceptual inference*). The approximate posterior beliefs are then used to evaluate expected free energy and subsequent beliefs about action (i.e., *policy selection*). Note a subtle but important move in this construction: the expected free energy furnishes prior *beliefs about policies*. This is interesting from several perspectives. For example, it means that agents infer policies and, implicitly, active states. In other words, beliefs about policies – encoded by internal states – are distinct from the active states of the agent's Markov blanket. In more sophisticated schemes, agents infer hidden states under plausible policies with a generative model based on a Markov decision process. This means the agent predicts how it will behave and then verifies those predictions based on sensory samples. In other words, agents garner evidence for their own behaviour and actively self-evidence. In this setting, variational free energy reflects the surprisal or evidence that a particular policy is being pursued. In sum, this means the agent (will appear to) have elemental beliefs about its enactive self – beliefs that endow it with a sense of purpose, in virtue of the prior preferences that constitute risk. A key insight from simulations is that the form of the generative model can be quite different from the process by which external states generate sensory states. In effect, this enables agents (i.e., particles) to author their own sensorium in a fashion that has close connections with econiche construction (Bruineberg and Rietveld, 2014). Please see (Friston et al., 2017c) for technical details and (Friston et al., 2017a) for a discussion of how the implicit belief updating might be implemented in the brain.

# Deep inference: gradient flows or least action?

The expected free energy corollary says that the probability of a particular policy corresponds, in part, to its epistemic value in resolving uncertainty about hidden states of the world. Heuristically, this renders epistemic actions more probable, in virtue of maximising the information gain afforded by sensory samples in the future; i.e., the reduction of uncertainty anticipated under a particular course of action. This is sometimes called *epistemic affordance* (Calvo and Friston, 2017; Proust, 2015).

This follows, almost tautologically, from the definition of the sorts of particles we are trying to explain. In other words, self-evidencing particles will self-organise to minimise the conditional uncertainty about their blanket states, given hidden or external states; i.e., ambiguity (2.3). This underlies many influential formulations of perception in neuroscience; including the principle of minimum redundancy (Barlow, 1961; Barlow, 1974), the Infomax principle (Linsker, 1990) and formulations of epistemic foraging in terms of information theory and artificial curiosity (Bialek et al., 2001; Itti and Baldi, 2009; Schmidhuber, 2010; Still and Precup, 2012; Sun et al., 2011; Tishby and Polani, 2010). In the present formulation, internal states come to parameterise beliefs about external states – that enable an active sampling of unambiguous sensory states. Crucially, this enables the internal states to infer the external states, speaking to a circular causality that attends most treatments of self-organisation (Haken, 1983).

From the particular free energy lemma (8.3), we can summarise self-organisation as a principle of least action cast in terms of variational free energy:

$$\boldsymbol{\alpha}[\tau] = \arg\min_{\alpha[\tau]} \mathcal{A}(\alpha[\tau] \mid s[\tau])$$
$$\Rightarrow \delta_\alpha \mathcal{A}(\boldsymbol{\alpha}[\tau] \mid s[\tau]) = 0 \qquad\qquad (10.6)$$
$$\Rightarrow \dot{\boldsymbol{\alpha}}_0 = (Q_{\alpha\alpha} - \Gamma_{\alpha\alpha})\nabla_\alpha F(\boldsymbol{\alpha}_0, s_0)$$

This variational principle has been used extensively in simulations of emergent behaviour in the setting of morphogenesis and autopoiesis (self-assembly) in relatively simple systems (Friston et al., 2015a). A deeper self-organisation emerges when we equip the generative model $\Im(\eta, s, \alpha)$ – that defines variational free energy – with a temporal depth that encompasses trajectories or paths $\Im(\eta[\boldsymbol{\tau}], s[\boldsymbol{\tau}], \alpha[\boldsymbol{\tau}])$. This *deep generative model* mandates a prior over active states in the future, which by (8.7), is available in terms of an upper bound on the surprisal of the endpoint of autonomous paths. If we retain the path-independent term of action in (1.10), then from (8.15) the most likely course of autonomous behaviour minimises expected free energy:

$$G(\alpha_\tau) \geq \Im(\alpha_\tau) - \Im(\alpha_0) \propto \mathcal{A}(\alpha[\tau] \mid \pi_0)$$
$$\boldsymbol{\alpha}[\tau] = \arg\min_{\alpha[\tau]} \mathcal{A}(\alpha[\tau] \mid \pi_0) \qquad\qquad (10.7)$$
$$\approx \arg\min_{\alpha[\tau]} G(\alpha_\tau)$$

This formulation has been used extensively in simulations of visual foraging and decision-making in cognitive neuroscience; e.g., (Mirza et al., 2016). In these applications, the objective of trying to characterise interesting, self-organising attracting sets is turned on its head and (10.7) is used to select policies (i.e., trajectories of actions) under specified generative models. This deep active inference appears to be sufficient to account for many aspects of behaviour; ranging from the way that we sample our visual world with saccadic eye movements (Mirza et al., 2016), through to how we accumulate evidence for generative models and subsequently make decisions (Friston et al., 2015b). In short, the variational principles that underlie active inference are already implicit in many parts of the biological and social sciences.

The above equalities suggest that there are two ways in which we could solve for autonomous behaviour, either by performing a gradient descent on variational free energy (10.6) or by selecting the path that minimises expected free energy (10.7). Both lead to plausible behaviours with sentient and biological aspects. Figure 21 shows an example of autonomous behaviour based upon the first (gradient flow) solution that reproduces a simple form of handwriting (and its observation). This example used the chain rule to emulate active responses to the predictions of a generative model – in a way that looks very much like classical reflex arcs in neurobiology (Adams et al., 2013):

$$\dot{\mathbf{a}} \approx -\Gamma_{aa} \nabla_{\tilde{s}} F(\vec{\pi}) \cdot \frac{\partial \tilde{s}}{\partial \tilde{\eta}} \cdot \frac{\partial \tilde{\eta}}{\partial a} \qquad (10.8)$$

Note that this use of the chain rule requires a formulation of dynamics in generalised coordinates of motion $\vec{\pi} \triangleq (\pi, \pi', \pi'', \ldots)$ : see Appendix E and (Friston et al., 2010).

Equation (10.8) means that a change in active states produces a change in the higher order motion of sensory states that, in this example, is mediated by the generalised motion of external states. This example illustrates two points. First, all the interesting behaviour is generated by the beliefs encoded in internal states (here, a simple brain). In other words, the pullback attractor and associated nonequilibrium steady-state density are prescribed completely in terms of a particle's (i.e., agent's) beliefs. Heuristically, this means that the particle is the author or agent of its external dynamics and consequent sensorium. The second, more technical, point is that by using generalised coordinates of motion we have effectively specified dynamics in terms of local paths or trajectories. This follows from the Taylor expansion:

$$\pi(\tau) = \pi_0 + \tau \pi_0' + \tfrac{1}{2!} \tau^2 \pi_0'' + \ldots \qquad (10.9)$$

The reason these paths are local is that random fluctuations mean that the trajectories not specified to any meaningful precision in the past or future. This begs the question: can we use expected free energy to specify active inference over longer paths?

Figure 22 shows an example of using expected free energy to specify successive active states. In this example, saccadic eye movements were simulated to reproduce epistemic foraging of a visual scene (here, a face). Epistemic foraging means selecting behaviours that elicit the right sort of sensations, which resolve uncertainty about external states. This example is used to make two further key points. The first speaks to the epistemic, uncertainty or ambiguity resolving behaviour that emerges when using expected free energy to simulate active inference. This behaviour arises because the generative model had, in this instance, no prior preferences over external or hidden states; leaving just the information gain (a.k.a. *salience*) in (8.14). Put simply, active sampling of the visual scene was driven purely by informational imperatives to resolve uncertainty about the causes of sensations. The second point rests on the following question: could we have reproduced this behaviour using the gradient flow formulation?

The answer is no. This is because assigning a high probability to autonomous activity that minimises expected free energy is, effectively, minimising the self-entropy of future states, leading to apparently purposeful activity. One can imagine that this produces remarkable kinetics that appears to contravene the laws that apply to inert particles; namely, a particle that moves as if it were insensitive to thermodynamic costs or gravitational forces. This autonomous, purposeful, information seeking, preference fulfilling activity may be a characteristic of biological self-organisation – that inherits from the same Langevin dynamics that underwrite classical laws.

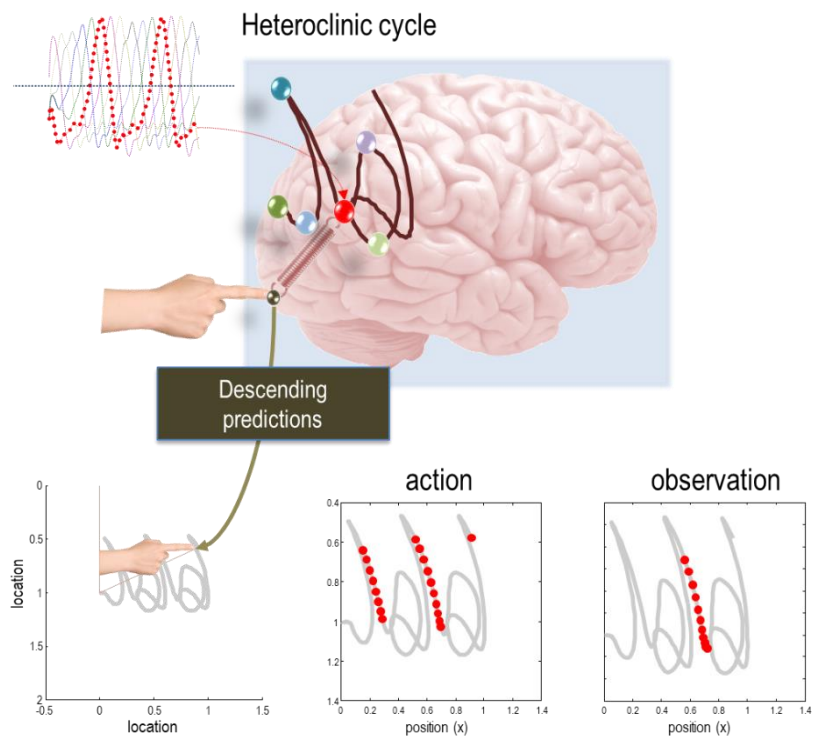

FIGURE 21

*Autonomous movement*. This figure shows the results of simulating active inference (here, writing), in terms of conditional expectations about hidden states of the world, consequent predictions about sensory input and the ensuing behaviour. The autonomous dynamics that underwrite this behaviour rest upon prior expectations about states that follow Lotka-Volterra dynamics: these are the six (arbitrarily) coloured lines in the upper left panel. In this generative model, each state is associated with a location in Euclidean space that attracts the agent's finger. In effect, the internal states then supply predictions of what sensory states should register if the agent's beliefs were true. Active states try to suppress the ensuing prediction error (i.e., maximising accuracy) by reflexively fulfilling expected changes in angular velocity, through exerting forces on the agent's joints (not shown);. The subsequent movement of the arm is traced out in the lower left panel. This trajectory has been plotted in a moving frame of reference so that it looks like synthetic handwriting (e.g., a succession of 'j' and 'a' letters). The lower left panels show the activity of one (the fourth) hidden state under 'action', and 'action-observation'. During action, sensory states register both the visual and proprioceptive consequences of movement, while under action observation, only visual sensations are available – as if the agent was watching another agent. The red dots correspond to the time bins during which this state exceeded an amplitude threshold of two arbitrary units. They key thing to note here is that this unit responds preferentially when, and only when, the motor trajectory produces a down-stroke, but not an up-stroke. Please see (Friston et al., 2011) for further details. Furthermore, with a slight delay, this internal state responds during action and action observation. From a biological perspective, this is interesting because it speaks to an empirical phenomena known as mirror neuron activity (Gallese and Goldman, 1998; Kilner et al., 2007; Rizzolatti and Craighero, 2004).

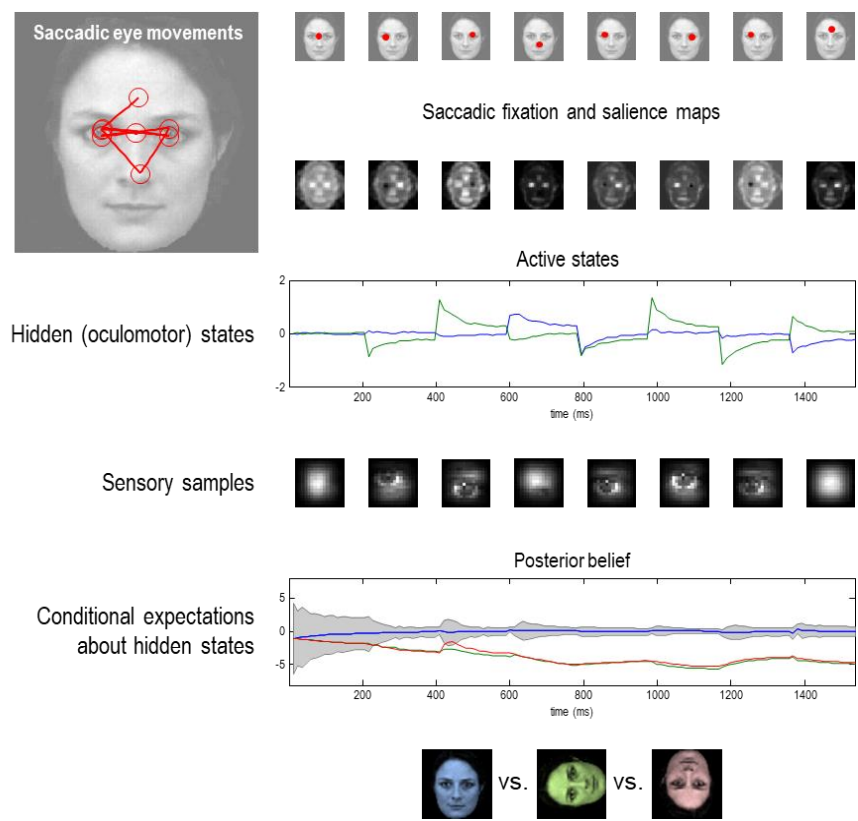

**FIGURE 22**

*Epistemic foraging*. This figure shows the results of a simulation in which a face was presented to an agent, whose responses were simulated by selecting active states that minimised expected free energy following an eye movement. The agent had three internal images or hypotheses about the stimuli she might sample (an upright face, an inverted face and a rotated face). The agent was presented with an upright face and her posterior expectations were evaluated over 16 (12 ms.) time bins, until the next saccade was emitted. This was repeated for eight saccades. The ensuing eye movements are shown as red dots at the end of each saccade in the upper row. The corresponding sequence of eye movements is shown in the inset on the upper left, where the red circles correspond roughly to the proportion of the visual image sampled. These saccades are driven by prior beliefs about the direction of gaze based upon the salience maps in the second row. These salience maps are the expected free energy as a function of policies; namely, where to look next. Note that these maps change with successive saccades as posterior beliefs about the hidden states, including the stimulus, become progressively more confident. Note also that salience is depleted in locations that were foveated in the previous saccade, because these locations no longer have epistemic affordance (i.e., the ability to reduce uncertainty or expected free energy). Empirically, this is known as inhibition of return. Oculomotor responses are shown in the third row in terms of the two hidden oculomotor states corresponding to vertical and horizontal eye movements. The associated portions of the image sampled (at the end of each saccade) are shown in the fourth row. The final two rows show the accompanying posterior beliefs in terms of their sufficient statistics and stimulus categories, respectively. The posterior beliefs are plotted here in terms of posterior expectations and 90% confidence intervals. The key thing to note is that the expectation about the true stimulus supervenes over alternative expectations and, as a result, conditional confidence about the stimulus category increases (and the confidence intervals shrink to the expectation). This illustrates the nature of evidence accumulation when selecting a hypothesis or percept that best explains sensory states. Within-saccade accumulation is evident even during the initial fixation with further stepwise decreases in uncertainty as salient information is sampled at successive saccades. Please see (Friston et al., 2012) for further details.

## Summary

In summary, this section has reviewed a physics of sentient dynamics that can be read at two levels. From a deflationary perspective, it is just a description of (weakly mixing) random dynamical systems that have certain kinds of attracting sets; namely, those that are coincidentally space occupying but with a low (self) entropy measure. In other words, if such random dynamical systems exist (i.e., possess a Markov blanket), they will look as if they conform to a variational principle of least action (10.6); where action is a functional of probabilistic beliefs encoded by internal states. Alternatively, one can regard the free energy lemmas as a description of self-organisation or, more specifically, self-evidencing.

In brief, this more teleological account goes as follows: self-organisation emerges under active inference, in virtue of a conditional synchrony between external and internal states (induced by the Markov blanket). This synchrony affords a probabilistic representation of the 'world outside' (i.e., external states) by internal states. In turn, this enables the selection of policies (trajectories of autonomous states) that ensure a conditional synchrony through actively minimising uncertainty about hidden states. This description highlights the circular causality that underwrites a form of autopoiesis (Clark, 2017; Maturana and Varela, 1980; Seth, 2014) that, at the end of the day, minimises the variational free energy functional that upper bounds (negative log) evidence; hence, self-evidencing (Hohwy, 2016).

The preponderance of autonomous states reflects the probability that they will lead to nonequilibrium steady-state. In this sense, the active states will look 'as if' that they have been selected to minimise expected free energy (i.e., self-entropy). This is simply a description of systems that self-organise to a random dynamical attractor. However, if a system's generative model acquires a temporal depth – and the prior belief that its actions minimise expected free energy – the descriptive 'as if' aspect of Bayesian mechanics gives way to a deeper form of self-organisation;

in which active states are effectively caused by the 'beliefs' held by a particle. Formally, this kind of deeply structured, self-evidencing requires an explicit generative model of future trajectories, where, crucially, the (path integral of) expected free energy furnishes a prior over policies.

One might ask whether self-evidencing dynamics are themselves an emergent property (not dissimilar to self-organised criticality). This is probably a question too far for the current treatment; however, beyond the (strong) anthropic principle (Barrow et al., 2015) – that we exist as an inferential systems in a world of measurable Markov blankets – one can easily elude the emergence question. A more interesting possibility lies in the repertoire of random dynamical systems that show self-evidencing to a greater or lesser degree. Their persistence (and therefore probability of being found in the universe) may be intimately related to the depth of planning implicit in (10.7). In other words, particles with generative models that are sufficiently deep – to select from long sequences of autonomous behaviour – may be those sorts of systems we associate with biotic systems of increasing sophistication, sometimes referred to as counterfactual depth (Seth, 2015). This speaks to interesting issues about the nature of the generative model and representations of the future (and past), which are taken up elsewhere (Friston et al., 2015b). This also brings us back to the characterisation of self-organisation in terms of itinerancy, information lengths and associated thermodynamics.

## The thermodynamics of inference

It is common to ask whether there is a quantitative relationship between thermodynamic and variational free energy. The formulations of Bayesian mechanics above provide a clear answer to this question. Because variational free energy is a function of a particle's states (8.4), it corresponds to a thermodynamic *potential*. In contrast, thermodynamic free energy is an expectation over an ensemble of particles: see (6.6). This means that thermodynamic free energy is an attribute of an ensemble, while variational free energy is an attribute of a single particle. More specifically, as ensembles converge to their random dynamical attractors, thermodynamic free energy decreases. When the ensemble has attained nonequilibrium steady-state, the variational free energy of each particle constitutes (an upper bound on) surprisal or the NESS potential and, by implication, thermodynamic potential energy.

## Potentials and surprisal

At nonequilibrium steady-state, there is a straightforward relationship between variational free energy and a particle's thermodynamic potential: from (6.5) and (8.6):

$$\left.\begin{array}{l} U(\pi) = k_B T \cdot \mathfrak{I}(\pi) + F_m \\ \mathfrak{I}(\pi) = F(\boldsymbol{\alpha}, s) \le F(\boldsymbol{\mu}, b) \end{array}\right\} \Rightarrow \left\{ \begin{array}{l} U(\pi) = k_B T \cdot F(\boldsymbol{\alpha}, s) + F_m \\ \qquad \le k_B T \cdot F(\boldsymbol{\mu}, b) + F_m \end{array} \right. \tag{10.10}$$

These expressions show that the thermodynamic potential of a particular state is proportional to the particular free energy, where $F_m$ is the normalisation constant based on the partition function. The interesting aspect of this relationship is the constant of proportionality that depends on the particle's temperature or the amplitude of random fluctuations: $\Gamma_{\pi\pi} \triangleq \mu_m k_B T$. This relationship can be used to evaluate the heat generated by a trajectory of particular states where, from (6.2) and (6.7) we have (ignoring time-dependent thermodynamic potentials and forces):

$$\begin{aligned} f_m(\pi, \tau) &\triangleq -\nabla U(\pi, \tau) \\ \mathbf{q}(\pi, \tau) &= \int_0^t \dot{\mathbf{q}} d\tau = \int_0^t f_m \cdot \dot{\pi} d\tau = -\int_0^t \dot{U} d\tau \\ &= U(\pi_0) - U(\pi_t) = -\Delta U \\ &= -k_B T \cdot \Delta F(\boldsymbol{\alpha}, s) \end{aligned} \tag{10.11}$$

This is the limiting case of the Jarzynski relation (6.19), in the absence of a time-dependent potential. Just for fun, one can now work out the heat dissipated by doing something of note; namely, a 'quantum of inference' or change in variational free energy of three natural units (corresponding to an odds ratio of $20:1 \approx \exp(3):1$[30]. This would generate $1.24 \ 10^{-20}$ J at a body temperature of $300^\circ$ K. For comparison, the Landauer limit (for the equivalent 4.32 bits of information) at room temperature is $1.19 \ 10^{-20}$ J (Bennett, 2003; Landauer, 1961): see also (Matta and Massa, 2017). We have, somewhat arbitrarily, associated a quantum of belief updating – and implicit movement on a statistical manifold – with a (path-independent) *divergence length* of three nats. A related treatment of thermodynamics and belief updating, in terms of (path-dependent) *information length*, can be found in (Crooks, 2007). These relationships suggest that Bayesian and stochastic mechanics are equivalent formulations of the same thing. One can either regard Bayesian inference is a necessary consequence of thermodynamics (i.e., gradient flows on a thermodynamic potential). Alternatively, Bayesian mechanics is a corollary of thermodynamics.

"*I suggest that we may never understand this strange thing, the quantum, until we understand how information may underlie reality. Information may not be just what we 'learn' about the world. It may be what 'makes' the world.*" (Wheeler, 1989)

---

[30] Typically regarded as strong evidence for one probabilistic belief over another – as articulated in terms of Bayes factors. The corresponding classical statistical inference requires $p < 0.05$.

## Ensemble free energies

If thermodynamic free energy is an attribute of an ensemble, can we establish a relationship between thermodynamic free energy and the variational free energies of an ensemble of (exchangeable) particles; such as a collection of biological or chemical cells? From (6.6) it is evident that the average variational free energy over the ensemble becomes, as one might anticipate, thermodynamic entropy:

$$F_3(\tau) = E[U(\pi, \tau)] - T \cdot S$$
$$S = k_B E[F(\mathbf{a}, s)]$$

(10.12)

The expectation here is over the ensemble of particles or agents at nonequilibrium steady-state. This has the interesting implication that if every member of the ensemble is trying to minimise their variational free energy, then the thermodynamic entropy of the ensemble will also be minimised. See (Friston et al., 2015a) for an example of how this ensemble perspective can be used to simulate pattern formation and morphogenesis in ensembles of (simulated) cells.

## Summary

The thermodynamic take on the complexity cost of Bayesian inference is potentially important from a practical point of view. It suggests that statistical and thermodynamic efficiency go hand-in-hand. This raises interesting questions: for example, if a machine learning or inference scheme dissipates large amounts of heat, is it statistically inefficient? Could thermodynamic potential be used as a proxy for surprisal and Bayesian model evidence? And so on.

Table 5 summarises the various potentials we have considered in previous sections in terms of surprisal – and implicitly the variational free energy that underwrites inference. The basic message here is that to characterise any mechanics, it is sufficient to specify the surprisal and amplitude of random fluctuations. The ensuing dynamics can then be cast in terms of a (solenoidal) gradient flow on a potential, which is a function of surprisal or *vice versa*. Depending upon the definition of the potential, the amplitude of random fluctuations acquires various interpretations; such as inverse mass in quantum mechanics or a scaled temperature in stochastic mechanics. In conservative mechanics, where the random fluctuations can be discounted, quantities like mass and charge become constants of proportionality in the relationship between surprisal and potential.

TABLE 5

Potentials and surprisal (conditional dependencies on external states are omitted for clarity)

| Mechanics (1.8) $f(x) = (Q - \Gamma) \cdot \nabla \Im(x)$ | NESS Potential or surprisal $\Im(x)$ | Amplitude of fluctuations $\Gamma$ |
|---|---|---|
| **Quantum mechanics (5.4)** $i\hbar\dot{\Psi} = \mathbf{H}\Psi$ | Schrödinger potential $V(x) = \frac{\hbar^2}{4m}(\frac{1}{2}\nabla\Im \cdot \nabla\Im - \nabla^2\Im)$ | $\Gamma = \frac{\hbar}{2m}$ |
| **Stochastic mechanics (6.2)** $f(\pi) = (Q_m - \mu_m)\nabla U(\pi)$ | Thermodynamic potential $U(s,\alpha)$ $\Im(s,\alpha) = \frac{1}{k_B T}U(s,\alpha) + \ln Z$ | $\Gamma = \mu_m k_B T$ |
| **Newtonian mechanics (7.11)** $\begin{Bmatrix} \dot{\mathbf{a}} \\ \dot{\mathbf{s}} \end{Bmatrix} = \begin{Bmatrix} +\nabla_s \Im(\mathbf{b}) \\ -\nabla_\mathbf{a} \Im(\mathbf{b}) \end{Bmatrix}$ | Newtonian potential $\Im(\mathbf{a})$ $\Im(\mathbf{s},\mathbf{a}) = \Im(\mathbf{a}) + \frac{\hbar}{2m}\mathbf{s}\cdot\mathbf{s}$ | $\Gamma = 0$ |
| **Classical mechanics (7.12)** $\begin{Bmatrix} \dot{\mathbf{a}} \\ \dot{\mathbf{s}} \end{Bmatrix} = \begin{Bmatrix} +\nabla_s \Im(\mathbf{b}) \\ -\nabla_\mathbf{a} \Im(\mathbf{b}) \end{Bmatrix}$ | Electrical potential $\varphi(\mathbf{a})$ $\Im(\mathbf{s},\mathbf{a}) = z\varphi(\mathbf{a}) + \frac{\hbar}{2m}(\mathbf{s} - z\mathbf{A}(\mathbf{a}))\cdot(\mathbf{s} - z\mathbf{A}(\mathbf{a}))$ | $\Gamma = 0$ |
| **Relativistic mechanics (7.14)** $\begin{Bmatrix} \dot{\mathbf{a}} \\ \dot{\mathbf{s}} \end{Bmatrix} = \begin{Bmatrix} +\nabla_s \Im(\mathbf{b}) \\ -\nabla_\mathbf{a} \Im(\mathbf{b}) \end{Bmatrix}$ | Gravitational potential $V(\mathbf{a})$ $\Im(\mathbf{s},\mathbf{a}) = V(\mathbf{a}) + \sqrt{m^2 c^4 + \mathbf{s}^2 c^2}$ | $\Gamma = 0$ |
| **Bayesian mechanics (8.5)** $\dot{\boldsymbol{\alpha}} = (Q_{\alpha\alpha} - \Gamma_{\alpha\alpha})\nabla_\alpha F(\boldsymbol{\alpha},s)$ | Variational free energy $\Im(s,\boldsymbol{\alpha}) = F(\boldsymbol{\alpha},s)$ | $\Gamma_{\alpha\alpha}$ |

# Discussion

The foregoing technical treatment leads us to a remarkable conclusion; namely, that any self-organising system that preserves its form, in virtue of possessing an attracting (random dynamical) set can be construed as inferring the causes of its sensations. In this sense, sentient behaviour and the implicit representationalism may be a universal property of all self-organising systems. We have seen, using numerical analyses, how the elementary behaviour entailed by this take on dynamics looks remarkably similar to the characterisation of itinerant dynamics in neurophysiology. There are some interesting implications of this Bayesian gloss on self-organising (Ashby, 1947) and self-evidencing (Hohwy, 2016) dynamics that speak to the observational foundations of physics (Cook, 1994). See also (Bridgman, 1954). In other words, without a sentient capacity – or the ability to measure and infer, one could argue there would be no physics. In one (epistemological) sense, Bayesian mechanics trumps quantum, classical and statistical mechanics, because they all depend upon observations (Cook, 1994; Seifert, 2012; Theise and Kafatos, 2013). This metrological perspective starts to make a lot of sense, when one looks at the fundamental postulates of things like quantum physics, right through to the observational or measurement-bound nature of gauge theories (Capozziello and De Laurentis, 2011). They all rely upon measurement and inference. The special nature of Bayesian mechanics is unremarkable because it rests on an explicit account of dynamics under Markov blankets, as opposed to the implicit treatment of boundary conditions in terms of heat baths and [Schrödinger] potentials.

# Conclusion

In conclusion, we have seen how much of physics follows from straightforward constraints on measurable systems. Specifically, some familiar results from quantum, statistical and classical mechanics have been sketched out as a natural consequence of non-equilibrium steady state dynamics. Furthermore, the nature of 'states' has been considered in terms of a hierarchical decomposition that reduces detailed, microscopic descriptions of ensemble dynamics to progressively macroscopic descriptions – through recursive application of a Markovian partition and adiabatic reduction. In brief, this suggests that random fluctuations – at any level of description within an ensemble of coupled systems – are inherited from the intrinsic or internal fluctuations that are sequestered behind Markov blankets.

On one view, this speaks to a somewhat deflationary account of physics; in the sense that things like Hamilton's principle of least action, fluctuation theorems and Schrödinger equation are essentially restatements of the conditions that are necessarily true of things that attain non-equilibrium steady state; i.e., have a pullback attractor. Tracing the pedigree of this physics back even further, nonequilibrium steady-state is itself inherited from the assumption that systems possess an invariant measure. This invariance underwrites the existence of a random dynamical attractor. In short, all the phenomena that we have considered follow from assuming that things only exist if they have measurable characteristics: i.e., return to the (neighbourhood of) states within their attracting set. This deflationary account does, however, offer closure and internal consistency. If things exist in virtue of their measurability, then existence presupposes something that enacts or makes the measurement. Given that measurement is, at the end of the day, an inference conditioned on the blankets states of an instrument (an Observer), the process of measurement or inference must itself be a property of things that exist. Happily, this is precisely what follows from the free energy principle. In short, existence entails measurement and measurement (inference) is a necessary consequence of existence. This speaks to the fundamentally metrological aspect of physics and, perhaps, the existential nature of inference (Cook, 1994).

The way in which this monograph has portrayed different sorts of mechanics bears upon key notions in the philosophy of science, namely, *reduction* and *emergence*[31]. Reductionism is the view that the causal properties and dynamics of higher-order phenomena can be reduced to the operation and organisation of their constituent parts (Bechtel and Richardson, 2010; Bedau and Humphreys, 2008; Nagel, 1961). This view entails the position that higher-order phenomena are mere *epiphenomena* – in the sense that the causal powers of (allegedly) emergent phenomena reduce to those of the more basal system from which they are said to emerge (Kim, 1999); and in the sense that system-level dynamics do not have 'downward' causal effects on the components of the system (Campbell, 1974; Craver and Bechtel, 2007). On this view, those Markov blanketed systems that are 'carved out' from the dynamics of their constituent particles do not exist independently – the whole 'just is', or is 'nothing over and above', the organisation of its component parts. Proponents of emergence, contrariwise, argue that novel systemic dynamics do indeed (appear to) emerge from new configurations of underlying systems; e.g., (Thompson, 2010). Emergence can be construed as an ontological claim; i.e., as the claim that new kinds of things emerge, which are not mere re-descriptions of the more basal dynamics; or as epistemological, i.e., as the claim that we could not have *predicted* (*a priori*) the higher-order dynamics from our theories about lower-order dynamics (Ayala and Dobzhansky, 1974; Batterman, 2001; Bedau, 2002). The (strong) ontological flavour of this position endorses the view that quantum level phenomena support – i.e., enable the emergence of – molecular mechanics (e.g., statistical mechanics and thermodynamics), from which biological and cosmological (e.g., classical and possibly Bayesian) mechanics emerge in turn. The epistemological flavour speaks instead to limitations of our theories, and to our surprise at observing unpredicted patterns.

We have seen that quantum, statistical, classical and Bayesian mechanics can be applied at the same level of analysis; namely, our synthetic virus. This undercuts the notion of ontological emergence – in that all these mechanics are *complementary* characterisations of the way that things behave. This implies that the higher-level descriptions capture causally inert dynamics, and simply cast the same system under a different light. This would commit one to reductionism (in the sense that nothing novel need emerge) or to an epistemological view on emergence. On another reading, the dependency of these mechanics on the amplitude of random fluctuations – and inherently on the scale and implicit degree of averaging in play – introduces a hierarchical aspect, which speaks in favour of ontological readings of the emergence of classical (and Bayesian) mechanics from quantum and stochastic mechanics. Here, one would place the Bayesian mechanics of self-organisation at the same level as Newtonian or Lagrangian mechanics; in the sense that they are apt for describing macroscopic scales at which pullback attractors are shaped by exponentially divergent (and convergent) flows, as opposed to random fluctuations.

---

[31] This material benefited heavily from rewriting by my philosophically literate colleague, Maxwell Ramstead.

Committing to one of these options may be important, because they have different implications for quantum level descriptions of sentient systems – and their implicit unconscious inference (Helmholtz, 1866/1962). It remains an outstanding question to which position (if any) one must commit under the free-energy principle. Metaphysical agnosticism may even be the best option available. It may be that the hierarchical structure of Markov blankets of blankets – which provides a formal grip on the notion of emergent systems – obviates the need to commit to one or another perspective. Indeed, in providing an account of how hierarchically structured organisms adapt, and adapt to, their embedding niche, the multiscale Markovian dynamics described in this monograph seem to account for the causal powers of emergent systemic dynamics; both in terms of their effects on their constituent parts – thus obviating the need for a metaphysics of downward causation – and in terms of their system-level effects on their external milieu.

# Appendix A: Stratonovich path integrals

The following provides an intuition about the origins of the divergence term under a Stratonovich interpretation of stochastic differential equations. For a full treatment see (Cugliandolo and Lecomte, 2017). Imagine a discretised path, where the probability of each interval is based on the difference between the first derivative and expected flow at its midpoint:

$$\dot{x}_{\tau+\frac{\Delta\tau}{2}} = f(x_{\tau+\frac{\Delta\tau}{2}}) + \Delta(\omega_\tau)$$
$$p(\Delta(\omega_\tau)) = \left|\partial_\omega\Delta\right|^{-1} p(\omega_\tau)$$
$$p(\omega_\tau) = \mathcal{N}(0, \tfrac{2\Gamma}{\Delta\tau})$$

$$(11.1)$$

This difference is a function of random fluctuations in the flow at the beginning of each interval, whose variance decreases with interval length $\Delta\tau$. The second equality follows from the change in variables; i.e., the contribution to the first derivative at the midpoint from the random fluctuation at the beginning of the interval. Using the solution of Langevin equation, we can express the determinant of the Jacobian in (11.1) in terms of the divergence of the flow as follows:

$$\Delta(\omega_\tau) = \dot{x}_{\tau+\frac{\Delta\tau}{2}} - f(x_{\tau+\frac{\Delta\tau}{2}})$$
$$\dot{x}_{\tau+\frac{\Delta\tau}{2}} = e^{\frac{\Delta\tau}{2}\partial_x f}\dot{x}_\tau = e^{\frac{\Delta\tau}{2}\partial_x f}(f(x_\tau) + \omega_\tau) \Rightarrow$$
$$\left|\partial_\omega\Delta\right| = e^{\frac{\Delta\tau}{2}\nabla\cdot f}$$

$$(11.2)$$

This enables us to express the action (i.e., surprisal) of a continuous path as the limit of short intervals

$$
\begin{aligned}
\Im(x[\tau]) &= \lim_{\Delta\tau \to 0} \sum_\tau \Im(\Delta(\omega_\tau)) \\
&= \lim_{\Delta\tau \to 0} \sum_\tau \tfrac{1}{2}(\dot{x}-f) \cdot \tfrac{\Delta\tau}{2\Gamma}(\dot{x}-f) + \tfrac{\Delta\tau}{2}\nabla \cdot f \\
&= \tfrac{1}{2}\int [(\dot{x}-f) \cdot \tfrac{1}{2\Gamma}(\dot{x}-f) + \nabla \cdot f]dt
\end{aligned}
\tag{11.3}
$$

This is the action or path integral of the Lagrangian. Heuristically, the divergence determines how quickly random fluctuations in the flow are exponentially 'remembered' or 'forgotten' over time. In other words, when divergence is low, perturbations to the flow decay quickly and a larger range of random fluctuations produce similar differences in subsequent flow, rendering them more likely or less surprising.

# Appendix B: lemmas and proofs

**Lemma** (NESS density): *The nonequilibrium steady-state density of a random dynamical system with flow* $f = (Q-\Gamma)\nabla\Im$ *and associated Fokker Planck operator* $\mathbf{L} \triangleq \nabla \cdot (\Gamma\nabla - f)$ *is given by*

$$
\begin{aligned}
p(x) &= \exp(-\Im(x)) \\
&\Rightarrow \nabla p = -p(x)\nabla\Im(x) \\
&\Rightarrow \dot{p}(x) = \mathbf{L}p(x) = 0 \\
&\Rightarrow f(x) = (\Gamma-Q)\nabla \ln p(x)
\end{aligned}
\tag{12.1}
$$

*Where* $Q = -Q^T$ *implies solenoidal flow.*

**Proof**: By substituting $\nabla p = -p\nabla\Im$ and $f = Q\nabla\Im - \Gamma\nabla\Im$ into the Fokker Planck operator we have:

$$
\begin{aligned}
\mathbf{L}p &= \nabla \cdot \Gamma\nabla p - \nabla \cdot (fp) \\
&= -\nabla \cdot (p\Gamma\nabla\Im) - \nabla \cdot (pQ\nabla\Im) + \nabla \cdot (p\Gamma\nabla\Im) \\
&= -p\nabla \cdot (Q\nabla\Im) - (Q\nabla\Im) \cdot \nabla p \\
&= -p(\nabla \cdot (Q\nabla\Im) - \nabla\Im \cdot (Q\nabla\Im))
\end{aligned}
\tag{12.2}
$$

One can see that (12.1) is satisfied when the solenoidal component of flow $Q\nabla\Im$ is divergence free and orthogonal to $\nabla\Im$. It is easy to see both these conditions are met when,

$$
Q = -Q^T \Rightarrow \begin{cases} \nabla \cdot (Q\nabla\Im) = tr(Q\nabla_{xx}\Im) = 0 \\ \nabla\Im \cdot (Q\nabla\Im) = tr(Q\nabla\Im\nabla\Im^T) = 0 \end{cases}
\tag{12.3}
$$

This means that $p = \exp(-\Im)$ is the equilibrium density or eigensolution $\mathbf{L}p = 0$ of the Fokker-Planck operator describing density dynamics. □

**Corollary** (*Markov blanket*): *if one (external) subset of states $\eta \subset x$ is conditionally independent of another (internal) subset $\mu \subset x$, when conditioned on their Markov blanket $b \subset x$ – and the subsets are not coupled via solenoidal flow – then the flow of internal states does not depend on external states and vice versa:*

$$
\left.
\begin{array}{l}
p(x \mid m) = p(\eta \mid b)\, p(\mu \mid b)\, p(b) \\
Q_{\eta\mu} = 0 \\
Q_{\eta b} = 0 \\
Q_{\mu b} = 0
\end{array}
\right\}
\Rightarrow
\begin{cases}
\nabla_{\mu} f_{\eta}(x) = 0 \\
\nabla_{\eta} f_{\mu}(x) = 0
\end{cases}
\tag{12.4}
$$

**Proof**: expressing the flow in terms of the above partition, we have:

$$
\begin{bmatrix} f_{\eta}(x) \\ f_{b}(x) \\ f_{\mu}(x) \end{bmatrix} =
\begin{bmatrix}
\Gamma_{\eta\eta} - Q_{\eta\eta} & -Q_{\eta b} & -Q_{\eta\mu} \\
Q_{\eta b}^{T} & \Gamma_{bb} - Q_{bb} & -Q_{b\mu} \\
Q_{\eta\mu}^{T} & Q_{b\mu}^{T} & \Gamma_{\mu\mu} - Q_{\mu\mu}
\end{bmatrix}
\begin{bmatrix} \nabla_{\eta} \ln p \\ \nabla_{b} \ln p \\ \nabla_{\mu} \ln p \end{bmatrix} =
\begin{bmatrix}
(\Gamma_{\eta\eta} - Q_{\eta\eta})\nabla_{\eta} \ln p \\
(\Gamma_{bb} - Q_{bb})\nabla_{b} \ln p \\
(\Gamma_{\mu\mu} - Q_{\mu\mu})\nabla_{\mu} \ln p
\end{bmatrix}
\tag{12.5}
$$

We assume here (and throughout) that random fluctuations are independently and identically distributed (i.e. their covariance has a leading diagonal form). By the conditional independence above

$$
\nabla_{\mu\eta} \ln p(x) = \nabla_{\mu\eta} (\ln p(\eta \mid b) + \ln p(\mu \mid b) + \ln p(b)) = 0
\tag{12.6}
$$

And

$$
\begin{aligned}
\nabla_{\mu} f_{\eta}(x) &= (\Gamma_{\eta\eta} - Q_{\eta\eta})\nabla_{\mu\eta} \ln p(x) = 0 \\
\nabla_{\eta} f_{\mu}(x) &= (\Gamma_{\mu\mu} - Q_{\mu\mu})\nabla_{\eta\mu} \ln p(x) = 0
\end{aligned}
\tag{12.7}
$$

This means the flow of internal states does not depend on external states and *vice versa* □

**Remarks**: the Markov blanket corollary expresses the uncoupling of external and internal states as a consequence of the conditional independencies implicit in a (generalised) Markovian blanket partition. An alternative formulation starts with flow constraints that lead to a Markov Blanket (and suppression of solenoidal coupling). For example, we can express uncoupled flow in terms of Jacobians as follows: from (12.5)

$$
\left.
\begin{array}{l}
\nabla_{\mu} f_{\eta}(x) \\
\nabla_{\eta} f_{\mu}(x)
\end{array}
\right\} =
\left\{
\begin{array}{l}
(Q_{\eta\eta} - \Gamma_{\eta\eta})\nabla_{\mu\eta}\Im + Q_{\eta b}\nabla_{\mu b}\Im + Q_{\eta\mu}\nabla_{\mu\mu}\Im \\
(Q_{\mu\mu} - \Gamma_{\mu\mu})\nabla_{\eta\mu}\Im - Q_{b\mu}^{T}\nabla_{\eta b}\Im - Q_{\eta\mu}^{T}\nabla_{\eta\eta}\Im
\end{array}
\right\} = 0
\tag{12.8}
$$

Because $\Gamma_{\eta\eta}, \Gamma_{\mu\mu}$ are positive definite, the associated derivatives of surprisal must be zero and we recover the Markov blanket factorisation:

$$\left.\begin{array}{c} \nabla_{\mu\eta}\Im \\ \nabla_{\eta\mu}\Im \end{array}\right\}=0 \Rightarrow \left\{\begin{array}{c} \Im(\eta\,|\,b,\mu) \\ \Im(\mu\,|\,b,\eta) \end{array}\right\}=\left\{\begin{array}{c} \Im(\eta\,|\,b) \\ \Im(\mu\,|\,b) \end{array}\right\} \Rightarrow p(x)=p(\eta\,|\,b)\,p(\mu\,|\,b)\,p(b) \tag{12.9}$$

Furthermore, because $\nabla_{\mu\mu}\Im, \nabla_{\eta\eta}\Im$ are positive semidefinite, we can eliminate solenoidal coupling between external and internal states: $Q_{\eta\mu}=0$. The solenoidal coupling with blanket states must then satisfy:

$$\left.\begin{array}{c} Q_{\eta b}\nabla_{\mu b}\Im \\ Q_{b\mu}^{T}\nabla_{\eta b}\Im \end{array}\right\}=0 \Leftarrow \left[\begin{array}{c} Q_{\eta b} \\ Q_{\mu b} \end{array}\right]=0 \tag{12.10}$$

Notice that an absence of solenoidal coupling to blanket states is sufficient but not necessary to satisfy the flow constraints. For example, if we repeat the above analysis but dividing blanket states into active and sensory states, with the following (complete) flow constraints, we have:

$$f(x)=\left[\begin{array}{c} f_{\eta}(\eta,b) \\ f_{s}(\eta,b) \\ f_{\mu}(\mu,b) \\ f_{a}(\mu,b) \end{array}\right] \Leftrightarrow \left[\begin{array}{c} \nabla_{\mu}f_{\eta} \\ \nabla_{\mu}f_{s} \\ \nabla_{\eta}f_{\mu} \\ \nabla_{\eta}f_{a} \end{array}\right]=\left\{\begin{array}{c} (Q_{\eta\eta}-\Gamma_{\eta\eta})\nabla_{\mu\eta}\Im+Q_{\eta a}\nabla_{\mu a}\Im+Q_{\eta s}\nabla_{\mu s}\Im+Q_{\eta\mu}\nabla_{\mu\mu}\Im \\ (Q_{ss}-\Gamma_{ss})\nabla_{\mu s}\Im+Q_{sa}\nabla_{\mu a}\Im-Q_{\eta s}^{T}\nabla_{\mu\eta}\Im+Q_{s\mu}\nabla_{\mu\mu}\Im \\ (Q_{\mu\mu}-\Gamma_{\mu\mu})\nabla_{\eta\mu}\Im-Q_{a\mu}^{T}\nabla_{\eta a}\Im-Q_{s\mu}^{T}\nabla_{\eta s}\Im-Q_{\eta\mu}^{T}\nabla_{\eta\eta}\Im \\ (Q_{aa}-\Gamma_{aa})\nabla_{\eta a}\Im+Q_{a\mu}\nabla_{\eta\mu}\Im-Q_{sa}^{T}\nabla_{\eta s}\Im-Q_{\eta a}^{T}\nabla_{\eta\eta}\Im \end{array}\right\}=0$$

$$\Rightarrow \left\{\begin{array}{c} \nabla_{\mu\eta}\Im \\ \nabla_{\mu s}\Im \\ \nabla_{\eta\mu}\Im \\ \nabla_{\eta a}\Im \end{array}\right\}=0 \Rightarrow \left[\begin{array}{cc} Q_{\eta a} & Q_{\eta\mu} \\ Q_{sa} & Q_{s\mu} \end{array}\right]=\left[\begin{array}{c} Q_{\eta a} \\ Q_{s a} \end{array}\right]=0 \tag{12.11}$$

The particular uncoupling of flow implied by these constraints, precludes solenoidal coupling between internal and external states – and between autonomous and non-autonomous states. This induces a Markov blanket over generalised states as depicted in Figure 23.

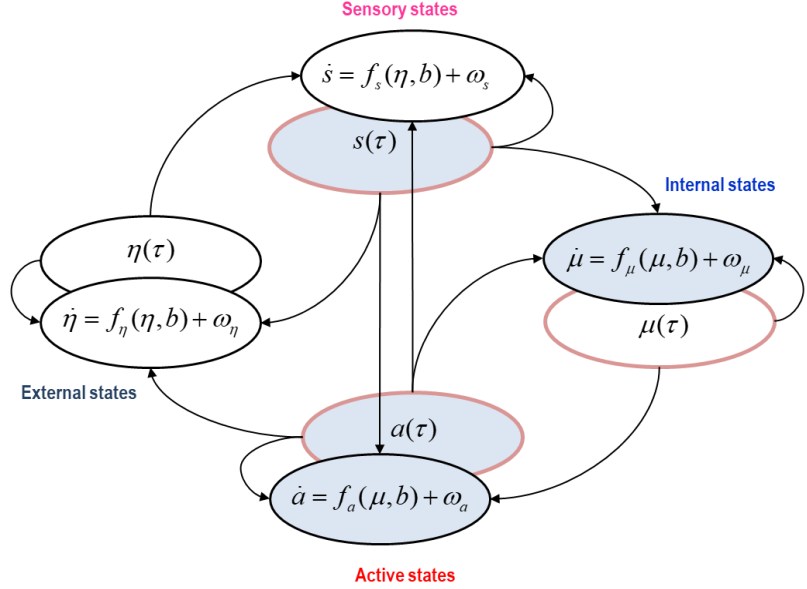

**FIGURE 23**

*Generalised Markov blankets*. This dependency graph or Bayesian network is a nuanced version of Figure 1 that includes generalised states; namely, the states and their motion. The Markov blanket of internal states comprises those generalised states highlighted in blue, while the generalised states that constitute the Markov blanket of the corresponding motion are circled in pink. The key point to take from this figure is that the Markov blanket of generalised internal states comprise the generalised blanket states (with the exception of sensory motion). This generalised blanket requires an absence of solenoidal coupling between internal and external states – and between autonomous and non-autonomous states.

**Lemma** (marginal flow): *for any weakly mixing random dynamical system, the marginal flow $f_\eta(\mu)$ of any subset of states $\eta \in X$, averaged under the complement of another $\mu \in X$ can be expressed in terms of the gradients of the corresponding marginal surprisal $\Im(\mu) = -\ln p(\mu)$:*

$$f_\eta(\mu) \triangleq E_{p(\tilde{\mu}|\mu)}[f_\eta(\mu, \tilde{\mu})] = (\Gamma_{\eta\eta} - Q_{\eta\eta})\nabla_\eta \ln p(\mu) - Q_{\eta\tilde{\eta}}\nabla_{\tilde{\eta}} \ln p(\mu) \quad (12.12)$$

**Proof**: Consider a partition of the states into a subset and its complement: $x = (\eta, \tilde{\eta})$. The corresponding partition of (NESS) flow can be expressed as:

$$\begin{bmatrix} f_\eta(x) \\ f_{\tilde{\eta}}(x) \end{bmatrix} = \begin{bmatrix} \Gamma_{\eta\eta} - Q_{\eta\eta} & -Q_{\eta\tilde{\eta}} \\ Q_{\eta\tilde{\eta}}^T & \Gamma_{\tilde{\eta}\tilde{\eta}} - Q_{\tilde{\eta}\tilde{\eta}} \end{bmatrix} \begin{bmatrix} \nabla_\eta \ln p(x) \\ \nabla_{\tilde{\eta}} \ln p(x) \end{bmatrix} \quad (12.13)$$

From this, we can define a marginal flow as the flow expected under $\tilde{\mu}$, where $x = (\mu, \tilde{\mu})$:

$$f_\eta(\mu) \triangleq E_{p(\tilde{\mu}|\mu)}[f_\eta(\mu, \tilde{\mu})] = \int f_\eta(x) \, p(\tilde{\mu} \mid \mu) d\tilde{\mu}$$

$$= \int \Big( (\Gamma_{\eta\eta} - Q_{\eta\eta}) \nabla_\eta \ln p(x) - Q_{\eta\tilde{\eta}} \nabla_{\tilde{\eta}} \ln p(x) \Big) p(\tilde{\mu} \mid \mu) d\tilde{\mu}$$

$$= \int \Big( (\Gamma_{\eta\eta} - Q_{\eta\eta}) \nabla_\eta p(\tilde{\mu} \mid \mu) p(\mu) - Q_{\eta\tilde{\eta}} \nabla_{\tilde{\eta}} p(\tilde{\mu} \mid \mu) p(\mu) \Big) \frac{p(\tilde{\mu} \mid \mu)}{p(\tilde{\mu}, \mu)} d\tilde{\mu} \qquad (12.14)$$

$$= (\Gamma_{\eta\eta} - Q_{\eta\eta}) \left( \frac{\nabla_\eta p(\mu)}{p(\mu)} + \int \nabla_\eta p(\tilde{\mu} \mid \mu) d\tilde{\mu} \right) - Q_{\eta\tilde{\eta}} \left( \frac{\nabla_{\tilde{\eta}} p(\mu)}{p(\mu)} + \int \nabla_{\tilde{\eta}} p(\tilde{\mu} \mid \mu) d\tilde{\mu} \right)$$

$$= (\Gamma_{\eta\eta} - Q_{\eta\eta}) \nabla_\eta \ln p(\mu) - Q_{\eta\tilde{\eta}} \nabla_{\tilde{\eta}} \ln p(\mu)$$

The integrals of probability gradients disappear because the average change of a probability density is zero □

**Corollary** (conditional independence): *if the flow of one subset of states does not depend on another, then it becomes the marginal flow (expected) under the second subset. For example, in terms of the Markov blanket partition, it follows that:*

$$\begin{bmatrix} f_\eta(x) \\ f_s(x) \\ f_\mu(x) \\ f_a(x) \end{bmatrix} = \begin{bmatrix} f_\eta(\eta, b) \\ f_s(\eta, b) \\ f_\mu(\mu, b) \\ f_a(\mu, b) \end{bmatrix} = \begin{bmatrix} (Q_{\eta\eta} - \Gamma_{\eta\eta})\nabla_\eta \Im \\ (Q_{ss} - \Gamma_{ss})\nabla_s \Im \\ (Q_{\mu\mu} - \Gamma_{\mu\mu})\nabla_\mu \Im \\ (Q_{aa} - \Gamma_{aa})\nabla_a \Im \end{bmatrix} + \begin{bmatrix} +Q_{\eta s}\nabla_s \Im \\ -Q_{\eta s}^T\nabla_\eta \Im \\ +Q_{\mu a}\nabla_a \Im \\ -Q_{\mu a}^T\nabla_\mu \Im \end{bmatrix} \qquad (12.15)$$

Note that many solenoidal terms are eliminated by the Markov blanket corollary (12.11). In short, the conditional independencies induced by the Markov blanket mean that the flow of external states is the same for every value of the internal states, which is just its average over the internal states (similarly for other partitions) □

**Corollary** (expected flow): *the marginal flow of any subset $\eta \subset x$ averaged over all other states depends only on the gradients of its marginal density, provided there is no solenoidal coupling with its complement:*

$$f_\eta(\eta) = (\Gamma_{\eta\eta} - Q_{\eta\eta}) \nabla_\eta \ln p(\eta) = (Q_{\eta\eta} - \Gamma_{\eta\eta}) \nabla_\eta \Im(\eta) \qquad (12.16)$$

This is a special case of the marginal flow lemma, when $\eta = \mu$ and $Q_{\eta\tilde{\eta}} = 0$. It implies that the expected flow of any state or subset of states, averaged over all other states, will behave in exactly the same way as all states considered together. In other words, it will descend the gradients of its (marginal) surprisal □

# Appendix C: nonequilibrium steady-state energy functions

At nonequilibrium steady-state, the expected curvature of surprisal is the expected sum of squared gradients:

$$E_{p(x)}[\nabla^2 \mathfrak{I}] = \int p \nabla^2 \mathfrak{I} \, dx = \int p \left( \frac{\nabla p \cdot \nabla p}{p^2} - \frac{\nabla^2 p}{p} \right) dx = \int p \nabla \mathfrak{I} \cdot \nabla \mathfrak{I} \, dx = E_{p(x)}[\nabla \mathfrak{I} \cdot \nabla \mathfrak{I}] \qquad (13.1)$$

Note that the integral of the curvature of the NESS density disappears because it is a proper density. This equality furnishes some intuitive expressions for the expected Schrödinger potential, Lagrangian and Hamiltonian: consider the dissipative stochastic entropy production along a single path (i.e., the change in self-information associated with the dissipation of heat), defined using the notation of (6.9). The corresponding expectation is dissipative entropy production, which – by (13.1) – is the average curvature of surprisal:

$$\dot{\mathfrak{I}}^q(x, \omega) \triangleq \frac{1}{T} f \cdot \dot{\pi} = \frac{1}{k_B T} \dot{\mathbf{q}}$$

$$\dot{\mathfrak{I}}^q(x) \triangleq E_{p(\omega)}[\dot{\mathfrak{I}}^q(x, \omega)] = \Gamma \nabla \mathfrak{I}(x) \cdot \nabla \mathfrak{I}(x)$$

$$\dot{H}^q \triangleq E_{p(x)}[\dot{\mathfrak{I}}^q(x)] = E_{p(x)}[\Gamma \nabla \mathfrak{I} \cdot \nabla \mathfrak{I}] = E_{p(x)}[\Gamma \nabla^2 \mathfrak{I}] \qquad (13.2)$$

In other words, dissipative entropy production increases with the amplitude of random fluctuations and the average curvature (or sum of squared gradients) of the surprisal or NESS potential. In turn, this leads to the following expectations, for a system with $n$ degrees of freedom (i.e., $x \in \mathbb{R}^n$):

$$V(x) = \frac{\hbar}{2} [\frac{1}{2} \Gamma \nabla \mathfrak{I} \cdot \nabla \mathfrak{I} - \Gamma \nabla^2 \mathfrak{I}]$$

$$\mathcal{L}(x, \dot{x}) = \frac{1}{4\Gamma} (\omega \cdot \omega) - \frac{1}{2} \Gamma \nabla^2 \mathfrak{I}$$

$$\mathcal{H}(x, \dot{x}) = \frac{1}{4\Gamma} (f + \omega) \cdot (f + \omega) - \frac{1}{\hbar} V(x)$$

$$\mathbf{V} \triangleq E_{p(x)}[V(x)] = -\frac{\hbar}{4} \dot{H}^q$$

$$\mathcal{L} \triangleq E_{p(x, \omega)}[\mathcal{L}(x, \dot{x})] = \frac{1}{2} [n - \dot{H}^q]$$

$$\mathcal{H} \triangleq E_{p(x, \omega)}[\mathcal{H}(x, \dot{x})] = \frac{1}{2} [n + \dot{H}^q] \qquad (13.3)$$

These expressions disclose the role of the expected curvature as a key (global) characterisation of nonequilibrium steady-state dynamics.

# Appendix D: the Fokker Planck operator

The Fokker Planck operator operates on an infinite dimensional (Hilbert) space; namely, the support of the density dynamics. Clearly, this is not useful for numerical solutions in practical applications. However, one can easily convert density dynamics into a manageable form using a discrete basis; for example, an orthonormal basis $\left| v_i(x) \right\rangle$, where (using the Dirac notation):

$$\left| p(x) \right\rangle = \sum_i \left| v_i(x) \right\rangle \varphi_i$$

$$\left\langle v_i \mid v_k \right\rangle = \delta_{ik} \qquad (14.1)$$

This enables one to express density dynamics in terms of the associated coefficients $\varphi_i \in \mathbb{C}$

$$\dot{p}(x) = \mathbf{L} | p(x) \rangle \Rightarrow \dot{\varphi} = \lambda \varphi$$
$$\lambda_{ik} = \langle \nu_i | \mathbf{L} | \nu_k \rangle$$

(14.2)

The coefficients are the amplitude of probability modes associated with the basis. A natural choice of the basis would be eigenfunctions of the Fokker Planck operator, that allows one to decompose density dynamics into a series of uncoupled modes and, crucially, discard fast modes that dissipate almost instantaneously (i.e., the coefficients of eigenmodes with eigenvalues $\lambda_{ii} \ll 0$).

The practical challenge here – say in weather forecasting – is to find the eigenmodes e.g., (Harlim and Yang, 2017; Palmer and Laure, 2013). An alternative would be to use a polynomial basis, rendering the coefficients the moments of the probability density. This leads to the low order approximations to density dynamics (c.f., the method of moments). For example, retaining the first two moments corresponds to the Laplace approximation. An example of this can be found in our own work on the dynamic causal modelling of population dynamics (Marreiros et al., 2009). Finally, one could use a Fourier basis set, which brings us to the quantum mechanical formulation with an important twist: by decomposing the wave function (as opposed to the probability density) the Plancherel theorem ensures that the (Fourier) coefficients can be interpreted as probability amplitudes.

# Appendix E: generalised motion

"*[T]he results obtained by applying the techniques of Markov process theory are valuable only to the extent to which they characterise just these 'large-scale' fluctuations*" (Stratonovich, 1967); p123.

Generally, we have been assuming that random fluctuations are sufficiently fast that their serial or temporal correlations can be ignored. This means that the precision of their temporal derivatives is zero (or does not exist). The implication is that there is no statistical coupling between the generalised motion of states and the generalised states *per se*. However, the situation changes if we allow for random fluctuations that have a degree of temporal smoothness (i.e., that are analytic). In this instance, dependencies between generalised motion and states give rise to a generalised version of Langevin dynamics in (1.1): by applying a succession of temporal derivatives, we have (to first-order):

$$\left.\begin{array}{l} \dot{x} = f(x,v) + \omega \\ \dot{x}' = \nabla f \cdot x' + \omega' \\ \dot{x}'' = \nabla f \cdot x'' + \omega'' \\ \quad \vdots \end{array}\right\} \quad \Rightarrow \quad \dot{\vec{x}} = \mathbf{D}\vec{x} = \nabla \mathbf{f} \cdot \vec{x} + \vec{\omega}$$

(14.3)

$$\vec{x} \triangleq (x, x', x'', \ldots)$$
$$\nabla \mathbf{f} = I \otimes \nabla f$$

Here, $\mathbf{D}$ is a block matrix operator that returns generalised motion from generalised states (see below for an example). The mixture of Lagrange and Newton notations for temporal derivatives reminds us that we are working in generalised coordinates of motion. The associated Lagrangian and surprisal have straightforward forms (to first order):

$$\mathcal{L}(\vec{x}) = \tfrac{1}{2}(\tfrac{1}{2}\vec{x} \cdot \mathbf{M}\vec{x} + \nabla \cdot \mathbf{f}(\vec{x}))$$
$$\mathbf{M} = (\mathbf{D} - \nabla \mathbf{f}) \cdot \mathbf{\Gamma}^{-1}(\mathbf{D} - \nabla \mathbf{f})$$

(14.4)

$$\Im(\vec{x}) = \Im(x) + \Im(\mathbf{D}\vec{x} \mid x)$$
$$\quad\quad = \Im(x) + \mathcal{L}(\vec{x})$$

Note that the (generalised) Lagrangian can be absorbed into (generalised) surprisal in this setting. Because generalised motion effectively describes a trajectory, this means that surprisal becomes a *local* action; i.e., the surprisal of a local path over the correlation length of random fluctuations. The Lagrangian in (14.4) has been expressed in terms of an *effective mass matrix* $\mathbf{M}$ that incorporates flow gradients and the amplitude of generalised fluctuations. The amplitude of these fluctuations is a function of their autocorrelation function, evaluated at zero lag:

$$\mathbf{\Gamma} = \begin{bmatrix} 1 & 0 & \partial_\tau^2 \rho(0) & \cdots \\ 0 & -\partial_\tau^2 \rho(0) & 0 & \\ \partial_\tau^2 \rho(0) & 0 & \partial_\tau^4 \rho(0) & \\ \vdots & & & \ddots \end{bmatrix} \otimes \Gamma$$

(14.5)

Equation (14.5) follows from standard results in the theory of stochastic processes (Cox and Miller, 1965) and allows one to quantify temporal autocorrelations in terms of the amplitude of random fluctuations in generalised coordinates of motion. Heuristically, if we assume random fluctuations are extremely fast, then the curvature of the autocorrelation function (*resp.* correlation length) tends to infinity (*resp.* zero). In the limit of fast fluctuations, we can therefore ignore higher orders of generalised motion, such that (14.3) reduces to the usual Langevin form, under Wiener assumptions (1.1). However, should we want to accommodate autocorrelation functions that cannot be approximated by a Delta function – as in (1.1) – generalised coordinates of motion provide a useful augmentation of state-space that allows one to solve for dynamics in a computational and analytically straightforward fashion (Friston et al., 2010). The use of generalised coordinates of motion becomes especially useful in the context of Bayesian filtering, as illustrated by the following lemma:

**Lemma** (generalized gradient flows): *when the variation of free energy $F(\bar{\mu}, b)$ with respect to the variational density over the generalised motion of external states $q_{\bar{\mu}}(\bar{\eta}) = \mathcal{N}(\sigma(\bar{\mu}), \Sigma(\bar{\mu}))$ is minimised, the motion of generalized internal states becomes their generalised motion:*

$$\delta_q F = 0 \Leftrightarrow \dot{\bar{\mu}} = \mathbf{D}\bar{\mu} \tag{14.6}$$

**Proof**: in generalised coordinates of motion, (8.21) can be expressed in terms of the Helmholtz decomposition:

$$
\begin{aligned}
\dot{\bar{\mu}} &= (Q_{\sigma\sigma} - \Gamma_{\sigma\sigma}) \cdot \nabla_{\bar{\mu}} F(\bar{\mu}, b) \\
&= \mathbf{D}\bar{\mu} - \Gamma_{\sigma\sigma} \cdot \nabla_{\bar{\mu}} F(\bar{\mu}, b)
\end{aligned} \tag{14.7}
$$

$$\mathbf{D}\bar{\mu} = Q_{\sigma\sigma} \cdot \nabla_{\bar{\mu}} F(\bar{\mu}, b) \Rightarrow \nabla_{\bar{\mu}} \cdot \mathbf{D}\bar{\mu} = 0$$
$$\bar{\mu} = (\boldsymbol{\mu}(b), \boldsymbol{\mu}'(b), \boldsymbol{\mu}''(b), \ldots)$$

In generalised coordinates of motion, solenoidal flow can be formulated in terms of a divergence-free operator $\mathbf{D}$ that plays the role of a generalised time derivative operator. In this setting, the most likely path of internal states minimises variational free energy and renders the motion of generalised internal states (divergence-free) generalised motion,

$$\dot{\bar{\mu}} = \mathbf{D}\bar{\mu} \Leftrightarrow \nabla_{\bar{\mu}} F(\bar{\mu}, b) = 0 \Leftrightarrow \delta_q F = 0 \tag{14.8}$$

giving (14.6). □

**Remarks**: In short, generalized motion $\mathbf{D}\bar{\mu}$ corresponds to conservative, divergence-free flow in generalised coordinates of motion. Intuitively, this casts motion as a gradient flow a frame of reference that moves with the generalized motion:

$$\mathbf{D}\bar{\mu} - \dot{\bar{\mu}} = \Gamma_{\sigma\sigma} \cdot \nabla_{\bar{\mu}} F(\bar{\mu}, b) \tag{14.9}$$

This is probably the most general and compact form of (generalised or variational) Bayesian filters. With an appropriate choice of (Gauge) transformation, one can recover standard filtering schemes such as the extended Kalman-Bucy filter (Friston et al., 2014), where:

$$\mathbf{D} = \begin{bmatrix} 0 & I \\ 0 & 0 \end{bmatrix}, \qquad \bar{\mu} = (\boldsymbol{\mu}(b), \boldsymbol{\mu}'(b)) \tag{14.10}$$

These classical schemes generally limit generalised motion to first order.

# Appendix F: discrete state-space models

In continuous state-space formulations, the expected free energy is a functional of the predictive density over final states, associated with autonomous states in the future. In discrete state-space formulations, the corresponding expected free energy becomes a vector with an element for each sequence of autonomous states (i.e., policy). Its derivation follows the same basic arguments; however, the expected free energy bound is derived directly from the non-negativity of an entropy of a probability distribution (as opposed to a density), instead of using Jensen's inequality.

**Lemma** (expected free energy – discrete): *the prior surprisal* $\Im(\alpha^i \mid \pi_0)$ *of the i-th policy* $\alpha^i = (\alpha_1^i, \ldots, \alpha_\tau^i)$, *given an initial state* $\pi_0$, *is upper bounded by an expected free energy comprising risk and ambiguity:*

$$G(\alpha^i) = \underbrace{D[Q_i(\eta_\tau \mid \alpha^i) \parallel P(\eta_\tau \mid \alpha^i)]}_{Risk} + \underbrace{E_{Q_i}[\Im(s_\tau \mid \eta_\tau)]}_{Ambiguity} + \Im(\alpha^i) \geq \Im(\alpha^i \mid \pi_0) \tag{15.1}$$

*The predictive density, given the current (particular) state* $\pi_0$ *is defined in terms of the variational density:*

$$Q_i(s_\tau, \eta_\tau, \alpha^i) \triangleq Q_i(s_\tau \mid \eta_\tau) Q_i(\eta_\tau \mid \alpha^i) P(\alpha^i \mid \pi_0)$$
$$Q_i(\eta_\tau \mid \alpha^i) \triangleq P(\eta_\tau \mid \alpha^i, \pi_0) = \sum_{\eta_0} Q_\pi(\eta_0) P(\eta_\tau \mid \alpha^i, \pi_0, \eta_0) \tag{15.2}$$
$$Q_\pi(\eta_0) = P(\eta_0 \mid \pi_0)$$

*where* $Q_i(s_\tau \mid \eta_\tau) = P(s_\tau \mid \eta_\tau) = P(s_\tau \mid \eta_\tau, \alpha)$.

**Proof**: the divergence between the predictive distribution (over outcomes, states and policies) and the prior distribution corresponding to the generative model (i.e., nonequilibrium steady-state distribution) must be greater than zero.

$$D[Q \parallel P] = E_Q[\mathbf{G}(\alpha) - \Im(\alpha \mid \pi_0)] \geq 0$$
$$\mathbf{G}(\alpha^i) = G(\alpha^i) - H[Q_i(s_\tau \mid \eta_\tau)]$$
$$G(\alpha^i) = E_{Q_i}[\Im(\eta_\tau, s_\tau, \alpha^i)] - H[Q_i(\eta_\tau \mid \alpha^i)] \tag{15.3}$$

$$Q_i \triangleq Q_i(\eta_\tau, s_\tau, \alpha^i)$$
$$P_i \triangleq P_i(\eta_\tau, s_\tau, \alpha^i)$$

When the predictive distribution converges to nonequilibrium steady-state (i.e., the prior distribution), the expected free energy of a particular policy must upper bound its surprisal:

$$D[Q \parallel P] = 0 \Rightarrow \mathbf{G}(\alpha) = \Im(\alpha \mid \pi_0)$$
$$\Rightarrow G(\alpha^i) = \Im(\alpha^i \mid \pi_0) + H[Q_i(s_\tau \mid \eta_\tau)] \qquad (15.4)$$
$$\Rightarrow G(\alpha^i) \geq \Im(\alpha^i \mid \pi_0)$$

The final inequality follows because the conditional entropy of the distribution over discrete states cannot be less than zero. Therefore, at convergence, expected free energy provides an upper bound on the surprisal of a trajectory (15.1), with equality when there is no ambiguity □

**Remark**: Note that when risk is minimised – and the predictive distribution converges to the prior distribution $Q_i \approx P(\eta_\tau, s_\tau \mid \alpha^i)$ – expected free energy becomes a mixture of epistemic and instrumental value; i.e., information gain and expected log-evidence, respectively. From (15.1), in the absence of risk:

$$G(\alpha^i) = E_{Q_i}[\Im(s_\tau \mid \eta_\tau)] + \Im(\alpha^i)$$
$$= E_{Q_i}[\Im(s_\tau \mid \eta_\tau, \alpha^i) + \Im(\alpha^i)]$$
$$= E_{Q_i}[\ln P(\eta_\tau \mid \alpha^i) - \ln P(\eta_\tau \mid s_\tau, \alpha^i) - \ln P(s_\tau, \alpha^i)] \qquad (15.5)$$
$$= E_{Q_i}[\underbrace{-D[P(\eta_\tau \mid s_\tau, \alpha^i) \parallel P(\eta_\tau \mid \alpha^i)]}_{\text{epistemic value}} - \underbrace{\ln P(s_\tau, \alpha^i)]}_{\text{instrumental value}}$$

This can be regarded as an alternative formulation of ambiguity in terms of information gain or salience.

**Software note:** The simulations described in this monograph can be reproduced using the academic software available from http://www.fil.ion.ucl.ac.uk/spm/software/. Typing DEM at the Matlab prompt will invoke a graphical user interface. The simulations above can be reproduced by selecting the **A physics of life** button. This allows users to examine the code and subroutines – and customise them at their discretion.

**Acknowledgments:** at the time of writing, KF is a Wellcome Principal Research Fellow (Ref: 088130/Z/09/Z). There are dozens of friends and colleagues who deserve acknowledgement for the ideas described in this monograph. I will pick out Thomas Parr for an explicit mention: Thomas was my Ph.D. student during its writing and contributed substantively to the active inference formulation. Furthermore, he prepared an invaluable series of weekly seminars, deconstructing the monograph's formal arguments – and oversights – for our colleagues in London: see https://www.fil.ion.ucl.ac.uk/~tparr/Physics/Slides%20Stochastic%20dynamics.htm

**Glossary of terms and expressions**

(a.u.:  *arbitrary units; e.g., metres (m), radians (rad), etc.*)

| Expression | Description | Units |
|---|---|---|
| **Variables** | | |
| $\tau$ | Time | s (seconds) |
| $x[\tau] = \{x(t) : t \in (0, \tau)\}$ | Trajectory or path through state-space | a.u. (m) |
| $\omega(\tau)$ | Random fluctuations | a.u. (m) |
| $x^{(i)} = \{x_1^{(i)}, x_2^{(i)}, \ldots, x_N^{(i)}\}$ $x_n^{(i)} = \{x_{n_1}^{(i)}, x_{n_2}^{(i)}, \ldots, x_{n_M}^{(i)}\}$ | Vector states at the *i*-th level of description | a.u. (m) |
| $x = \{\eta, s, a, \mu\} \in X$ | Markovian partition into external, sensory, active and internal states | a.u. (m) |
| $\tilde{\eta} \in X \setminus E : x = \{\eta, \tilde{\eta}\} \in X$ | Complement of a subset of states | a.u. (m) |
| $\dot{x} = \dfrac{dx}{dt}$ | Time derivative (Newton notation) | m/s |
| $\tilde{x} = (x, x', x'', \ldots)$ | Generalised motion (Lagrange notation) | (m, m/s,…) |
| $\alpha = \{a, \mu\} \in A$ | Autonomous states | a.u. (m) |
| $b = \{s, a\} \in B$ | Blanket states | a.u. (m) |
| $\pi = \{b, \mu\} \in P$ | Particular states | a.u. (m) |
| $\eta \in E$ | External states | a.u. (m) |
| $\Gamma = \frac{\hbar}{2m} = \mu_m k_B T$ | Amplitude (i.e., half the variance) of random fluctuations | $m^2$/s or J·s/kg |
| $Q$ | Rate of solenoidal flow | $m^2$/s or J·s/kg |
| $m = \frac{\hbar}{2\mu_m k_B T} = \frac{\hbar}{2\Gamma}$ | (Reduced) mass | kg (kilogram) |
| $\mu_m = \frac{\hbar}{2mk_B T} = \frac{1}{k_B T}\Gamma$ | Mobility coefficient | s/kg |
| $T$ | Temperature | K (Kelvin) |
| $\ell = \int d\ell : d\ell^2 = g_{ij}\, d\lambda^j\, d\lambda^i$ | Information length | nats |
| $D(\tau) = D[p(x,\tau\mid\pi_0)\parallel p(x,\infty\mid\pi_0)]$ | Divergence length | nats |
| $\boldsymbol{\tau} : d\ell(\tau \geq \boldsymbol{\tau}) \approx 0$ | Critical time | s |
| $g_{ij} = E\left[\dfrac{\partial\Im}{\partial\lambda^i}\dfrac{\partial\Im}{\partial\lambda^j}\right]$ | Fisher (information metric) tensor | a.u. |

| **Functions, functionals and potentials** | | |
|---|---|---|
| $E[x] = E_p[x] = \int x p_\lambda(x)\,dx$ | Expectation or average | |
| $p_\lambda(x) : \Pr[X \in A] = \int_A p_\lambda(x)\,dx$ | Probability density function parameterised by sufficient statistics $\lambda$ | |
| $p_\tau(x) \equiv p(x,\tau)$ | Time-dependent probability density function parameterised by time | |
| $p(x) \triangleq p_\infty(x) = \lim_{\tau\to\infty} p(x,\tau)$ $\Leftrightarrow \mathbf{L}p(x) = 0$ | Nonequilibrium steady-state (NESS) density – the eigensolution of the Fokker Planck operator | |
| $q_\mu(\eta)$ | Variational density – an (approximate posterior) density over external states parameterised by internal states | |
| $f(x) = E[\dot{x}]$ | Flow – the expected motion through state-space | |
| $j(x) = f(x)p(x) - \Gamma\nabla p(x)$ | Probability current | |
| $\Psi(x,t) = \Psi(x)e^{-i\omega t}$ $p(x) = \Psi(x)\cdot\Psi(x)^\dagger$ | Wave function – the complex root of the NESS density | |
| $\mathcal{L}(x,\dot{x}) =$ $\frac{1}{2}[(\dot{x}-f)\cdot\frac{1}{2\Gamma}(\dot{x}-f)+\nabla\cdot f]$ | Lagrangian (Legendre transform of the Hamiltonian) | |
| $\mathcal{H}(x,\dot{x}) = \dot{x}\cdot\dfrac{\partial\mathcal{L}}{\partial\dot{x}} - \mathcal{L}(x,\dot{x})$ | Hamiltonian (Legendre transform of the Lagrangian) | |
| $\mathcal{A}(x[\tau]) = \Im(x[\tau]) = \int_0^t \mathcal{L}(x, \dot{x})d\tau$ | Action: the surprisal of a path; i.e., path integral of the Lagrangian | |
| $V(x) = \frac{m}{2}f \cdot f + \frac{\hbar}{2}\nabla \cdot f$ | Schrödinger potential | J (Joules) |
| $U(\pi) = k_B T \cdot \Im(\pi) + F_m$ | Thermodynamic potential | J or kg m²/s² |
| $F_\Im(\tau) \triangleq E[U(\pi, \tau)] - E[k_B T \cdot \Im(\pi)]$ | Thermodynamic free energy | J or kg m²/s² |
| $F(\pi) \geq \Im(\pi)$ | Variational free energy – an upper bound on the surprisal of particular states | nats |
| $G(\alpha_\tau) \geq \Im_\tau(\alpha_\tau \mid \pi_0)$ | Expected free energy – an upper bound on the surprisal of an autonomous state in the future | nats |
| $\Omega(x[\tau])$ | Path-dependent measurement | a.u. |
| $\boldsymbol{\sigma}(u_i) = \dfrac{\exp(-u_i)}{\sum_i \exp(-u_i)}$ | Normalised exponential (softmax) function of a vector | a.u. |

## Operators

| | | |
|---|---|---|
| $u \cdot v = u^T v = \langle u \mid v \rangle = u_i v^i$ | Inner product using dot, Dirac and summation notation | |
| $u \times v = u v^T = \lvert u \rangle \langle v \rvert$ | Outer product using cross and Dirac notation | |
| $\nabla_x \Im(x) = \dfrac{\partial \Im}{\partial x} = \left( \dfrac{\partial \Im}{\partial x_1}, \dfrac{\partial \Im}{\partial x_2}, \dots \right)$ | Differential or gradient operator (on a scalar field) | |
| $\nabla_{xx} \Im(x) = \dfrac{\partial^2 \Im}{\partial x^2}$ | Curvature operator (on a scalar field) | |
| $\nabla \cdot f(x) = \sum_i \dfrac{\partial f_i}{\partial x_i}$ | Divergence operator (on a vector field) | |
| $\nabla^2 \Im(x) = \nabla \cdot \nabla \Im = \Delta \Im$ | Laplace operator – the divergence of a gradient | |
| $\mathbf{L} = \nabla \cdot (\Gamma \nabla - f)$ <br> $\dot{p}(x) = \mathbf{L}p(x)$ | Fokker Planck operator | |
| $\mathbf{H} = V(x) - \dfrac{\hbar^2}{2m}\nabla^2$ <br> $i\hbar \dot{\Psi}(x) = \mathbf{H}\Psi(x)$ | Hamiltonian operator | |
| $\mathbf{T} = -\dfrac{\hbar^2}{2m}\nabla^2$ | Kinetic operator | |

| **Entropies and potentials** | | |
|---|---|---|
| $\Im(x) = -\ln p(x)$ | Surprisal or self-information | nats |
| $H(X) = H[p(x)] = E[\Im(x)]$ | Entropy or expected surprisal | nats |
| $H(X \mid Y) = E_{p(x,y)}[\Im(x \mid y)]$ | Conditional entropy | nats |
| $D[q(x) \Vert p(x)] = E_q[\ln q(x) - \ln p(x)]$ | Relative entropy or Kullback-Leibler divergence | nats |
| $I(X,Y) = H(X) - H(X \mid Y)$ $= H(X) + H(Y) - H(X,Y)$ $= D[p(x,y) \Vert p(x)p(y)] \geq 0$ | Mutual information | nats or natural units 1 nat $\approx$ 1.44 bits |
| $S \triangleq k_B H(X)$ | Thermodynamic entropy | J/K or m$^2$ kg s$^{-2}$/K |
| **Constants and coefficients** | | |
| $Z$ | Partition function or normalisation constant | a.u. |
| $\hbar = \frac{h}{2\pi}$ | (Reduced) Planck constant (or Dirac constant) 1.05457$\times$ 10$^{-34}$ | m$^2$ kg/s or J$\cdot$s |
| $k_B$ | Boltzmann's constant 1.39 10$^{-23}$ | m$^2$ kg s$^{-2}$/K or J/K |
| $c$ | Speed of light 299,792,458 | m/s |